\documentclass{aa}  
\usepackage{caption}
\usepackage{adjustbox}
\usepackage{longtable}
\usepackage{geometry}
\usepackage{subcaption}
\usepackage{balance}
\usepackage{graphicx}
\usepackage{adjustbox}
\usepackage{rotating}
\usepackage{ulem}
\usepackage[varg]{txfonts}
\usepackage{appendix}
\usepackage{pgffor}
\usepackage{xcolor}
\usepackage{placeins}
\usepackage{afterpage}
\usepackage{multirow}
\usepackage{alphalph}
\usepackage{hyperref}
\hypersetup{
    colorlinks=true,
    linkcolor=blue,
    citecolor=blue,
    urlcolor=blue,
    pdftitle={},
    pdfpagemode=FullScreen,
    }
\usepackage{siunitx}
\usepackage{booktabs}

\usepackage{xcolor}

\newcommand{\xmm}{{\it XMM-Newton}}
\newcommand{\chandra}{{\it Chandra}}
\newcommand{\SRG}{{\it SRG}}

\usepackage{pgffor}
\usepackage{float}

\usepackage[format=plain, labelsep=period, labelfont=bf]{caption}
\DeclareSIUnit{\erg}{erg}
\begin{document} 

   \title{First Study of the Supernova Remnant Population in the Large Magellanic Cloud with eROSITA}

   \subtitle{}

   \author{Federico Zangrandi
          \inst{1}
          \and
          Katharina Jurk\inst{1}      
          \and
          Manami Sasaki\inst{1}
          \and
          Jonathan Knies\inst{1}
          \and
          Miroslav D. Filipovi\'c\inst{2}
          \and
          Frank Haberl\inst{3}
          \and
          Patrick Kavanagh\inst{4}
          \and
          Chandreyee Maitra\inst{3}
          \and
          Pierre Maggi\inst{5}
          \and
          Sara Saeedi\inst{1}
          \and
          Dominic Bernreuther\inst{1}
          \and
          B\"arbel S. Koribalski\inst{6,2}
          \and
          Sean Points\inst{7}
          \and
          Lister Staveley-Smith\inst{8}
          }

   \institute{Dr. Karl Remeis Observatory, Erlangen Centre for Astroparticle Physics (ECAP), Friedrich-Alexander-Universität Erlangen-Nürnberg,
Sternwartstraße 7, 96049 Bamberg, Germany \label{inst1}\\
              \email{federico.zangrandi@fau.de}
        \and
        Western Sydney University, Locked Bag 1797, Penrith South DC, NSW 2751, Australia \label{inst2}
        \and
        Max-Planck-Institut f\"ur extraterrestrische Physik,  Gie{\ss}enbachstra{\ss}e 1, 85748 Garching, Germany \label{inst3}
        \and
        Department of Experimental Physics, Maynooth University, Maynooth, Co. Kildare, Ireland \label{inst4}
        \and
         Université de Strasbourg, CNRS, Observatoire astronomique de Strasbourg, UMR 7550, 67000 Strasbourg, France \label{inst6}
        \and
        Australia Telescope National Facility, CSIRO, Space and Astronomy, P.O. Box 76, Epping, NSW 1710, Australia 
        \and 
        Cerro Tololo Inter-American Observatory, NOIRLab, Cassilla 603, La Serena, Chile \label{inst7}
        \and
        International Centre for Radio Astronomy Research (ICRAR), University of Western Australia, 35 Stirling Highway, Perth, WA 6009, Australia \label{inst8}
        \\
             }

   \date{Received ; accepted }

 
  \abstract
   {}
   { 
The all-sky survey carried out by the extended Roentgen Survey with an Imaging Telescope Array (eROSITA) on board {\it Spektrum-Roentgen-Gamma (Spektr-RG, SRG)} has provided us with spatially and spectrally resolved X-ray data of the entire Large Magellanic Cloud (LMC) and its immediate surroundings in the soft X-ray band down to \SI{0.2}{keV} with an average angular resolution of 26\arcsec\ in the field of view. In this work, we have studied the supernova remnants (SNRs) and candidates in the LMC using data from the first four all-sky surveys (eRASS:4).  
From the X-ray data in combination with results at other wavelengths, we  obtain information about the SNRs, their progenitors, and the surrounding interstellar medium (ISM). The study of the entire population of SNRs in a galaxy helps us to understand the underlying stellar populations, the environments, in which the SNRs are evolving, and the stellar feedback on the  ISM. 
}
   {The eROSITA telescopes are the best instruments currently available for the study of extended soft sources like SNRs in an entire galaxy due to their large field of view and high sensitivity in the softer part of the X-ray band. We applied the Gaussian gradient magnitude (GGM) filter to the eROSITA images of the LMC to highlight the edges of the shocked gas in order to find new SNRs. We visually compared the X-ray images with those of their optical and radio counterparts to investigate the true nature of the extended emission. The X-ray emission is evaluated using the contours with respect to the background, while for the optical we used line ratio diagnostics, and non-thermal emission in the radio images. We used the Magellanic Cloud Emission Line Survey (MCELS) for the optical data. For the radio comparison, we used data from the Australian Square Kilometre Array Pathfinder (ASKAP) survey of the LMC. Using the star formation history (SFH) derived from the near-IR photometry of the VISTA survey of the Magellanic Clouds (VMC) we have investigated the possible progenitor type of the new SNRs and SNR candidates in our sample. 
   }
   {We present the most updated catalogue of SNRs in the LMC. Previously known SNRs and candidates were detected with 1$\sigma$ significance down to a surface brightness of $\Sigma$ [0.2--5.0\,keV] $= 3.0 \times 10^{-15}$ erg s$^{-1}$ cm$^{-2}$ arcmin$^{-2}$ and were examined. 
   The eROSITA data have allowed us to confirm one of the previous candidates as an SNR. We confirm three newly detected extended sources as new SNRs, while we propose 13 extended sources as new X-ray SNR candidates.
   We also present the analysis of the follow-up \xmm\ observation of MCSNR J0456--6533  discovered with eROSITA. Among the new candidates, we propose J0614--7251 (4eRASSU J061438.1$-$725112) as the first X-ray SNR candidate in the outskirts of the LMC.
}
   {}


   \maketitle

\section{Introduction}
Some stars end their life with a supernova (SN) explosion, which can be of two types. Massive stars with initial main-sequence mass above $8 \, \mathrm{M}_{\odot}$ explode as core-collapse (CC) supernovae, which enrich the interstellar medium (ISM) mainly with $\alpha$-elements (i.e., O, Ne, Mg, Si, S). Less massive stars finish their life as white dwarfs (WDs). In binary systems, WDs can accrete mass from their companion star and can result in a thermonuclear explosion (SN Ia), which mainly releases Fe-group elements into the ISM. 
Supernovae are responsible for the chemical enrichment of galaxies but also release a great amount of energy (${\sim} 10^{51} \, \mathrm{erg}$) at once into the ISM. 

By the shock waves of the supernova, objects called supernova remnants (SNRs) are created. The explosion ejects stellar material into the ISM with high velocities (${\sim} 10^4$ \SI{}{\km \ \s^{-1}}). The shock of the blast wave propagates into the ISM increasing its temperature and ionising it. 
The blast wave shock is decelerated during its expansion, which causes a second shock running into the ejecta called a reverse shock. The high-temperature plasma (${>}10^6$ \SI{}{\K}) in the shocked ISM and the shocked ejecta makes the SNRs emit in the X-ray regime.
In addition, the shock fronts are responsible for accelerating particles through the diffusive shock acceleration process making the SNRs one of the main sources of cosmic rays \citep{1934PNAS...20..259B, 1997A&A...324..641Z}.

Supernova remnants in X-rays are diffuse thermal sources due to the high-temperature plasma in its interior with an electron temperature of  $0.2$ to $\SI{5.0}{\keV}$. The youngest SNRs can also show non-thermal X-ray emission due to synchrotron processes. The X-ray synchrotron emission however diminishes rapidly since it is produced by the most energetic electrons, which radiate and lose their energy quickly \citep{vink2020physics}. In radio, the synchrotron radiation is visible for the entire lifetime of the remnant. In addition to its remnant, the CC explosion leaves a compact object, which can also radiate in X-rays.

Studying an SNR's X-ray spectrum allows us to infer the properties of the hot plasma such as its temperature, ionisation state, and chemical composition. These quantities are connected to the progenitor star of the remnant, its evolutionary stage, and the properties of the ISM in which the explosion occurred. Combining all this information, it is possible to further comprehend the role of SNRs in the dynamical and chemical evolution of galaxies.

Galactic absorption complicates the study of SNRs in our own Galaxy, the Milky Way (MW). The absorption is particularly dramatic for soft X-ray sources, preventing the detection of obscured or faint SNRs.  So far the number of confirmed Galactic SNRs is 303\footnote{Green D. A., 2022, `A Catalogue of Galactic Supernova Remnants (2022 December version \url{http://www.mrao.cam.ac.uk/surveys/snrs/})}\citep{2019JApA...40...36G}, less than what is expected from the star-formation rate and stellar evolution in the Milky Way. The measured Galactic SNe rate is between $0.02\mbox{--}0.03 \,\mathrm{yr}^{-1}$ \citep{1994ApJS...92..487T}. The number of detectable SNRs also depends on the visibility time of SNRs.
Predictions of the number of visible Galactic SNRs are quite uncertain, as the visibility time depends on the distance of the SNR, the density of its environment, and the absorption on the line of sight. For SNRs in M33, \citet{10.1093/mnras/stw2566} estimated a radio visibility lifetime of 20-30 kyr. Another limitation on the visibility time of an SNR is the crowdedness of the observed region. Assuming that the visibility time of the MW is similar to M33 we expect between 1000 and 4000 detectable SNRs. Instead, the Large Magellanic Cloud (LMC) is a perfect target for the study of the entire population of SNRs in a galaxy. The LMC is located outside of the Galactic plane, which means that the absorption along the line of sight is reduced. The average Galactic column density on the line of sight to our sources is $N_\text{H}^\text{Gal} \sim 6.30 \times 10^{20}$ \SI{}{\cm^{-2}}, while the absorption in the LMC corresponds to a column density in the range of $(0.0\mbox{--}1.1) \times 10^{20}$  \SI{}{\cm^{-2}} \citep{Maggi16}. In addition, the LMC is the nearest \citep[${\sim} 50$ kpc,][]{Pietrzy19} star-forming galaxy, viewed almost face-on \citep{van_der_Marel_2001} where we expect to obtain a more complete sample of SNRs. 

Several population studies of SNRs in the LMC have been conducted in the past using multi-wavelength data. In X-rays, e.g., \citet{Maggi16} studied the population using \xmm\ data. In radio,
\citet{Badenes10}, \citet{Bozzetto_2017}, and \citet{Filipovich23} presented results from surveys carried out at Parkes Observatory, Molonglo Observatory Synthesis Telescope (MOST), and the Australian Square Kilometer Array Pathfinder \citep[ASKAP,][]{Johnston08, Pennock21}. Optical SNRs were studied \citep{Yew21} using data obtained from the Magellanic Cloud Emission Line Survey \citep[MCELS,][]{MCELS99}.
It is worth pointing out that the data at different wavelength ranges are useful to identify SNRs in different phases of their evolution. Synchrotron emission in the radio is visible for SNRs at any age. In X-rays synchrotron emission from the shell is only seen in a few young SNRs, while a possible pulsar wind nebula (PWN) can also cause non-thermal emission. The shocked thermal plasma inside the SNR makes it a bright X-ray source in general. While SNRs are not yet radiatively cooling, they are usually not optically bright, unless there is emission from dense ejecta. Therefore, if emission from the SNR shell is prominent in the optical, the SNR tends to be more evolved.
\citet{Maggi16} studied $59$ confirmed SNRs using \xmm\ X-ray observations, obtaining $51$ high quality spectra, while \citet{Bozzetto_2017} used 
radio data from MOST and optical data from the Advanced Technology Telescope (ATT) at the Siding Springs Observatory \citep[see also][]{Payne07, Payne08, Filipovic05} and discussed $15$ SNR candidates, one of which was confirmed by \cite{Maitra19} using \xmm\ data. 
\citet{Yew21} proposed and confirmed two SNRs and proposed $16$ new SNR candidates using MCELS data. Another candidate was confirmed in \citet{Maitra21} using \xmm. \citet{2022A&A...661A..37S} confirmed another previous optical candidate using eROSITA data during the performance verification phase. Recently, \citet{Kavanagh22} confirmed seven SNR candidates using \xmm\ data while \citet{Filipovich23} proposed $13$ new SNR candidates and confirmed two SNRs using the most recent ASKAP data.

\citet{2022MNRAS.512..265F} found a possible SNR in the outskirts of the LMC, which belongs to the new category of sources called "Odd Radio Circle" (ORC J0624--6948) due to its circular shape in the radio. In summary, there were 73 confirmed SNRs and 35 SNR candidates. 
Using the luminosity function of the SNR population in the LMC, \citet{Maggi16} pointed out the incompleteness of the sample, especially in the low-luminosity regime. Given the LMC stellar mass of $2.7 \times 10^9 \,\mathrm{M}_{\odot}$ \citep{2006lgal.symp...47V} and the star formation rates, we expect to have $0.2\mbox{--}0.4$ SNe per century. Assuming a life time of $50 \times 10^4 \ \mathrm{yr}$ we  expect $100\ \text{to}\ 200$ SNRs in the LMC \citep{van2006local, vink2020physics}. 

As the eROSITA all-sky survey (eRASS) provides data from the entire sky, we can investigate the entire LMC and surroundings in X-rays. Compared to the all-sky survey performed by ROSAT, the angular and energy resolution of the eROSITA survey is significantly improved. Furthermore, eROSITA is sensitive in a broader energy band. Using data from eROSITA, we want to find the missing SNRs and improve the statistical study of the SNR population in the LMC.
In this paper, we present the latest catalogue of  SNRs and candidates in the LMC.   
A detailed eROSITA study of the spectra of the brightest SNRs will be presented in a second paper (Zangrandi et al., in prep.).

\section{Data}
\subsection{X-rays}
\subsubsection{eROSITA}
We used data from the extended Roentgen Survey with
an Imaging Telescope Array (eROSITA) in the all-sky survey mode (eROSITA all-sky survey, eRASS). 
eROSITA is part of the {\it Spektrum-Roentgen-Gamma} (\SRG)  observatory \citep{Sunyaev21}, which was launched in July 2019 and started scanning the entire sky in December 2019. So far four all-sky surveys (eRASS1--4, the sum called eRASS:4) have been completed, giving us an unprecedented deep and uniform X-ray view of the entire sky. The full description of the first eRASS:1 survey, data processing, and source detection are discussed in \citet{Merloni24}. eROSITA is composed of seven telescope modules (TMs). Each TM consists of Wolter-1 mirror modules with 54 nested mirrors and a CCD detector \citep[for more details on eROSITA see][]{Predehl21}. The on-axis half energy width (HEW) of eROSITA is about 18\arcsec. As predicted by \citet{2020SPIE11444E..4QD}, there is an off-axis degradation of the angular resolution, which means that HEW at the edge of the field of view (FOV) is the largest, i.e., the resolution is the poorest. At $1.5 \,\mathrm{keV}$ the HEW at the edge is around 69\arcsec\ (see Fig.\,4 in \citet{2020SPIE11444E..4QD}). According to \citet{Merloni24} the most important angular resolution in the survey mode is the average over the entire FOV, which is approximately 26\arcsec\ (0.4 \arcmin) \citep{Predehl21}.

Data processing was performed on the eRASS:4 data with the standard eROSITA Science Analysis Software System (eSASS) software \citep{Brunner22}, version 211214. The pipeline configuration 020 was used to pre-process the data presented in this paper.
We used \texttt{evtool} to create the cleaned event files, selecting good time intervals and valid detection patterns (\texttt{PATTERN=15}). To extract the spectra and create the redistribution matrix file (RMF) and ancillary response file (ARF) we used the \texttt{srctool} task.
We combined the data of eRASS:4 to obtain a mosaic image of the LMC.  
The exposure map of the entire LMC was produced with the \texttt{expmap} command, correcting for the vignetting in the energy band $0.2\mbox{--}\SI{5.0}{keV}$, which is the energy band used for the image analysis. The exposure time varies strongly across the LMC, and the exposure time of the sources analysed in this paper spans from \SI{1.5}{ks} to \SI{16.8}{ks}. 

For the entire analysis, we only used data from TM1, 2, 3, 4, and 6 (TM 12346) due to the light leak found in TM 5 and 7 \citep{Predehl21}. The light leak particularly affects the soft part of the X-ray spectrum where most of the SNR emission is expected. 

\subsubsection{\xmm}

We have identified new SNR candidates using eROSITA data as will be described in Sect.\,\ref{eROSITACandidate} and applied for follow-up observations with \xmm. The source MCSNR J0456--6533 was observed with \xmm\ on May 5, 2022 (obs.ID 0901010101) with the European Photon Imaging Camera \citep[EPIC,][]{2001A&A...365L..18S, 2001A&A...365L..27T} using the medium filter\footnote{\url{https://xmmweb.esac.esa.int/cgi-bin/xmmobs/public/obs_view_cosmos.tcl?action=Get+Selection&search_instrument=0&search_order_by=3&search_obs_id=0901010101}}.  
\xmm\ Extended Source Analysis Software (ESAS, version 20.0.0)\footnote{\url{https://heasarc.gsfc.nasa.gov/docs/xmm/xmmhp_xmmesas.html}} was used to produce filtered event files and to create one merged image of the  EPIC-pn, MOS1, and MOS2 data in the $0.2 \mbox{--} \SI{4.5}{\kilo\eV}$ energy band. To reduce the data, the procedure described in the \xmm\ ESAS cookbook\footnote{\url{https://heasarc.gsfc.nasa.gov/docs/xmm/esas/cookbook/xmm-esas.html}} was followed.
After filtering out bad time intervals caused by soft proton flares, the resulting exposure times were between $42\mbox{--}\SI{44}{\kilo\s}$ for the EPIC detectors. 
Apart from MOS1-CCD3 and CCD6 which were lost due to micro-meteorite hits and hence were excluded from the analysis, no other CCDs were observed to be in an anomalous state.
The source detection task \texttt{cheese} was performed to remove the contribution of point sources in the entire energy band stated above. 
The point sources were masked by using a point spread function (PSF) threshold of 0.5, which means that the point source emission is removed down to a level where the surface brightness of the source is 0.5 of that of the local background, and a minimum separation of 40\arcsec.
Using the tasks \texttt{mos-spectra} and \texttt{pn-spectra}, 
spectra and response files for the entire field of view of the observation for the energy interval $0.2\mbox{--}\SI{10.0}{\kilo\eV}$ were created from the filtered event files. 
Quiescent particle background (QPB) spectra were created with the \texttt{mos-back} and \texttt{pn-back} tasks. 
To determine the level of residual soft proton (SP) contamination, spectral fits to the data were performed. 
The count-rate, exposure, model QPB, and SP background images from the single instruments were combined with the \texttt{comb} task. The background-subtracted and exposure-corrected images are then adaptively smoothed with the \texttt{adapt} task using a binning factor of two and a minimum of 50 counts. Finally, the \texttt{bin\_image} task produced binned count-rate images with a binning factor of two.
Using a three-colour composite image in the energy bands $0.2\mbox{--}0.7\,\mathrm{keV}$, $0.7\mbox{--}1.1\,\mathrm{keV}$, and $1.1\mbox{--}5.0\,\mathrm{keV}$, regions for the spectral analysis were defined based on the X-ray colour (see Sect.\ref{section:J0456-6533}).

For the spectral analysis, the task \texttt{evselect} was used to select single to quadruple pixel events ($\mathrm{PATTERN} \leq 12$) for EPIC-MOS1 and MOS2  and single to double pixel events ($\mathrm{PATTERN}\leq4$) for EPIC-pn. Point sources were detected by \texttt{edetect\_chain} and after checking the extent likelihood, proper point sources were removed from the extraction regions for the source and the local background. To rescale the background spectrum to the source spectrum, areas of the extraction regions were calculated by the task \texttt{backscale} (in arcmin$^2$) to take CCD gaps and bad pixels into account. Finally, the spectra were binned with a minimum of 30 counts and grouped with the respective RMF and ARF files.

\subsection{Optical}
For multi-wavelength comparison, we used optical images from the Magellanic Clouds Emission Line Survey \citep[MCELS, ][]{MCELS99}. These images were taken at the University of Michigan (UM) Curtis Schmidt telescope at Cerro Tololo Inter-American Observatory (CTIO). The angular resolution of the images is about 4.6\arcsec. We supplement our study using the narrow-band filters $\mathrm{H}\alpha \ (\lambda_c = 6563 \, \AA, \ \mathrm{FWHM} = 30 \, \AA)$, $\mathrm{[\ion{S}{II}]} \ (\lambda_c = 6724 \, \AA, \ \mathrm{FWHM} = 50 \, \AA)$ and $\mathrm{[\ion{O}{III}]} \ (\lambda_c = 5007 \, \AA, \ \mathrm{FWHM} = 50 \, \AA)$. We use continuum-subtracted images around the emission lines.

\subsection{Radio}
We have also used radio continuum data from the Australian Square Kilometre Array Pathfinder (ASKAP), in particular, the publicly available four-pointing mosaic of the LMC. The radio-continuum image covers $\SI{120}{deg^2}$ at $888 \, \mathrm{MHz}$, i.e., the entire LMC and its outskirts. The sensitivity of the map at $888 \, \mathrm{MHz}$ is 58\,$\mu \text{Jy beam}^{-1}$ \citep{10.1093/mnras/stac210}.
 For more details see \citet{Pennock21}.
 
\section{X-ray analysis}
\subsection{\textbf{Identification of supernova remnants}}
\label{IdentificationOfRemnants}
The identification of the SNRs in the LMC was performed by visual inspection of the count rate image of eRASS:4. To support the visual investigation, we apply the Gaussian gradient magnitude (GGM) filter \citep[][see Sect.\,\ref{section:GGM}]{2016MNRAS.457...82S}, which allows to see extended emission by enhancing the edges, and visually inspect the GGM image to find regions of interest for further investigation. The confirmation of the remnants relies on multi-wavelength data as described in Sect.\,\ref{multi-wavelength}.

We summarize the number of the SNRs and candidates in Table \ref{tab:Summary}.
\begin{table}[h]
        \renewcommand\arraystretch{1.1}
        \centering
        \caption{Number of sources per category and references to the subsection, table, and figures, in which the sources are discussed. }
        \begin{tabular}{@{} p{.22\textwidth}@{} 
        p{.09\textwidth}@{} p{.07\textwidth}@{} p{.06\textwidth}@{}
        p{.06\textwidth}@{}  @{}}\toprule\midrule
            Category 
            & Number 
            & Section
            & Table  
            & Figures
            \\\midrule
            Previously known SNRs  
             & \textbf{73}
              & \ref{KnownSNRnotMaggi}
             & \ref{Tab:SNRNotInMaggi}
             &  \ref{appendix:SNRNotInMaggi}\\[0.2cm]
             Previous candidate \\ confirmed in this work
             & \textbf{1}
             & \ref{previousCandConf}
             & \ref{Tab:PreviousCandConfi}  
             & \ref{J0454-7003}\\[0.2cm]   
            SNRs found and \\ confirmed in this work 
             & \textbf{3} 
             & \ref{erositasnrs}
             & \ref{Tab:eROSITASNR}   
             & \ref{CommonMainText}
             \\[0.2cm]
             Previous candidates, not  \\ yet confirmed in this work 
             & \textbf{34}
             &\ref{canNotConf}
             & \ref{Tab:PreviousCandRemainCand}
             & \ref{CandiRemainCandi_images} \\[0.2cm]
            New candidates proposed \\ in this work 
             & \textbf{13}
              & \ref{eROSITACandidate}
             & \ref{Tab:eROSITACandidate} 
             & \ref{eROSITACandi_images}\\
            \\\midrule
              Total number of SNRs 
             & \textbf{77}
             &
             & \ref{Tab:TotalMCSNR}\\ [0.2cm]
              Total number of \\
              SNR candidates
             & \textbf{47}
             &
             & 
             \\
             \bottomrule
        \end{tabular}
       
        \label{tab:Summary}
    \end{table}

\begin{figure*}[h]
\centering
\includegraphics[width=\textwidth]{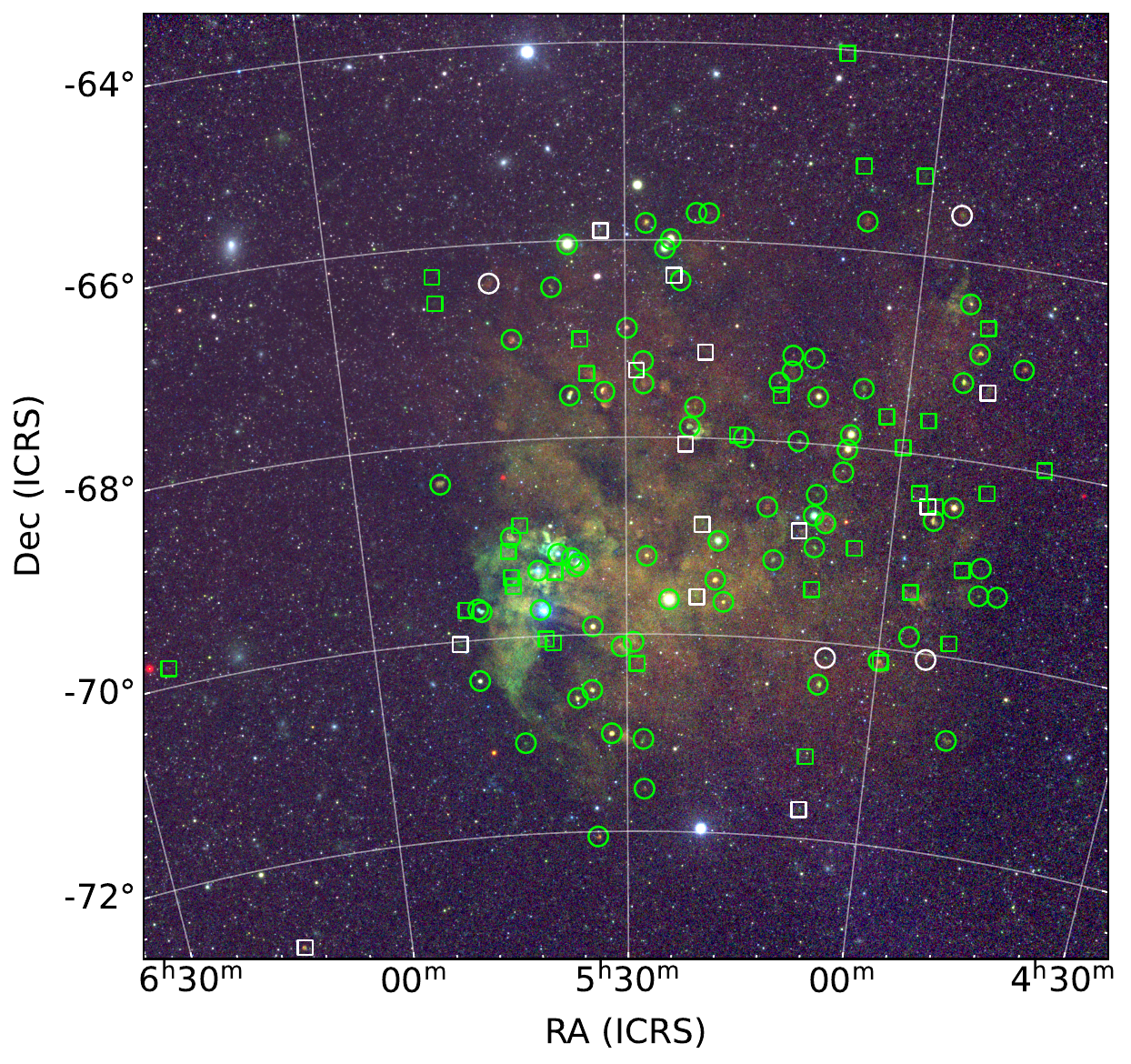}
\caption{Exposure corrected three-colour image of the entire LMC observed by eROSITA during the eRASS:4 in the colours red: $0.2\mbox{--}\SI{0.7}{\keV}$, green: $0.7\mbox{--}\SI{1.1}{\keV}$, and blue: $1.1\mbox{--}\SI{5.0}{\keV}$. 
The circles show the positions of confirmed SNRs while the squares show the position of SNR candidates. The green colour indicates that the sources were known in previous studies while in white colour we show the sources newly detected with eROSITA.
\label{FigLMC}
}
\end{figure*}

\subsection{Images}
\label{ImageCreationSection}
The eROSITA survey data are divided into separate sky tiles, which in total cover the entire sky. We used the eSASS package to generate a mosaic event list of the LMC by combining sky tiles including the LMC observed with TM1-TM4 and TM6. 
We created event maps in three different energy bands: $0.2\mbox{--}\SI{0.7}{\keV}$, $0.7\mbox{--}\SI{1.1}{\keV}$, and $1.1\mbox{--}\SI{5.0}{\keV}$, which are appropriate to detect and identify X-ray emission from SNRs \citep{Kavanagh16}.
We binned 80 sky pixels obtaining an image with a pixel size of 4$\arcsec/ \mathrm{pixel}$. 

The exposure map was obtained using the task \texttt{expmap} and the same binning and energy ranges as for the event maps, with vignetting correction applied. We divided the event map by the exposure map in order to acquire an exposure-corrected image. The final image was smoothed with a Gaussian kernel of $3$ pixels. 

For the point source identification, we used the point source catalogue obtained by the eSASS team using eRASS:4. The pipeline to obtain such a catalogue is described in \citet{Merloni24}.
To exclude the point sources we selected sources in the catalogue with at least a detection likelihood $\texttt{DETLIKE} > 10$ if the extension likelihood $\texttt{EXTLIKE} = 0$, and $\texttt{DETLIKE} > 20$ if the extent $\texttt{EXT} > 0$. We excluded a circular region centred on the point sources with a radius of 28\arcsec\ which is slightly larger than the average half energy width over the field of view \citep{Predehl21}. We noticed that the brightest SNRs, which are typically young and have a rather small extent, are detected as point sources.
If a point source is detected inside a source (SNR or candidate) but there is no clear evidence of a point source in the image, we did not exclude the point source, as it can also be a structure inside an SNR. We have 27 such cases. In total, we excluded $77$ point sources from the examined sources.
We removed the events from the point sources from the original mosaic event file and recreated an exposure-corrected image.
The images shown in the paper are the original exposure-corrected images. For the analysis, we used the point-source subtracted images.

Figure\,\ref{FigLMC} shows the three-colour image of the LMC for the three energy bands $0.2\mbox{--}0.7\,\mathrm{keV}$, $0.7\mbox{--}1.1\,\mathrm{keV}$, and $1.1\mbox{--}5.0\,\mathrm{keV}$.  We have marked the positions of known SNRs and SNR candidates with green symbols (circles and boxes, respectively) and new SNRs and candidates with white symbols. 

\subsection{Size of supernova remnants}
\label{section:Sizes}
To study the sources and to derive their count rate, and hence, flux, it is necessary to define their extent. To include the entire X-ray emission, we drew a circular or elliptical region around each SNR or SNR candidate by eye in the eRASS:4 image. For elongated sources, we used an ellipse while for the others we used circles. The count rates are obtained using the counts within these regions collected by TM 12346. We defined a background region close to each source in order to estimate the net count rate and we used the net count rates to calculate the hardness ratios. For the brighter sources, the same regions were used to extract the spectra, from which the luminosity is calculated.

\subsection{Luminosity}
\label{section:luminosity}
As eROSITA is a new X-ray telescope, first we verify that almost all the previously known SNR were detected with eROSITA. In this case, we consider a 1$\sigma$ emission above the local background as a sufficient detection since we are considering already known SNRs. There are five known sources, for which there is no 1$\sigma$ detection in eRASS:4 image or the 1$\sigma$ detection appears only in a few pixels but not in the entire SNR. 
Among them, there is MCSNR J0522--6543, which was classified as a bona-fide SNR by \citet{Filipovich23} based on optical and radio detections. It was not detected in X-rays, although it was in the FOV of the \xmm \ observation of MCSNR J0521--6543 (ID Obs: 0841320101, PI P. Maggi). 
The other sources (MCSNR J0447--6918, MCSNR J0449--6903, MCNSR J0456--6950, MCSNR J0510--6708) were confirmed in \citet{Kavanagh22} and \citet{Kavanagh16} using \xmm\ data, which means that they are detectable in X-rays.
The difficulty in detecting these sources with eROSITA is most likely related to the intrinsic X-ray faintness of the sources and their low exposure time in the eROSITA observations, which varies over the entire LMC. The average exposure time for these sources is $1.9\,\mathrm{ks}$ which is lower compared to the average exposure time of the other eROSITA detected SNRs of $3.1\,\mathrm{ks}$. The eROSITA exposure times are about one order of magnitude lower than the exposure time of the \xmm \ observations used in \citet{Kavanagh16} and \citet{Kavanagh22}.  
Although these five sources are not significantly detected in the images, we calculated the count rates using the regions as described in Section \ref{section:Sizes}. For MCSNR J0522--6543 we could only derive an upper limit (see Table \ref{Tab:TotalMCSNR}).

The source MCSNR J0524--6624 is the faintest known SNR in our sample in terms of X-ray surface brightness. MCSNR J0524--6624 was identified by \citet{1985ApJS...58..197M} from optical and radio observations. \citet{Maggi16} reported that no X-ray data were available at that time. The region was later observed with \xmm \ in 2019 (Obs ID: 0841320201, PI P. Maggi), where a faint X-ray emission was detected. 
With eROSITA we can detect MCSNR J0524--6624 with 1$\sigma$ emission above the local background. 
We estimated its flux in the energy range of 0.2--5.0\,keV using the energy conversion factor (ECF) calculated assuming a single temperature plasma in non-ionization equilibrium. A detailed explanation of the models used to calculate the ECF will be provided in a subsequent paper (Zangrandi et al. in prep.). The estimated flux is $F$ [0.2--5.0\,keV] ${\sim} 4.3 \times 10^{-14}$ erg s$^{-1}$ cm$^{-2}$ obtained by multiplying the count rates collected by TM 12346 with the ECF. 
Considering the X-ray size of the remnant with a diameter of $D \sim 4 \,\mathrm{'}$, we can estimate a surface brightness of $\Sigma$ [0.2--5.0\,keV] ${\sim} 3.0 \times 10^{-15}$ erg s$^{-1}$ cm$^{-2}$ arcmin$^{-2}$. Since the surface brightness is independent of distance we can compare it with a faint Galactic remnant such as the Monogem Ring. From the measurements by \citet{Knies24} we  estimate a surface brightness $\Sigma \sim 1.2 \times 10^{-14}$ erg s$^{-1}$ cm$^{-2}$ arcmin$^{-2}$ for the Monogem Ring in the same energy band. 

As a more quantitative test, we compare the luminosity of SNRs measured with eROSITA with the luminosity in the literature to check the reliability of the flux measurements. As a reference, we considered the luminosities in the catalogue of \citet{Maggi16} who performed a detailed spectral analysis of the SNRs in the LMC using \xmm\ observations. In order to determine the luminosity of the sources in our sample we performed spectral analyses of the sources with at least 400 net counts, combining data from TM1-TM4 and TM6. A detailed explanation of the spectral analysis and further studies of the population of SNRs in the LMC will be presented in a future paper (Zangrandi et al., in prep.). In this work, we only compare the luminosities measured with eROSITA to those obtained with \xmm\ to check for consistency. For the 13 brightest eROSITA sources with at least 1000 net counts per TM, we could fit the spectra with the same spectral models as  \citet{Maggi16}. We started from the same parameter values and performed a combined fit with the data from TM 12346. To calculate the luminosity $L$, we determined the flux $F$ in the $0.3\mbox{--}\SI{8.0}{\keV}$ energy interval using XSPEC. The eROSITA telescope is sensitive in the energy range $0.2\mbox{--}\SI{10.0}{\keV}$ but we expect that most of the SNR flux comes from the soft band. We used the relation $L=4 \pi d^2 F$, where we assumed $d = 50 \ \mathrm{kpc}$ as the distance to the LMC for all sources, as assumed in \citet{Maggi16}. We note that the distance of the LMC has been updated by \citet[][]{Pietrzy19}. Even though the recent value is more accurate it is still consistent with the approximation of $50 \ \mathrm{kpc}$. Since we want to check the consistency of the flux measurement with eROSITA with that of \citet{Maggi16}, we decided to assume the same value for the distance.
Since we have five different spectra for each source (one for each TM) we calculated the average of the luminosity estimated for each spectrum, and we compared the mean luminosity with  \citet{Maggi16}. The luminosities of \cite{Maggi16} were calculated mainly using  \xmm, except for the source J0550--6823 where \chandra\ data were used to evaluate $L_\text{X}$. 

\begin{figure}
\centering
\includegraphics[width=0.49\textwidth]{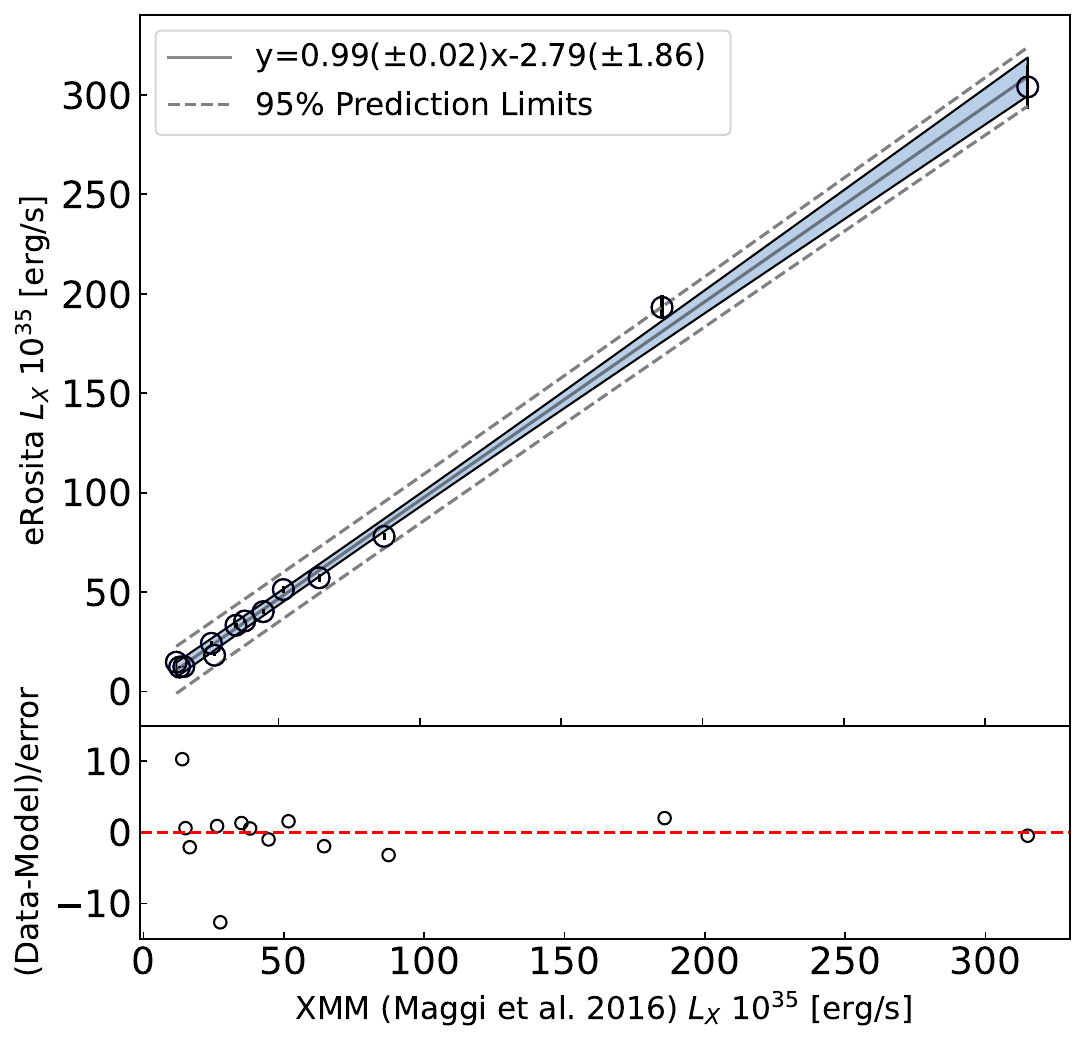}
    \caption{Comparison between the luminosity measured with eROSITA and the luminosity reported in \citet{Maggi16} based on data from \xmm, in the energy range $0.3\mbox{--}{8.0}\,\mathrm{keV}$. The shaded region shows the 95\% confidence interval around the best-fit line. The dashed line shows the 95\% prediction interval. The bottom panel shows the residuals between the data and the best fit. From the plot the agreement between the luminosities of SNRs obtained by the two different instruments is evident.}
\label{LumComp}
\end{figure}

In Fig.\,\ref{LumComp} we compare the luminosities measured in this work with the luminosities reported in \cite{Maggi16}. 
We fitted a linear relation. In Fig.\,\ref{LumComp} we plotted the best-fit line, the confidence interval, and the prediction interval, both at 95\% confidence, and the residuals. While the confidence interval gives the interval, in which the correlated data are located with 95\% confidence, the prediction interval is the interval, in which new data will be found with a 95\% probability.
 
The slope and the intercept of the best-fit line are $s=0.99 \pm 0.02$ and $q= (-2.79  \pm 1.86) \times 10^{35}$ erg s$^{-1}$, respectively. The intercept indicates an offset between eROSITA and \xmm \  luminosities. The negative value despite the uncertainty shows a slight underestimation of the luminosities with eROSITA.

We also evaluated the consistency for the 38 fainter sources, which are not included in the above fit as a direct comparison to \citet{Maggi16} was not possible, but for which we performed a spectral analysis assuming a simple thermal emission model. For these sources, we calculated the difference in the luminosity with respect to the values reported in \citet{Maggi16}. We measured the median and the standard deviation of the difference $L_\text{XMM}-L_\text{eROSITA}$. We obtain a median of $0.04$ and a standard deviation of $0.47$ both in units of $10^{35} \text{erg}\, \text{s}^{-1}$. The analysis confirms the consistency of luminosities obtained with eROSITA and by \citet{Maggi16}.

\subsection{Gaussian gradient magnitude filter}
\label{section:GGM}
By visually inspecting the eROSITA LMC images we searched for new SNR candidates. 
To enhance the diffuse emission, we applied a Gaussian gradient magnitude (GGM) filter \citep{2016MNRAS.457...82S} on the eRASS:4 images. The filter calculates the magnitude of the gradient of an image using Gaussian derivatives. Firstly, the input image is smoothed with a Gaussian filter of a certain $\sigma$. Secondly, the derivative along the $x$- and $y$-axis is taken. The magnitude of the gradient is then determined by summing the squared derivatives under the square root. 
Where the intensity of the image changes rapidly over the pixels the magnitude has a greater value, which can be used to highlight regions of rapid change in intensity. 
Usually, the edges of objects are characterized by such a change in intensity over pixels, which will be shown as maxima in the filtered image. Therefore, the GGM filter can act as an edge detection algorithm. 
The resulting image depends on the choice of $\sigma$ for the GGM filter, which is measured in pixels. For a certain $\sigma$ the filter will highlight the edges in the image with a certain pixel scale. Thus, we exploited various values of $\sigma$ ($\sigma$ = 1, 2, 4, 8, and 10 pixels) and combined the resulting filtered images into one. In order to reduce the noise resulting from point sources we applied the GGM on the point source subtracted count rate image.

\begin{figure}
\centering
\includegraphics[width=0.49\textwidth]{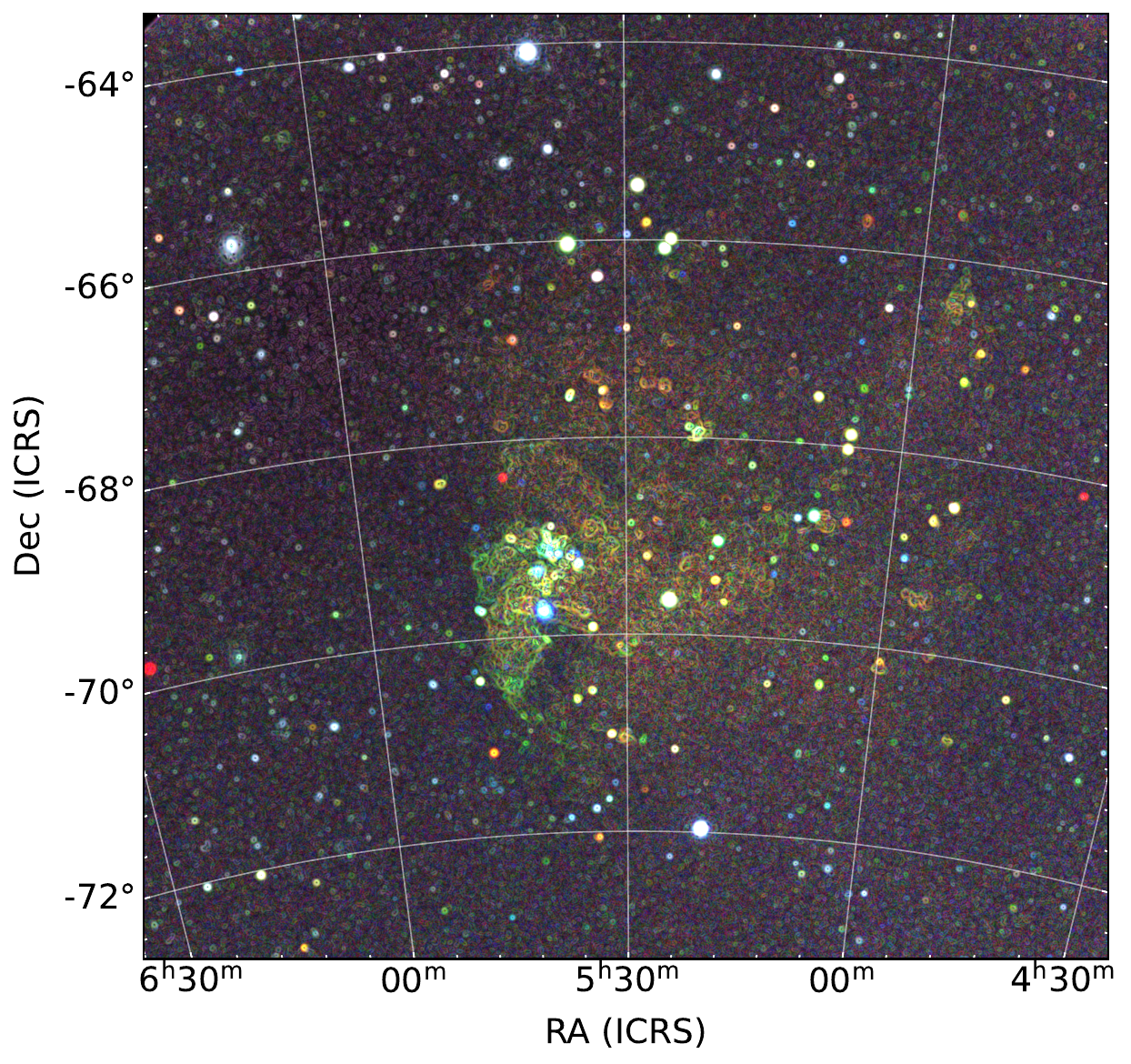}
\caption{eROSITA three-colour image after point sources were removed and GGM filter was applied. This GGM-filtered image shows the magnitude of the gradient of the input image. The filter was applied to each energy band separately and the resulting images were combined into a three-colour image.}
\label{GGM}%
\end{figure}

\begin{figure}[h]
        \centering
         \includegraphics[width=.5\textwidth]{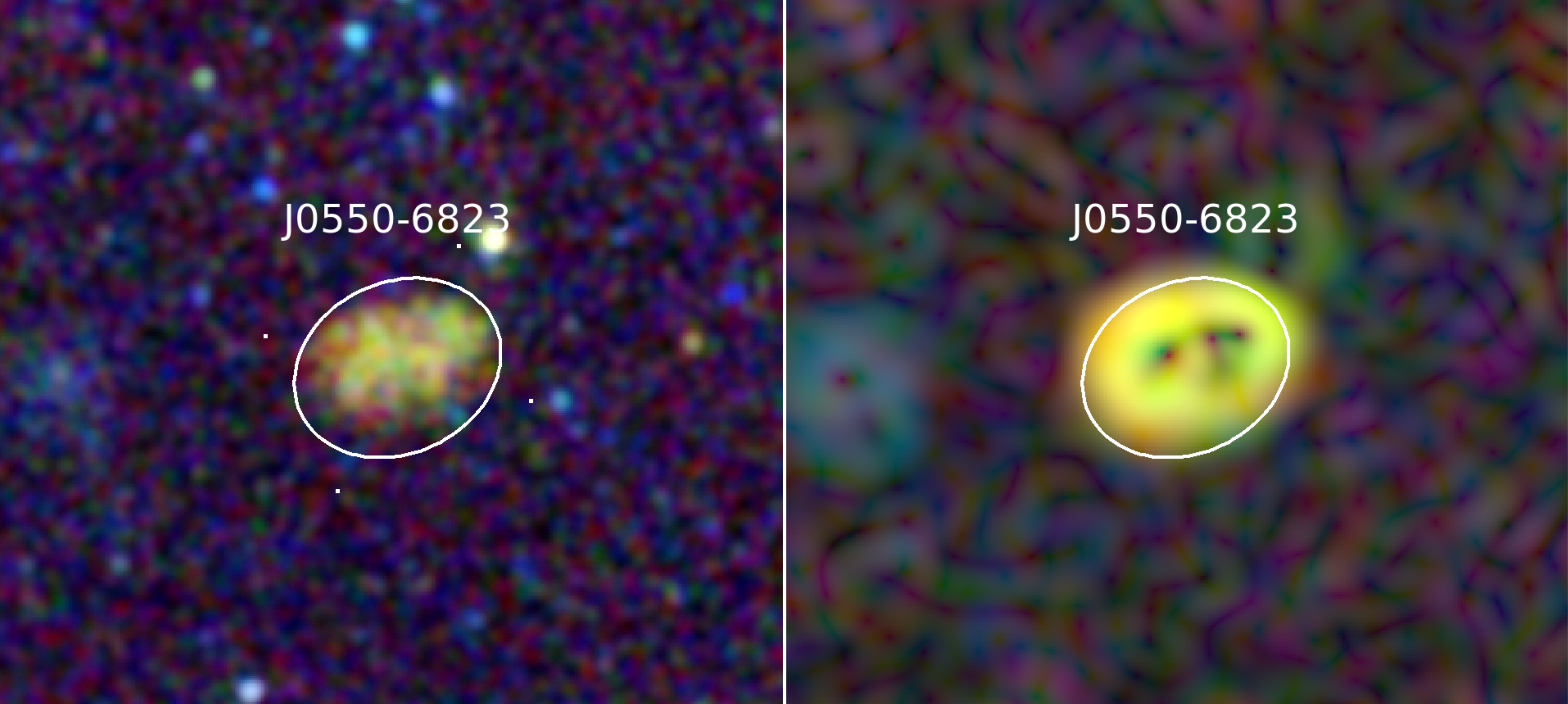}
    \caption{Three-colour eROSITA image of SNR J0550--6823 (left) and with GGM filter applied.}
          \label{exampleGGM}
\end{figure}

We repeated the procedure described above for each energy band.
The result of this technique is shown in Fig.\,\ref{GGM}.
This image was useful to detect faint sources or, also to check if at the position of a known SNR candidate an edge structure is detectable. In Fig.\,\ref{exampleGGM} we show the effect of the GGM filter on a relatively bright, known SNR. The GGM filter highlights the edges of the emission. This technique was particularly useful in identifying SNR candidates in crowded regions. We point out that this procedure was used to find interesting regions for further investigation.
For the identification, the sources detected with the GGM filter were further investigated using the X-ray count rate image and images at other wavelengths as described in Sect.\,\ref{multi-wavelength}.

\subsection{Hardness ratio}
\label{HR}
The faintness of many of the sources prevented us from performing a detailed spectral analysis for all sources. Therefore, we calculated the hardness ratios (HR) for all sources. 
We defined four energy bands: soft $=0.3\mbox{--}\SI{0.7}{\keV}$, medium $=0.7\mbox{--}\SI{1.1}{\keV}$, hard $=1.1\mbox{--}\SI{2.3}{\keV}$, harder $=2.3\mbox{--}\SI{4.0}{\keV}$ and determined the net count rates in each energy band. We then computed three hardness ratios according to equation \ref{equationHR} for two adjacent energy bands:
\begin{equation}
   \label{equationHR}
    \mathrm{HR}_{i} = \frac{R_\text{i+1} - R_\text{i}}{R_\text{i+1} + R_\text{i}}   
    \text{ \ \ \ }
    \mathrm{EHR}_{i} = 2 \frac{\sqrt{(R_\text{i+1}\ eR_\text{i})^2+(R_\text{i} \ eR_\text{i+1})^2}}{({R_\text{i+1} + R_\text{i})^2}}
   \end{equation}
where EHR$_{i}$ is the error of each HR$_{i}$ \citep{2018A&A...620A..28S}.
$R_i$ is the net count rate in each energy band,
$t_i$ is the average exposure time inside the emission region for each energy band,
and
$eR_{i}$ is the error of the net count rate calculated as $eR_i=\sqrt{R_i/t_i}$. The average exposure time was calculated using the exposure map obtained from the \texttt{exmap} task corrected for the vignetting effect.

\begin{figure*}[h]
\centering
\includegraphics[width=1.0\textwidth]{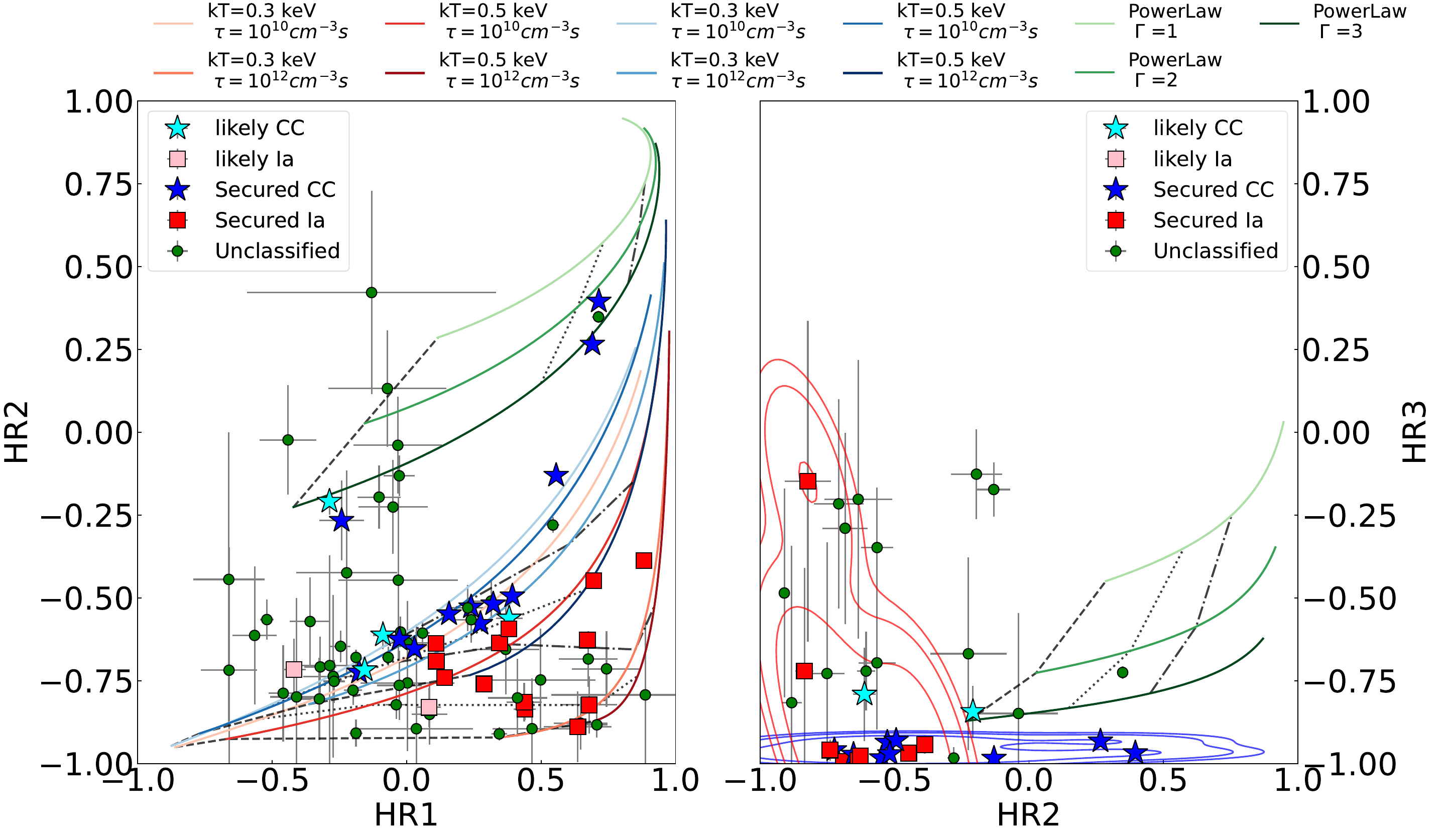}
\caption{Hardness ratios of candidates and confirmed SNRs in our sample. The energy bands used to calculate the hardness ratios are for HR1: soft ($=0.3\mbox{--}0.7\,\mathrm{keV}$) and medium ($0.7\mbox{--}1.1\,\mathrm{keV}$), for HR2: medium ($0.7\mbox{--}1.1\,\mathrm{keV}$) and hard ($1.1\mbox{--}2.3\,\mathrm{keV}$), and for HR3: hard ($1.1\mbox{--}2.3\,\mathrm{keV}$) and harder ($2.3\mbox{--}4.0\,\mathrm{keV}$). The sources are separated according to the explosion type as reported in \citet{Maggi16}. The SNRs tend to be separated according to the progenitor type of the remnant. The lines describe the expected HR values assuming different models. In particular we considered Fe-rich thermal plasma emission (VNEI with $\mathrm{Fe} = 2 \,\mathrm{Z}_{\odot}$) plotted with red lines and O-rich thermal plasma emission (VNEI with $\mathrm{O} = 2 \,\mathrm{Z}_{\odot}$) plotted with blue lines with different ionization time scales $\tau$ ($\tau = 10^{10} \,\mathrm{s}\,\mathrm{cm}^{-3}$ and $\tau = 10^{12} \,\mathrm{s}\,\mathrm{cm}^{-3}$) and different temperatures ($kT = 0.3\,\mathrm{keV}$ and $kT = 0.5\,\mathrm{keV}$). In addition, we also plotted power law models in green with different photon indices $\Gamma$ ($\Gamma =1, 2, 3$). The dashed and dotted black lines indicate different column densities $N_\mathrm{H} = 0.0, 0.5 \ \mathrm{and} \ 1.0 \,\mathrm{cm}^{-2}$. The contours in HR2--HR3 represent the Gaussian kernel-density estimation for core-collapse (CC) and type Ia SNRs. The contours indicate 68\%, 90\%, and 95\% probability.
}
\label{FigureHR}
\end{figure*}

We only include sources with a net count rate greater $0.01\,\mathrm{cts}\,\mathrm{s}^{-1}$ in the energy band $0.2\mbox{--}\SI{5.0}{\keV}$. Among them, we selected the sources with a net count rate greater than $0.001\,\mathrm{cts}\,\mathrm{s}^{-1}$ in each band (soft, medium,  hard, and harder). 
We plotted $\mathrm{HR}_{1}$ vs. $\mathrm{HR}_{2}$ and $\mathrm{HR}_{2}$ vs. $\mathrm{HR}_{3}$ in Fig.\,\ref{FigureHR} with SNR types from the classification proposed in \citet{Maggi16}. In the plot, we also exclude the data points with an error larger than $0.5$. The energy bands are chosen in a way that will allow us to separate core-collapse (CC) SNRs and type Ia SNRs. 
For type Ia SNRs, we expect enhanced Fe-L emission between 0.7 and 1.1\,keV. Therefore, HR1 is positive, while HR2 is negative. Core-collapse SNRs will be also brighter at higher energies, hence positive HR2. As the sensitivity of eROSITA drops significantly above 2.3\,keV, HR3 is unfortunately not so useful, which can be also seen in the large errors. 
\par We also compare the HRs with predicted values assuming different typical spectral models. We assume a thermal plasma in non-ionisation equilibrium (NEI) with different ionization time scales $\tau$ and different temperatures. We use the model VNEI, which allows us to vary the element abundances, and assume an enhanced Fe abundance by setting $\mathrm{Fe} = 2 \,\mathrm{Z}_{\odot}$ to simulate a type Ia SNR, while in the other case we assume an enhanced O abundance ($\mathrm{O} = 2 \,\mathrm{Z}_{\odot}$ as the ejecta of CC SNRs is enriched with $\alpha$-elements. 
In addition, we assume a power-law spectrum with different photon indices $\Gamma$. These models take into account the hard sources in which there can be non-thermal emission from the shell in young SNRs or from a pulsar wind nebula.

The Fe-rich models tend to occupy the lower right corner of the HR1--HR2 plot in Fig.\,\ref{FigureHR}, while the O-rich models tend to result in higher HR2. This trend is also visible in the distribution of the data points where we can clearly see a separation between type Ia (red squares) and CC (blue stars) SNRs. 
In the HR2$-$HR3 diagram where the separation is much more difficult to see, we instead estimated the probability density function of the two distribution of CC and type Ia SNRs using the python package \texttt{gaussian\_kde} \footnote{\url{https://docs.scipy.org/doc/scipy/reference/generated/scipy.stats.gaussian_kde.html}}, 
which allows us to better display the separation between CC and type Ia SNRs.
We note that although SNRs with different progenitors seem to have different HRs we cannot determine the origin of the SNR by only considering the HRs.  In this work, we discuss the possible progenitor type by combining the HRs with information about the underlying stellar population at the location of the SNR (see Sect.\,\ref{section:SFH}).

\section{Multi-wavelength analysis}
\label{multi-wavelength}

Supernova remnants are multi-wavelength objects and can be observed from radio to X-rays, in some cases also in gamma-rays. Morphologically, SNRs have mainly bubble- or shell-like structures and can easily be confused with \ion{H}{II} regions and planetary nebulae (PNe). To find new SNRs, a multi-wavelength investigation is mandatory. The candidates were first identified in the eRASS:4 data by visually inspecting the count rate image and the GGM image (see Sect.\ \ref{GGM}). If there is X-ray emission with a surface brightness 1$\sigma$ above the local X-ray background
with a diffuse or shell-like emission, we classify the source as an SNR candidate. 
In addition, we also inspected the optical narrow-band images and the radio image.
Only if we observe emission indicative of an SNR in at least one other band, either in optical or in radio, and the X-ray emission is at 3$\sigma$ level, we confirm the source as an actual SNR. 

In the optical, we look for an excess in the ratio of two emission line fluxes ${\mathrm{[\ion{S}{II}]/H}\alpha}$. For ${\mathrm{[\ion{S}{II}]/H}\alpha < 0.4}$ the gas is photo-ionised, a process which typically occurs in \ion{H}{II} regions around young and hot stars or in super-bubbles.
For $\mathrm{[\ion{S}{II}]/H}\alpha > 0.4$ the ionisation is likely caused by a shock \citep{Mathewson73, Dodorico80, Fese80, Blair97, Matonick97, Dopita10, Lee14, Vucetic19a, Vucetic19b, Lin20}.
The presence of a shock wave is a strong indication of the presence of an SNR.
In the radio band, the presence of an SNR is usually identified by measuring the spectral index $\alpha$ defined as $S_{\nu} \propto \nu^{\alpha}$, where $S_{\nu}$ is the flux density and $\nu$ the frequency. For SNRs, we expect $\alpha \sim -0.5$, which indicates non-thermal emission in the radio band \citep{Filipovic98, 2011A&A...525A.138G}.

\subsection{Optical}
To investigate the optical counterpart we use the MCELS data, which is very useful for studying the SNR as it provides us with narrow-band images of H$\alpha$, [\ion{S}{II}], and [\ion{O}{III}]. \par 
Essential for the detection and classification of SNRs is the ratio ${\mathrm{[\ion{S}{II}]/H}\alpha}$ as described above. For ${\mathrm{[\ion{S}{II}]/H}\alpha}$ higher than 0.4, ionisation and excitation of S is enhanced due to a radiative shock as suggested by several radiative shock models \citep{1979ApJS...39....1R, 1987ApJ...316..323H, 1995ApJ...455..468D, 2008ApJS..178...20A}. 
However, several studies of SNRs and SNR candidates have shown, that this separation can be less clear especially for sources with low surface brightness \citep[see, e.g.,][for a study of sources in M33]{2018ApJ...855..140L}. For this reason, in this work, we use ${\mathrm{[\ion{S}{II}]/H}\alpha > 0.67}$, which is a more reliable lower limit for identifying shock emission \citep{1985ApJ...292...29F}.
In addition, it is necessary that the emission of H${\alpha}$ is significant, as a low H${\alpha}$ flux might cause an artificial enhancement of the ${\mathrm{[\ion{S}{II}]/H}\alpha}$ ratio. 
As pointed out also by \cite{Yew21} the line ratio is an effective way to identify radiative SNRs but it will miss the young non-radiative and Balmer-dominated SNRs \citep{1980ApJ...235..186C}.

\subsection{Radio}
\label{radioSection}
From SNRs, we expect predominantly non-thermal emission in radio via synchrotron radiation. In young SNRs, synchrotron emission is also observed in the X-ray band. The more energetic electrons, emitting synchrotron radiation in X-rays, lose energy faster than the less energetic electrons, which will stay relativistic and emit in the radio band much longer. We used the public data from the ASKAP interferometer at $888$ MHz \citep{Pennock21}. The ASKAP LMC survey covers the entire LMC.  

In order to highlight the non-thermal emission in the radio images we used the same approach as in \citet{Filipovich23} and in \citet{Ye91}. 
If there was no supernova explosion inside an emission nebula we expect a correlation between the H${\alpha}$ emission and the radio continuum. This is due to the fact that the free electrons that produce thermal radio emission via Bremsstrahlung are the same as those that recombine with the protons to produce the H${\alpha}$ lines. We can use this proportionality to highlight the non-thermal emission in the radio images. After subtracting a scaled H${\alpha}$ image from the radio continuum what remains is just the non-thermal emission. In order to subtract the optical image from the radio image, we need a normalisation factor. This factor can be determined from the correlation between the pixel values in H${\alpha}$ and the radio continuum in regions where we expect that the emission is just thermal. In order to measure the correlation, we selected different  \ion{H}{II} regions in the entire LMC. We measure the [\ion{S}{II}]/H${\alpha}$ ratio and select only those emission nebulae with a ratio [\ion{S}{II}]/H${\alpha} <\ 0.4$ in order to make sure that there are no SNRs hidden inside the selected \ion{H}{II} regions. We measured the intensity of H${\alpha}$ and radio continuum from the same region and compared the values. We performed a linear fit in the H$\alpha$-radio diagram and calculated the Pearson coefficient to evaluate the goodness of the fit. The slope of the plot is the normalisation factor, which can be used to normalise the H${\alpha}$ before subtracting it from the radio continuum. We averaged the different slopes using a weighted average using the Pearson coefficients as the weight. 

\begin{figure*}[h]
     \centering
     \begin{subfigure}[]{0.24\textwidth}
         \centering
         \includegraphics[width=\textwidth]{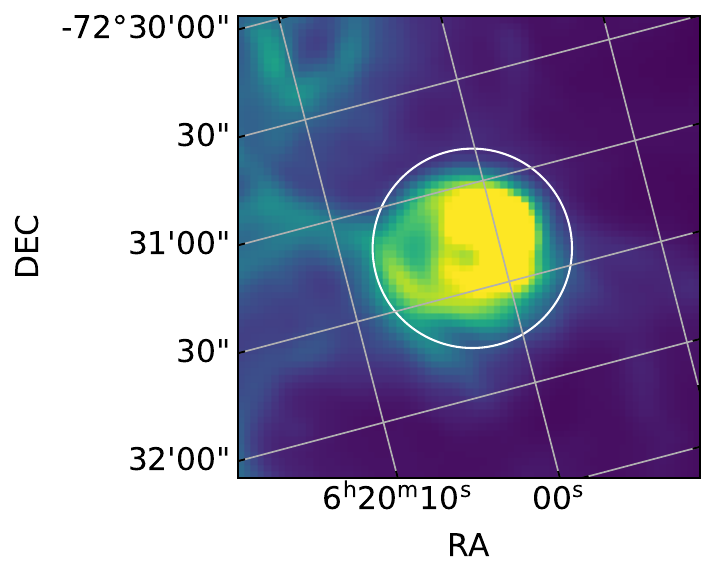}
         \caption{}
         \label{L140_Halpha}
     \end{subfigure}
     \begin{subfigure}[]{0.24\textwidth}
         \centering
         \includegraphics[width=\textwidth]{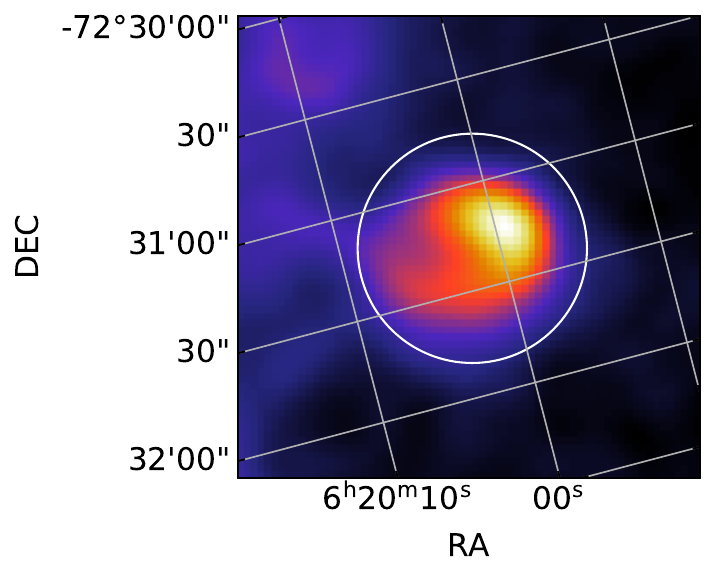}
         \caption{}
         \label{L140_radio}
     \end{subfigure}
     \raisebox{1mm}{\begin{subfigure}[]{0.27\textwidth}
         \centering
         \includegraphics[width=\textwidth]{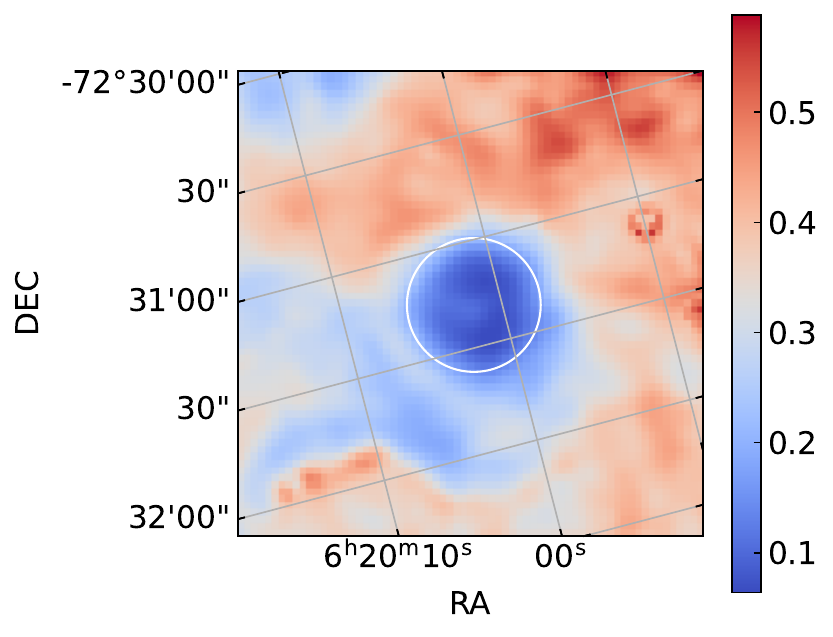}
         \caption{}
         \label{L140_sii2ha}
     \end{subfigure}
    } 
     \begin{subfigure}[]{0.23\textwidth}
         \centering
         \includegraphics[width=\textwidth]{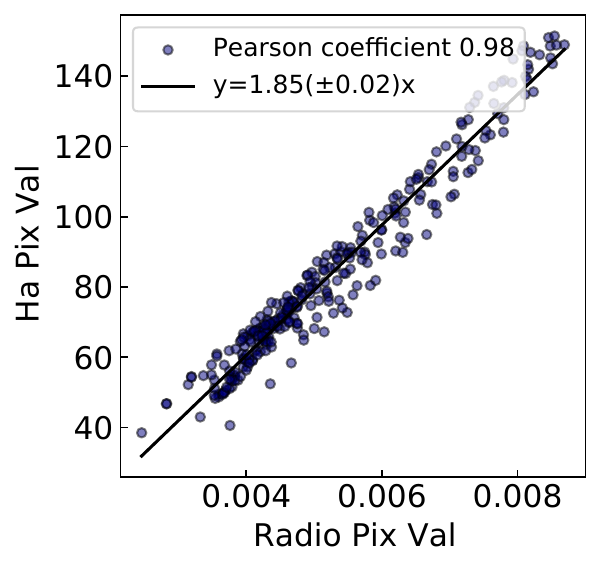}
         \caption{}
         \label{L140Linearfit}
     \end{subfigure}
     \caption{A subregion of the \ion{H}{II} region DEM-L140 chosen to highlight the non-thermal emission in the radio images. In particular Fig. \ref{L140_Halpha} and \ref{L140_radio} show the emission of the selected region in H$\alpha$ and radio respectively. Figure\,\ref{L140_sii2ha} displays the image of the ${\mathrm{[\ion{S}{II}]/H}\alpha}$ ratio. The latter image is used to ensure that the selected region is not affected by any SNR which would increase the ${\mathrm{[\ion{S}{II}]/H}\alpha}$ ratio to values larger than $0.4$. In Fig.\,\ref{L140Linearfit} the linear correlation between the H$\alpha$ and radio emission in the selected area is reported. The black solid line represents the best linear fit for this correlation. \label{L140}
     }
\end{figure*}

In Fig.\,\ref{L140} we show the example of the \ion{H}{II} region DEM-L140. Figure\,\ref{L140_Halpha} shows the H${\alpha}$ emission, Fig.\,\ref{L140_radio} shows the radio continuum of DEM-L140, and Fig.\,\ref{L140_sii2ha} shows the [\ion{S}{II}]/H${\alpha}$ ratio. In order to avoid outliers, which could contaminate the linear fit we recursively cleaned the data until the standard deviation of the pixel value vector stopped to decrease. 
The linear relation and the linear fit are shown in Fig.\,\ref{L140Linearfit}.
We adopted the average slope as the normalization factor for the H${\alpha}$ image and subtracted the scaled H${\alpha}$ emission from the radio-continuum emission. 
The \ion{H}{II} regions used are: DEM-L111, DEM-L140, DEM-L194, DEM-L196,LHA-120-N44J, LHA-120-N70, MCELS-L401, N11, N44C, NGC-1899. From these \ion{H}{II} regions, we selected several smaller regions and obtained a total of $61$ sub-regions. To calculate the average slope we only kept the regions which show a Pearson coefficient greater than $0.9$. Using this criterion we have a final number of $11$ regions that contribute to the averaged slope. In Table \ref{tab:HIIRegionUsed} we report the selected \ion{H}{II} sub-regions used to calculate the average slope with the relative fitted slopes and the Pearson coefficients for each sub-region. At the end of the table, we give the final average slope used to normalise the H$\alpha$ emission. The error on the average slope was calculated using the formula to propagate the error on the weighted average: 

\begin{equation}
\label{errorAverage}
    \sigma = \frac{1}{\sum_i w_i}\sqrt{\sum_i(w_i\sigma_i)^2},
\end{equation}
where $w_i$ are the Pearson coefficients and $\sigma_i$ are the errors on the single fitted slope. The error on the single slope has been calculated using the python package used to fit the slope of the correlation \texttt{scipy.optimize.curve\_fit}\footnote{\url{https://docs.scipy.org/doc/scipy/reference/generated/scipy.optimize.curve\_fit.html}}.

\begin{table}[h]
        \renewcommand\arraystretch{1.1}
        \centering
        \caption{ \ion{H}{II} regions used to calibrate the normalization factor for the H$\alpha$ images.}
        \begin{tabular}{@{} p{.16\textwidth}@{} 
        p{.17\textwidth}@{} 
        p{.15\textwidth}@{}  @{}}\toprule\midrule
            Region 
            & Fitted Slope [$\times 10^4$]
            & Pearson coefficient  \\\midrule
             DEM-L140 
             & $1.85 \pm 0.02$ 
             & $0.98$  \\
             LHA-120-N44J 
             & $1.75 \pm 0.03$  
             & $0.97$  \\
             LHA-120-N70 (a)  
             & $1.81 \pm 0.02$ 
             & $0.95$  \\
             LHA-120-N70 (b)  
             & $1.50 \pm 0.02$  
             & $0.91$ \\
             LHA-120-N70 (c) 
             & $1.45 \pm 0.02$ 
             & $0.93$\\
             N11 (a) 
             & $1.04 \pm 0.02$ 
             & $0.98$\\
             N11 (b)  
             & $1.10 \pm 0.04$ 
             & $0.91$\\
             N11 (c)  
             & $1.50 \pm 0.05$ 
             & $0.93$\\
             N11 (d) 
             & $1.70 \pm 0.05$  
             & $0.93$\\
             N11 (e)  
             & $0.87 \pm 0.02$  
             & $0.92$\\
             N11 (f)  
             & $1.61 \pm 0.02$  
             & $0.99$\\\midrule
             Average  
             & $1.48 \pm 0.01$  
             & $0.95$\\
             \bottomrule
        \end{tabular}
         \tablefoot{The first column reports the name of the \ion{H}{II} region used to extract the H$\alpha$ and radio pixel values. The second column indicates the fitted slope of the linear relation between radio and H$\alpha$. The third column documents the Pearson coefficient of the linear fit. The latter was used as a weighting factor in the calculation of the averaged normalisation factor.
        }
        \label{tab:HIIRegionUsed}
    \end{table}
    
After the subtraction of the scaled H$\alpha$ from the radio continuum image only the non-thermal radio emission remains. We used it to draw contours in the radio continuum images at 3 and $5\sigma$ levels above the background to search for significant emission in an SNR. 

\section{SFH-based progenitor classification} \label{section:SFH}
A necessary (but not sufficient) condition for a CC SNR is the presence of recent star formation (SF) activity near the SNR.
In order to find the possible origin of SNRs and SNR candidates, we estimated the number of OB stars in the proximity of each source. We used the star formation history (SFH)  values measured by \citet{SFHpaper} using the near-infrared photometry from the VISTA survey of the Magellanic Clouds (VMC). The VMC data covers a sky area of ${\sim} 96\ \text{deg}^2$ and consists of infrared observations in the $Y$, $J$ and $K_\text{s}$ bands.
\citet{SFHpaper} calculated the SFH by comparing the synthetic Hess diagrams with the observed Hess diagram. A Hess diagram shows how many stars are located in a sub-region of a color-magnitude diagram (CMD). The CMDs, and therefore the Hess diagrams, can be constructed using different colours, in particular:
$Y - K_\text{s}$ vs. $K_\text{s}$ and $J-K_\text{s}$ vs. $K_\text{s}$, called Y$\text{K}_\text{s}$ and J$\text{K}_\text{s}$ respectively in the figures showing the SFH for the eROSITA SNRs and SNR candidates (in Fig.\,\ref{CommonMainText} and Fig.\,\ref{eROSITACandi_images}). The synthetic CMDs were calculated using a library of isochrones, which depend on the initial mass function (IMF), distance, extinction, binary fraction and SFH. A set of isochrones constitutes a model. In general, it is possible to independently measure the distance, extinction, and binary fraction and to assume an IMF. In this case, the best fit model can be used to calculate the SFH. For more details on the models see \citet{SFHpaper}, \cite{2001ApJS..136...25H}, and references within.
\par 
We report the SFH around each source within a radius of $100 \, \mathrm{pc}$. This corresponds to the maximal projected distance that a star with a typical velocity of $10\ \mathrm{km} \, \mathrm{s}^{-1}$ travels in $10^7 \, \mathrm{yr}$. For the eROSITA candidates and SNRs we plotted the SFH (see Fig. \ref{CommonMainText} and Fig.\,\ref{eROSITACandi_images}). In \cite{Maggi16} the authors calculated the number ratio of the two SNR types $N_\text{CC}/N_\text{Ia} = 1.35^{+0.11}_{-0.24}$. This number seems to be lower than the results of the local SN survey \citep{Li11} and the ratio derived from the abundance pattern of the intra-cluster medium chemically enriched by supernovae \citep{2007ApJ...667L..41S} and indicates a higher relative number of type Ia SN in the LMC than in other galaxies. As discussed in \citet{Maggi16}, however, this lower $N_{CC}/N_{Ia}$ could be caused by observational bias, as some CC SNRs can be missed because they occur inside superbubbles. 
However, in our study, we confirm the same tendency, as many of the newly confirmed SNRs and new candidates show indications to be thermonuclear SNRs.

\section{ Classifications of SNRs and candidates}
\label{ClassificationsOfCandidates}
As summarised in Section \ref{IdentificationOfRemnants} the first step in the identification of the new candidates was the visual inspection of the eROSITA count rate image in three energy bands and the GGM filtered images.
In order to understand if there is emission associated with an SNR, we compared the X-ray, optical and radio images (see Sect.\,\ref{multi-wavelength}). 
The images are shown in Figs.\,\ref{CommonMainText}, \ref{J0614Rate}, and in Fig\,\ref{eROSITACandi_images}. For each SNR or candidate, we show the three-colour eROSITA image of the source in the bands $0.2\mbox{--}0.7$ keV, $0.7\mbox{--}1.1$ keV, and $1.1\mbox{--}\SI{5.0}{\keV}$ in the left image. The contours show the source surface brightness with  $1\sigma$, and $3\sigma$ above the local X-ray background. 
The second image is the GGM image of the X-ray emission, described in Sect.\,\ref{section:GGM}. The GGM was used in order to visually find possible region of interest, but it was not used to confirm any of the sources. The third panel shows the MCELS image, which shows the emission of H${\alpha}$, [\ion{S}{II}], and [\ion{O}{III}]. The contours in the optical image represent regions with [\ion{S}{II}]/H$_{\alpha} > 0.67$. In the right panel, we show the radio continuum image from the ASKAP survey of the LMC at $888 \, \mathrm{MHz}$. The contours indicate non-thermal emission at $2\sigma$, $3\sigma$, and $5\sigma$ above the background as described in Sect.\,\ref{radioSection}. For the first criterion to classify the source we used the X-ray contours. If at least $1\sigma$ emission (cyan contours in the eROSITA images) is detected we propose the source as a candidate. The classification of a source as a confirmed SNR requires a  $3\sigma$ emission (red contours) in the X-ray image and, in addition, a confirmation in at least one other wavelength band (radio or optical): either the optical line ratio is [\ion{S}{II}]/H$_{\alpha} > 0.67$ or there is non-thermal emission in the ASKAP radio data.
Optical or radio emissions, which satisfy these criteria are shown with red contours in the MCELS and ASKAP images. The most promising SNRs and candidates have been proposed and accepted for follow-up \xmm\  observations. Section \ref{erositasnrs} presents 3 newly discovered and confirmed SNRs, Sect.\,\ref{previousCandConf} reports the confirmation of one previous candidate as an SNR, Sect.\,\ref{eROSITACandidate} presents 13 new candidates, Sect.\,\ref{KnownSNRnotMaggi} discusses the eROSITA data for 14 previously known SNRs, which were not in the catalogue of \citet{Maggi16} and out of which two are not detected, and Sect.\,\ref{canNotConf} presents 34 previous candidates, which are still not confirmed as SNRs after the eRASS:4 analysis.
All confirmed (both previously known and newly confirmed in this work) SNRs in the LMC are listed in Table \ref{Tab:TotalMCSNR}.

\subsection{SNRs discovered and confirmed with eROSITA}\label{erositasnrs}

\def\namesOne{{J0456-6533}/{MCSNR J0456--6533},{J0506-7009}/{MCSNR J0506--7009},{J0543-6624}/{MCSNR J0543--6624}}
\def\labelname{{J0456-6533},{J0506-7009},{J0543-6624}}

\begin{figure*}[h]
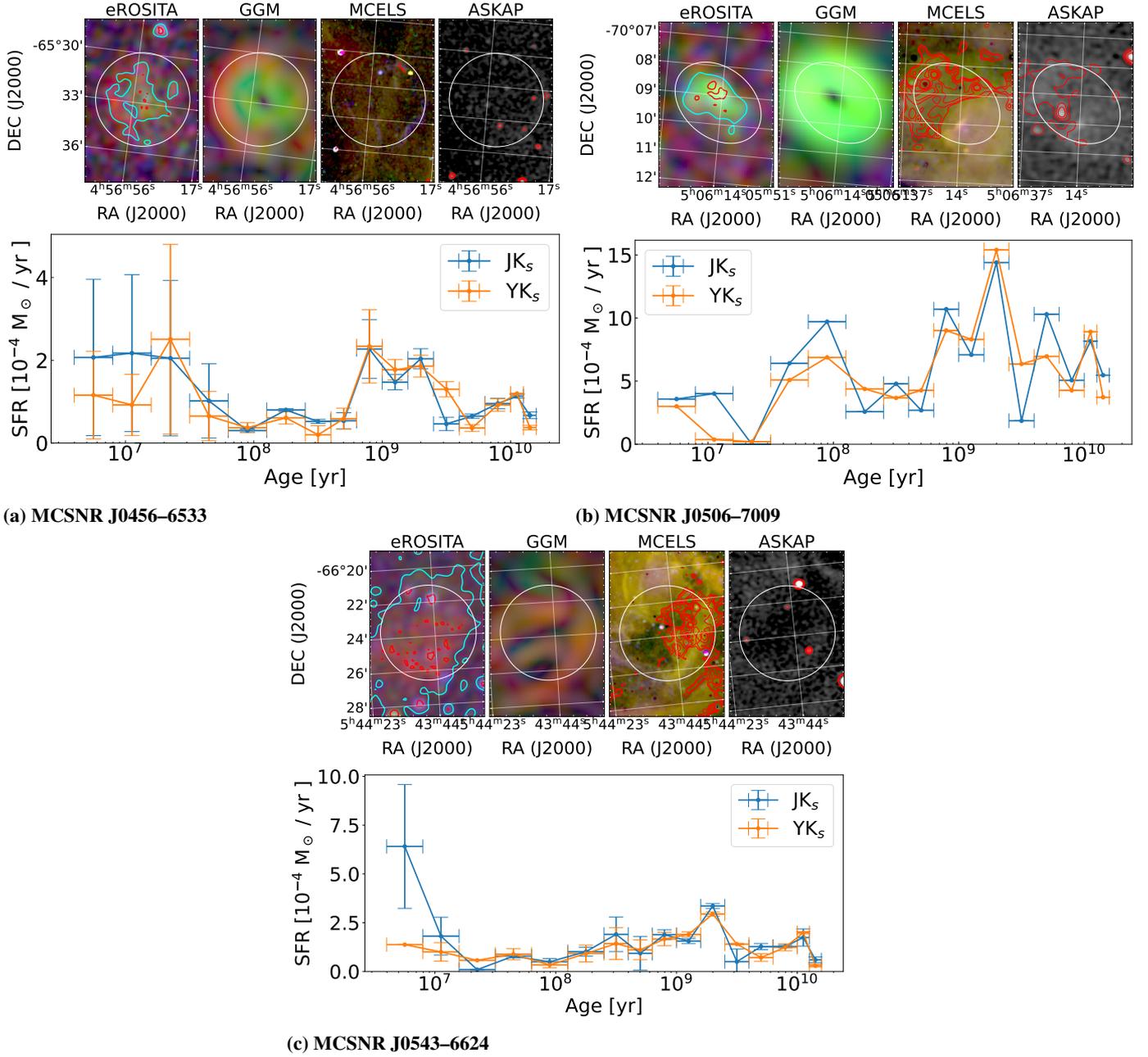

        \centering
        \foreach \i/\j in \namesOne { 
            \begin{subfigure}{0.49\textwidth}
            \centering
                 \includegraphics[width=\textwidth]{Images/Subplots/Ima\i.pdf}
                 \includegraphics[width=\textwidth]{Images/SFH/SFH_\i_100pc.pdf}
                 \caption{\textbf{\j}}
                 \label{\i}
        \end{subfigure}
         }
    \caption{For each source we show the eROSITA count rate three-colour image (left) with red: $0.2\mbox{--}\SI{0.7}{\keV}$, green: $0.7\mbox{--}\SI{1.1}{\keV}$, and blue: $1.1\mbox{--}\SI{5.0}{\keV}$, the GGM filter image (see Sect.\,\ref{section:GGM}) applied to the eROSITA count rate (middle left),  the MCELS survey three-colour image (middle right) with red: H$\alpha$, green: [\ion{S}{II}], and blue: [\ion{O}{III}], and the ASKAP radio continuum image (right) in the upper panel. The white circles (ellipses) show the extraction region for determining the count rates (see Sect.\,\ref{section:Sizes}). The cyan (red) contours in the eROSITA three-colour image show the detection at $1\sigma \ (3\sigma)$ over the background in the energy band $0.2\mbox{--}\SI{1.1}{\keV}$. The contours in the optical image represent [\ion{S}{II}]/H$\alpha > 0.67$. The contours in the radio image show the non-thermal emission calculated as described in Sect.\,\ref{radioSection}. In the lower panel we show the SFH as calculated in \cite{SFHpaper} using $J-K_\text{s}$ and $Y-K_\text{s}$ (see Sect.\,\ref{section:SFH}).}
    \label{CommonMainText}
\end{figure*}

The new SNRs first found in eROSITA and then confirmed thanks to the multiwavelength analysis are described below. The name of each source follows the convention for SNRs in the Magellanic Clouds (MC) while in brackets we report the name following the eROSITA convention. The newly found and confirmed SNRs with their X-ray properties obtained from eRASS:4 are listed in Table \ref{Tab:eROSITASNR}. 

\paragraph{\textbf{{MCSNR J0456--6533 (4eRASSU J045650.7–653244)}:}\label{text:J0456-6533}} The source is shown in Fig.\,\ref{CommonMainText} (upper left). This source was seen for the first time in the eROSITA data.  
There is a diffuse soft emission, while the emission is harder in the centre (with emission in the 0.7--1.1\,keV band) and thus appears green in the three-colour image. Even though the $3\sigma$  detection is present just in a small portion of the source we confirm the source as an SNR thanks to the \xmm\ follow-up observation described in Sect.\,\ref{section:J0456-6533}.
In the optical image, there is a shell-like structure in [\ion{O}{III}]  with a small enhancement of [\ion{S}{II}]/H$_{\alpha} > 0.67$ in the north of the shell. In the radio image, there is a very faint emission correlated with the optical shell, which surrounds the X-ray emission. 
We thus confirm the source as an SNR. 
We also show the plot of the SFH, in which there is an enhancement of star formation at around $10^9 \, \mathrm{yr}$ ago. The recent SFR has a large uncertainty, which prevents us from arguing for a recent star formation activity. Combining the information from the spectral analysis, discussed in Sect.\,\ref{section:J0456-6533} and the SFH we suggest a type Ia origin for the remnant.

\paragraph{\textbf{{MCSNR J0506--7009 (4eRASSU J050615.8–700920)}:}\label{text:J0506-7009}} 
The source is shown in Fig.\,\ref{CommonMainText} (upper right).
The source is located next to a molecular cloud known as LMC N J0506--7010 \citep{2008ApJS..178...56F}. It has a relatively small size of 98\arcsec\  $\times$ 72\arcsec.  The source shows a peak of emission in the energy band of $0.7\mbox{--}1.1\, \mathrm{keV}$.
The image shows a clear $3\sigma$ detection.
In the optical band, we can see a faint shell of [\ion{S}{II}] and a strong enhancement of [\ion{S}{II}]/H${\alpha}$ especially in the northeast. In the radio band, the continuum emission is very faint but with a non-thermal emission in the northeast, with the contours suggesting a semi-shell structure. We confidently confirm the source as a new SNR. 
The SFH shows a peak at around $10^9 \, \mathrm{yr}$ ago, but also a lower peak at around $10^8\, \mathrm{yr}$ ago. The SFH does not show a particular activity in the recent past. The hardness ratios $\mathrm{HR}_1 = 0.68 \pm 0.11$ and $\mathrm{HR}_2 = -0.68 \pm 0.08$ suggest that the source is located in the region populated by type Ia source in the $\mathrm{HR}_1-\mathrm{HR}_2$ diagram. Combining the HRs, the colour of the image, and the SFH we suggest a thermonuclear progenitor for the SNR.  

\paragraph{\textbf{MCSNR J0543--6624 (4eRASSU J054348.6--662351):}\label{text:J0543-6624}} The source is shown in Fig.\,\ref{CommonMainText} (bottom).
The source shows a diffuse soft X-ray emission with an irregular rectangular shape and has a  $3\sigma$ emission in the centre. In the optical band, we can see a rectangular shape similar to X-rays embedded in a \ion{H}{II} region with an enhancement of the ratio [\ion{S}{II}]/H${\alpha}$. In radio, there is also a faint rectangular shell in agreement with the optical shell, however, we do not detect any clear non-thermal emission. The SFH peaks at $10^7$ years ago, which suggests a possible CC origin for the remnant. Also in the HR diagrams, the source is in the region typical for CC SNRs. The CC origin is also in agreement with the fact that the source is embedded in a \ion{H}{II} region. 
Combining the X-ray and the optical information we confirm the source as an SNR.

\subsection{Previous MCELS candidate confirmed with eROSITA}
\label{previousCandConf}
\paragraph{\textbf{MCSNR {J0454--7003}:}\label{text:J0454-7003}} The source is shown in Fig.\,\ref{J0454-7003} and the X-ray property reported in Table \ref{Tab:PreviousCandConfi}. This source was proposed by \cite{Yew21} as an optical SNR candidate. It is located on the southeast edge of the \ion{H}{II} region LHA 120-N 185 overlapping half with the \ion{H}{II} region \citep{1976MmRAS..81...89D,2012ApJ...755...40P}.
In the optical band, the candidate shows a circular structure where the emission is dominated by [\ion{S}{II}] and H${\alpha}$. The ratio  [\ion{S}{II}]/H${\alpha}$ is $>$ 0.67 in the southern part of the shell and where the source coincides with the \ion{H}{II} region. In the radio image, there is diffuse emission at the position, however, there is no significant detection. In X-rays, there is diffuse emission in the \ion{H}{II} region and $3\sigma$ emission at the position of the optical SNR candidate. As this source fulfils the optical ([\ion{S}{II}]/H${\alpha} >$ 0.67) and X-ray ($3\sigma$ detection) criteria we confirm this source as an  SNR.

\subsection{SNR candidates detected with eROSITA}
In the eROSITA data, we detected 16 new diffuse sources in the X-ray images. Among them, we are able to confirm three as SNRs as presented in Sect.\,\ref{erositasnrs}. We present the 13 new SNR candidates detected with eROSITA for the first time in Appendix \ref{eROSITACandidate}. Among the new eROSITA candidates, we emphasize the peculiarity of 4eRASSU J061438.1--725112 (J0614--7251) described in the following section. This source is the first X-ray SNR candidate detected in the outskirts of the LMC. The X-ray properties are reported in Table \ref{Tab:eROSITACandidate}, while the images are shown in Fig.\,\ref{eROSITACandi_images}.

\begin{figure*}[h]
     \centering
     \begin{subfigure}[]{0.49\textwidth}
         \centering
         \includegraphics[width=\textwidth]{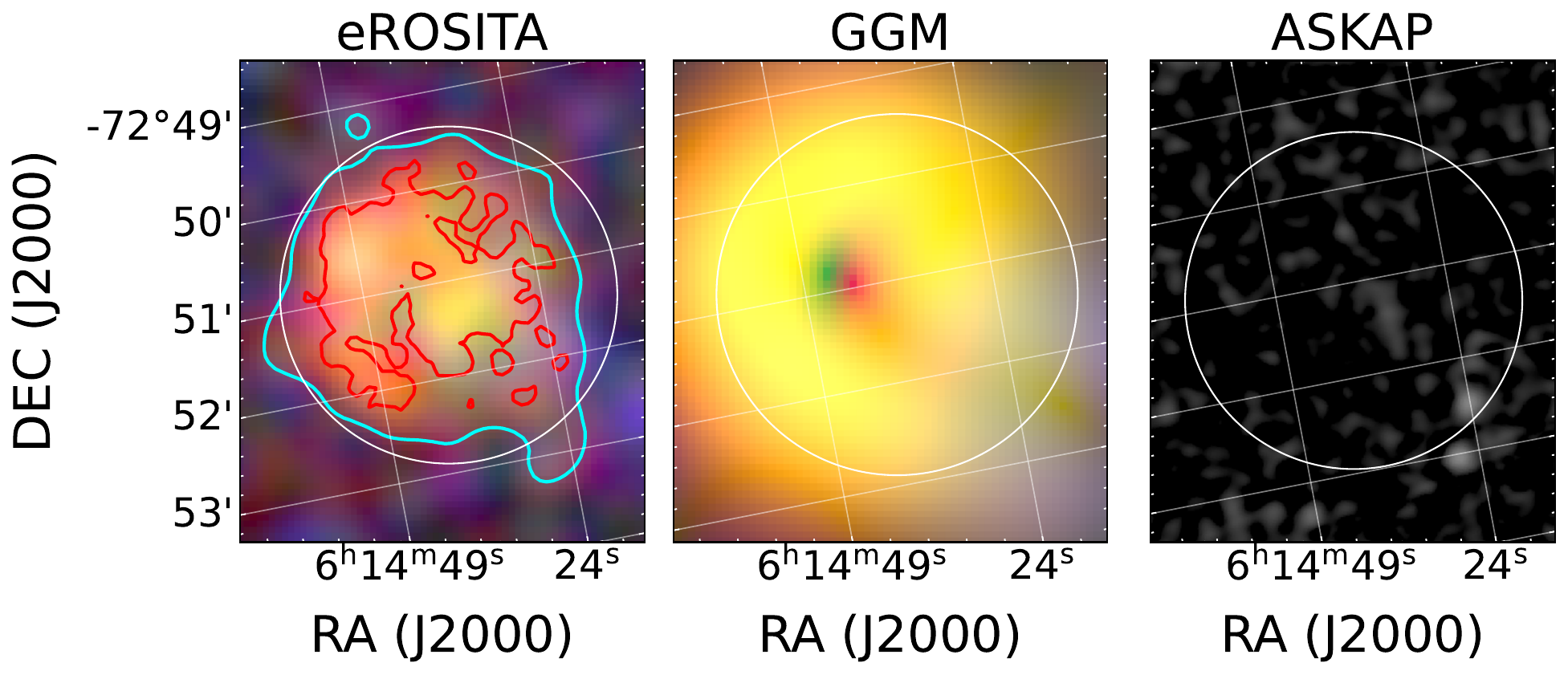}
         \caption{\textbf{J061438.1$-$725112}}
         \label{J0614Rate}
     \end{subfigure}
     \begin{subfigure}[]{0.49\textwidth}
         \centering
         \includegraphics[width=\textwidth]{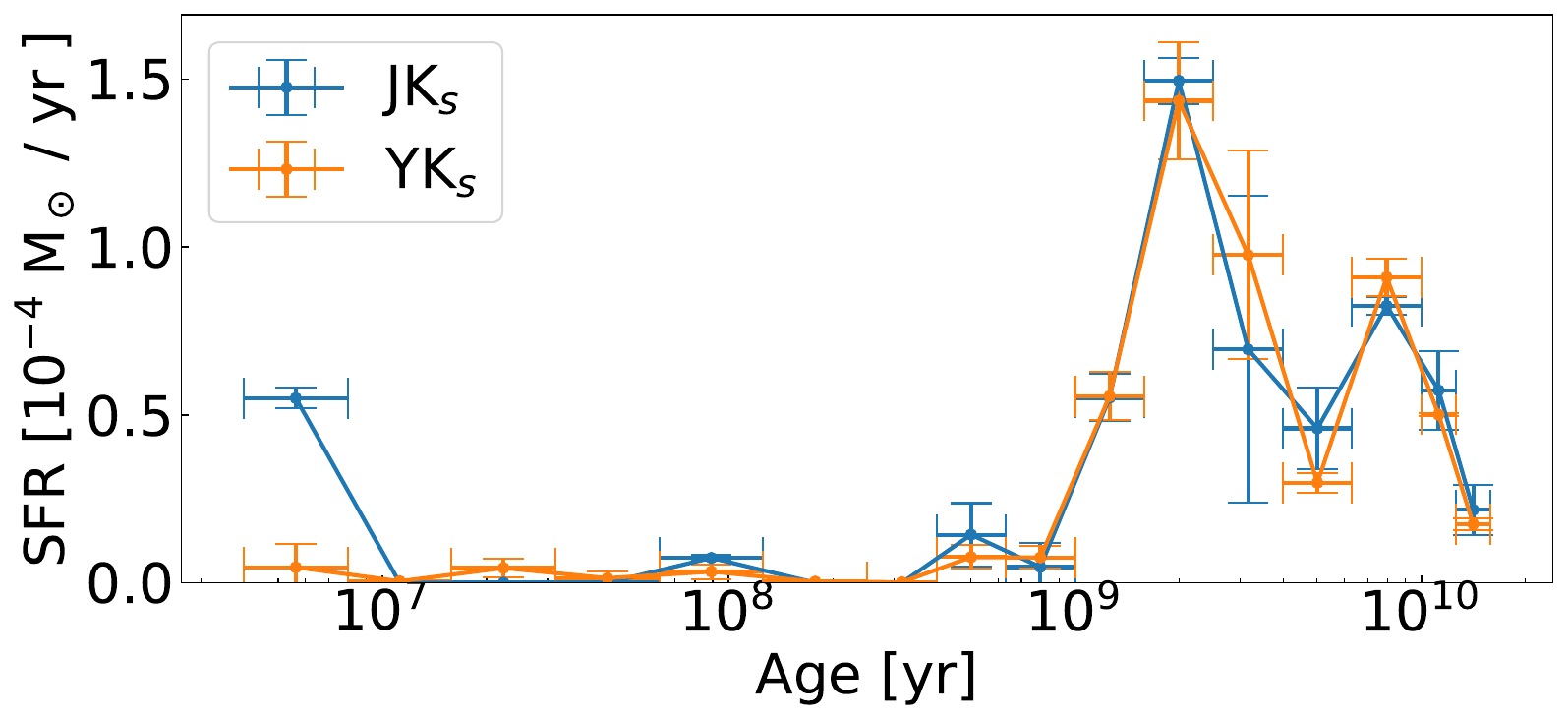}
         \caption{\textbf{SFH of J061438.1$-$725112 }}
         \label{J0614SFH}
     \end{subfigure}
     \caption{\textbf{(a)} eROSITA count rate three-colour image of J061438.1$-$725112 (left) with red: $0.2\mbox{--}\SI{0.7}{\keV}$, green: $0.7\mbox{--}\SI{1.1}{\keV}$, and blue: $1.1\mbox{--}\SI{5.0}{\keV}$, the GGM filter image (see Sect.\,\ref{section:GGM}) applied to the eROSITA count rate (middle left), and the ASKAP radio continuum (right). The white circle shows the extraction region for determining the count rates (see Sect.\,\ref{section:Sizes}) The cyan (red) contours in the eROSITA three-colour image show the detection at $1\sigma \ (3\sigma)$ over the background in the energy band $0.2\mbox{--}\SI{1.1}{\keV}$. The contours in the optical image represent [\ion{S}{II}]/H$\alpha > 0.67$. The contours in the radio image show the non-thermal emission calculated as described in Sect.\,\ref{radioSection}. The \textbf{(b)} panel shows the SFH as measured in \cite{SFHpaper} using $J-K_\text{s}$ and $Y-K_\text{s}$ (see Sect.\,\ref{section:SFH}).}
\end{figure*}

\paragraph{\textbf{{4eRASSU J061438.1--725112 (J0614--7251)}:}\label{text:J0614-7251}} The source is shown in Fig.\,\ref{J0614Rate}.
This source is of particular interest as it is not located in the LMC but far out in the southeast, also outside of the Magellanic Bridge (MB). The source was discovered in the eROSITA images because of its relatively bright appearance with an approximately spherical shape. The region is not covered by the MCELS survey, therefore we do not have any optical data. The region was covered in the ASKAP survey of the LMC, but it is on the edge of the image where the radio sensitivity is lower and does not allow us to detect anything in the radio band.  
In X-ray, the source has a shell-like structure in the soft X-rays in the east. The inner part appears slightly harder with an orange colour. 
The source shows a prominent emission above the $3\sigma$ level. The net count rate is $(2.76 \pm 0.12)\times 10^{-1}$ counts s$^{-1}$. As there is no optical or radio confirmation, this source is an SNR candidate.
In the SFH in Fig.\,\ref{J0614SFH} there are no recent peaks in star formation, therefore, ruling out the CC origin for this SNR candidate. The circular shape also suggests that the source could have originated from a Type Ia SN \citep{lopez2009astrophysical,Lopez_2011}.
Interestingly, the HR diagram suggests that the source might be related to a CC SN. Further investigations are needed to confirm or rule out the SNR nature of this source and, in the former case, to correctly classify the origin of the SNR candidate. We stress the peculiarity of the spatial location of the source. If this source is confirmed to be an SNR it will be the first SNR detected outside a galaxy in the Magellanic System. The lack of multi-wavelength information does not allow us to confirm the source as an SNR, but it remains a very promising SNR candidate.

\subsection{Known SNRs not included in \citet{Maggi16}, but observed with eROSITA}
The catalogue of \citet{Maggi16} contained 51 confirmed SNRs. Here we summarise the SNRs, which were confirmed in later studies. The multiwavelength description of these sources are given in Appendix \ref{KnownSNRnotMaggi}. The eROSITA X-ray properties are reported in Table \ref{Tab:SNRNotInMaggi}. The images are shown in Fig. \ref{appendix:SNRNotInMaggi}. 

\subsection{Previous candidates which remain candidates}
In Appendix \ref{canNotConf} we describe the previously known candidates which could not be confirmed using the eROSITA data. The respective X-ray properties are listed in Table \ref{Tab:PreviousCandRemainCand} and the images are shown in Fig.\,\ref{CandiRemainCandi_images}.

\section{\xmm\ observations of MCSNR J0456--6533} \label{section:J0456-6533}
\begin{figure}[h]
\centering
\includegraphics[scale=0.3]{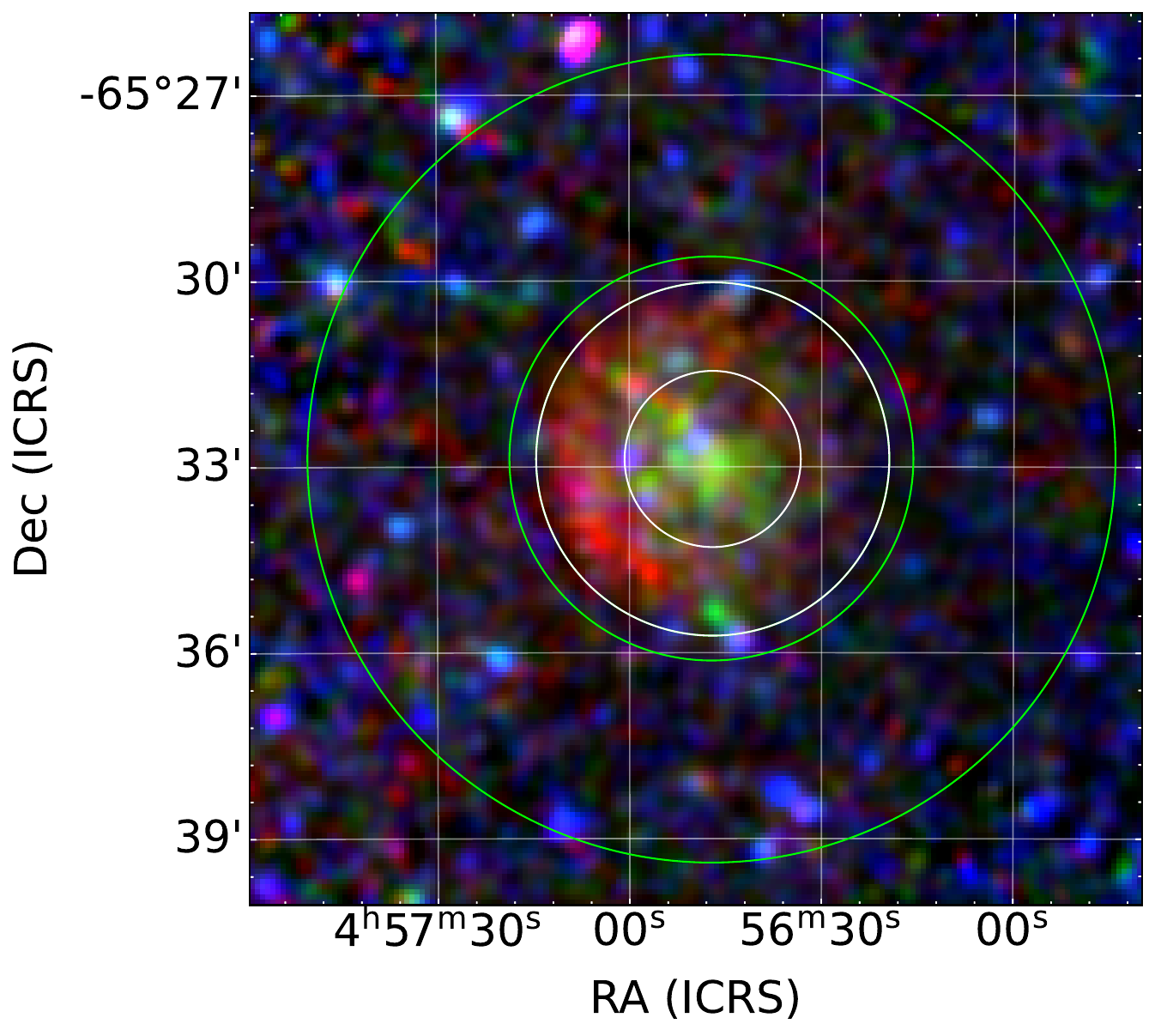}
\caption{Three-colour image of J0456--6533 with red for $0.3\mbox{--}0.7\,\mathrm{keV}$, green for $0.7\mbox{--}1.1\,\mathrm{keV}$, and blue for $1.1\mbox{--}4.5\,\mathrm{keV}$. The inner white circle and the white annulus around it mark the two source extraction regions. The green annular region around the SNR shows the background extraction region.}
\label{J0456-6533XMM}
\end{figure}

The source was detected for the first time in the eROSITA survey and proposed for an \xmm\ follow-up observation. The original exposure time was ${\sim} 45.0\,\mathrm{ks}$, while after the flare removal, the exposure time was reduced to 33.7 ks. Figure\,\ref{CommonMainText} shows the eROSITA X-ray image and the comparison with emission at other wavelengths. 
As described in Sect.\,\ref{text:J0456-6533}, in the optical there is a shell in  [\ion{O}{III}].
In the radio continuum, there is also a very faint shell which coincides with the shell in the optical. In the X-ray three-colour image the SNR has a green central region suggesting a peak in the emission around $\SI{1.0}{\keV}$ which is typically due to the presence of Fe L lines and characteristic for a type Ia SN. The remnant also presents a softer ring in the outer region, which is mainly visible in the east. 
The colours suggest that we are observing ejecta that have been heated by the reverse shock.
Figure\,\ref{J0456-6533XMM} shows the \xmm\ three-colour image in which the same structures as in the eROSITA image are visible. From the radius of the remnant (${\sim} \SI{37}{pc}$), assuming an pre-explosion number density of \SI{0.1}{\cm^{-3}} and an explosion energy of $10^{51} \, \mathrm{erg}$ we estimate the age to ${\sim} 42 \ [(E/10^{51} \,\text{erg})/(n/0.1\, \text{cm}^{-3}]^{-1/2} \,\mathrm{kyr}$, assuming a Sedov expansion \citep{Borkowski_2001}. \par 
We performed a spectral analysis of the source using the \xmm\ data. We chose an interior circular region covering the emission appearing in green in the three-colour image around ${\sim} 1 \, \mathrm{keV}$, an outer annulus covering the soft shell, and a circle covering the entire source (white circles in Fig.\,\ref{J0456-6533XMM}). In particular, we are interested in understanding the origin of the different colours in the X-ray three-colour image between the inner and outer regions. The spectrum of an additional background region is taken from a ring around the source  (green annulus in Fig.\,\ref{J0456-6533XMM}).\par

We used XSPEC (version 12.13.0 c) and AtomDB (version 3.0.9) to analyse the spectra.  
We performed a combined fit using data from EPIC-pn, -MOS1, and -MOS2 cameras. 
To model the background we used different contributions \citep{2008A&A...478..615S}: the Local hot bubble (LHB) emission modelled as a non-absorbed thermal plasma with a rather low temperature of $kT = \SI{0.1}{\keV}$,
the Galactic halo modelled as two absorbed thermal components and an extragalactic component caused by the unresolved AGN in the background modelled by a power law.
\par
For the thermal components of the background, we use the APEC model for thermal plasma in collisional ionisation equilibrium \citep{Smith_2001}. For the LHB, the temperature is fixed to $kT = \SI{0.1}{\keV}$. The Galactic halo emission consists of a `hot' component with a typical temperature of $kT = 0.3\mbox{--}\SI{0.8}{\keV}$ and a `cold' component with $kT = 0.1\mbox{--}\SI{0.3}{\keV}$.  
For the absorption of the halo emission, the Galactic column density in the direction of the source was assumed \citep{Dickey-Lackman90}. 
For the emission of the unresolved extragalactic background, we used an absorbed power-law model with a fixed photon index of $\Gamma = 1.46$ \citep{Lumb_2002,DeLuca,Moretti}. 
This component is absorbed not only by the Galactic absorption but also by the material inside the LMC.
Therefore, an additional absorption component was included and allowed to vary during the fit. The normalization of all the components was free to vary during the fit. We also considered Solar wind charge exchange (SWCX), which can have an important contribution at low energies and was modelled with six Gaussian functions with a line width of zero. The SWCX lines modelled are the following:
\ion{C}{VI} $(\SI{0.46}{\keV})$, \ion{O}{VII} $(\SI{0.57}{\keV})$, \ion{O}{VIII} $(\SI{0.65}{\keV})$, \ion{O}{VIII} $(\SI{0.81}{\keV})$, \ion{Ne}{IX} $(\SI{0.92}{\keV})$, \ion{Ne}{IX} $(\SI{1.02}{\keV})$ and \ion{Mg}{VI} $(\SI{1.35}{\keV})$ \citep{Snowden_2004}. \par
The particle background was modelled assuming a power law. Since the particle background does not interact with the mirrors of the telescope, the power law was not folded with the ancillary response file (ARF) of the instrument. The fit of the background spectrum yielded $\Gamma = 0.15$. 

To take the instrumental background into account, we added Gaussian lines for the instrumental lines Al K$\alpha$ at $\SI{1.48}{\kilo\eV}$ and Si K$\alpha$ at $\SI{1.75}{\kilo\eV}$ for EPIC-MOS1/MOS2. For EPIC-pn  the Al K$\alpha$ line as well as four additional lines at $\SI{7.49}{\keV}$, $\SI{8.05}{\keV}$, $\SI{8.62}{\keV}$, and $\SI{8.90}{\keV}$ need to be considered. \par
\begin{figure}[h]
\centering
\includegraphics[scale=0.4]{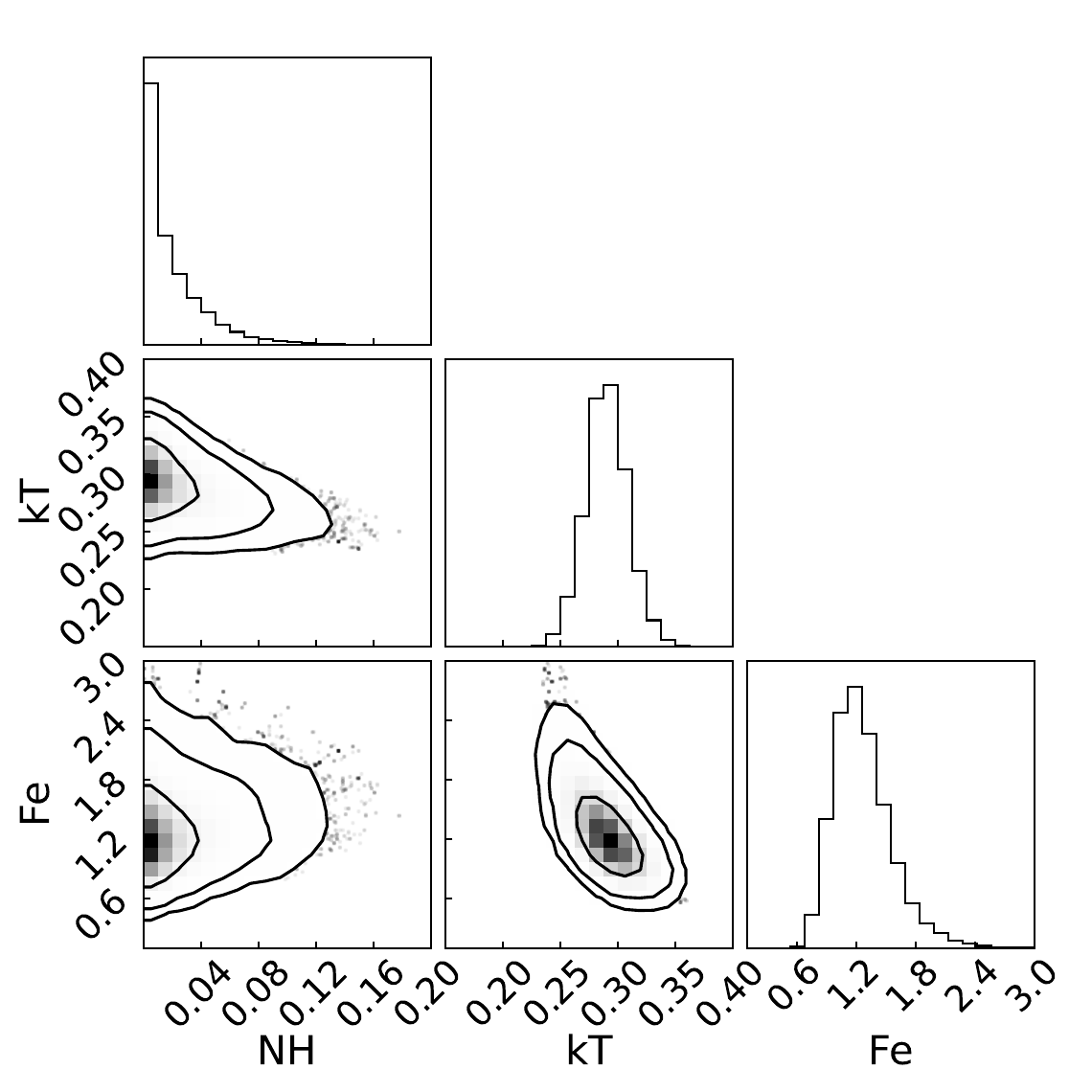}
\caption{Contour plots for the fitted parameters in the inner region of the remnant. The plotted contours are at 1, 2, and $3\sigma$. From the contours, we measure the enhancement of Fe, which is always larger than $0.5 \,\mathrm{Z}_{\odot}$, i.e., the average abundance of ISM elements in the LMC.}
\label{J0456-6533XMM_inner_contours}
\end{figure}
\begin{figure}[h]
        \centering
         \includegraphics[width=.59\textwidth, trim = 0 37 0 80, clip]{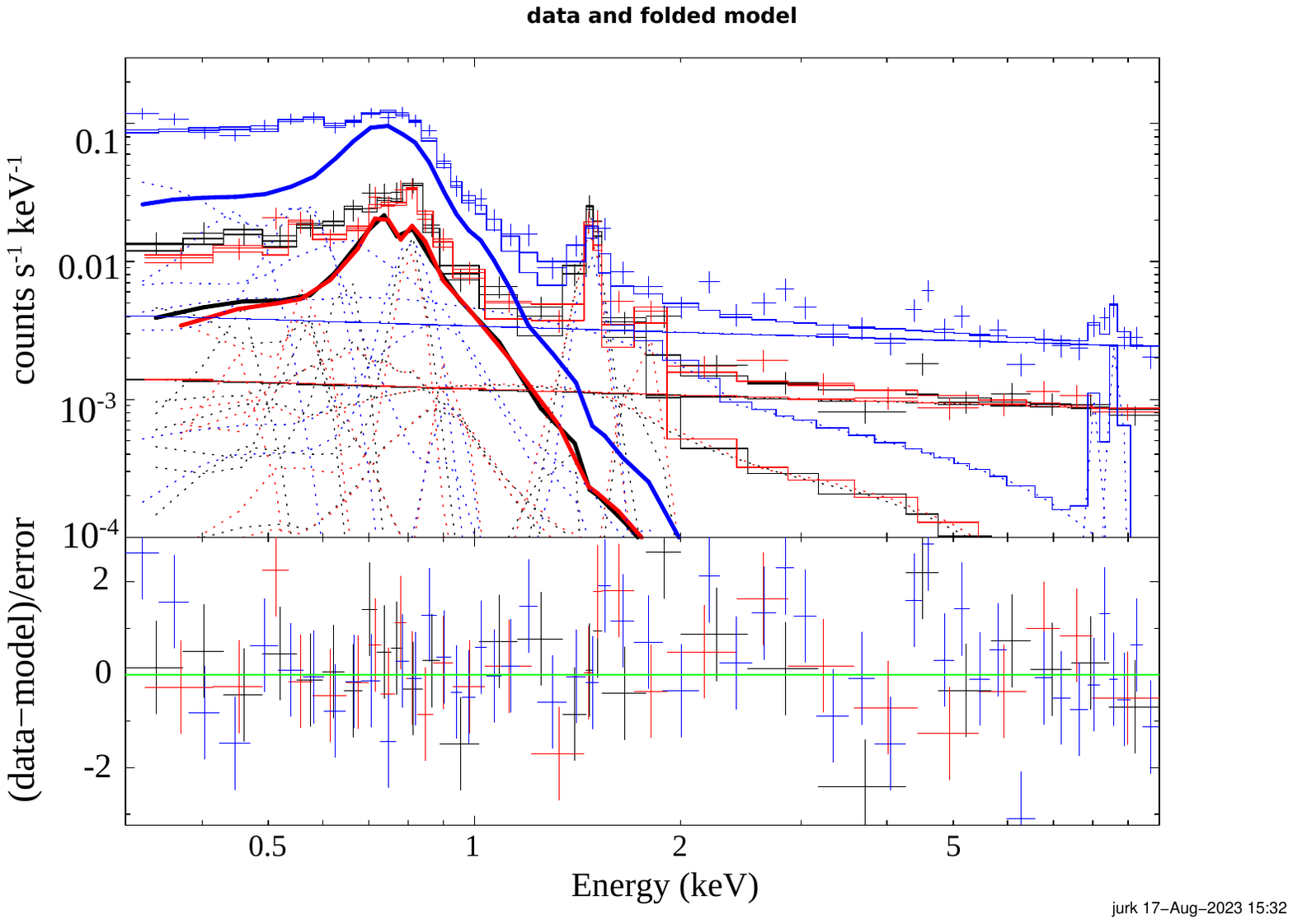}
        \includegraphics[width=.59\textwidth, trim = 0 20 0 80, clip]{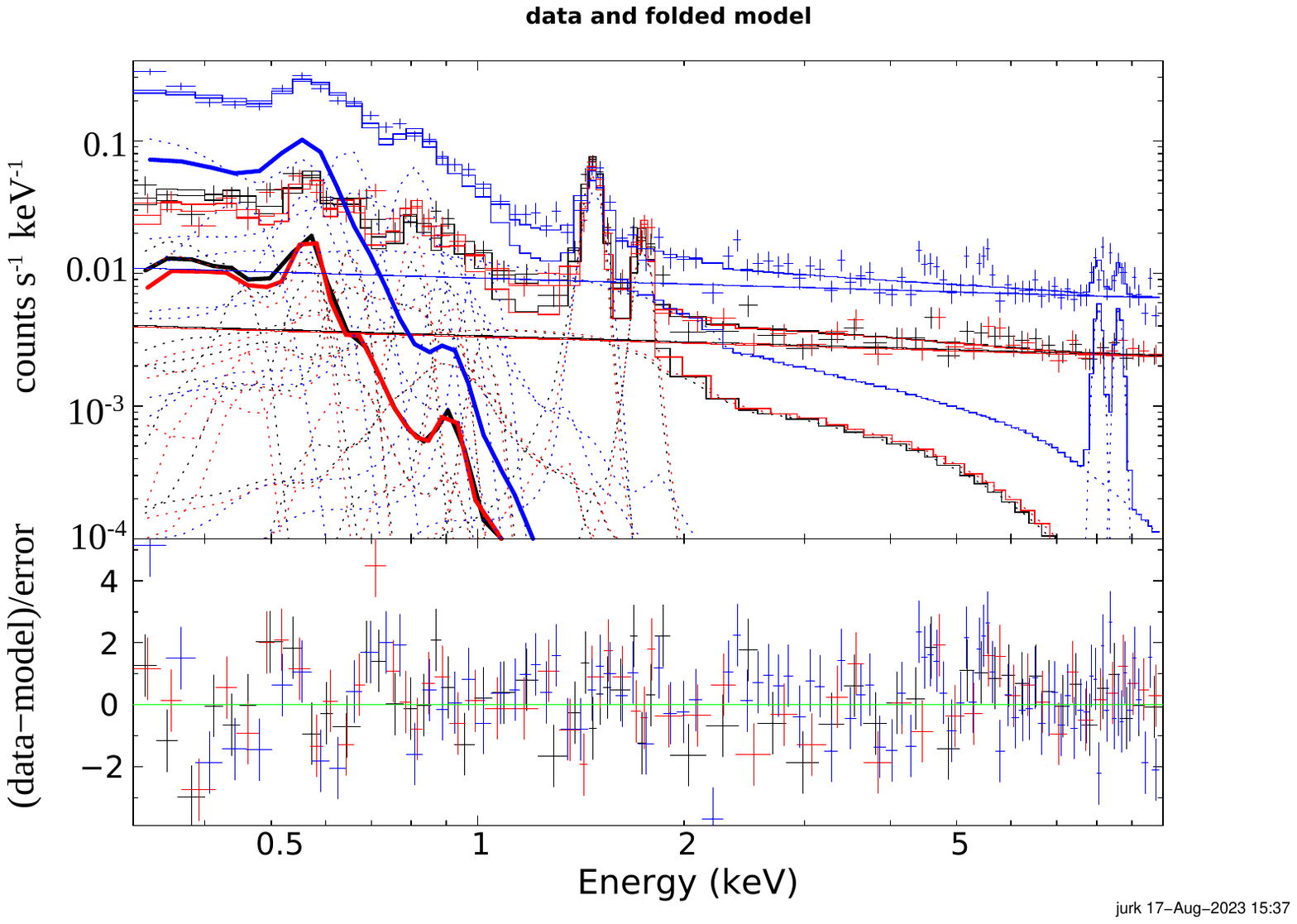}
    \caption{\xmm\ EPIC spectra of the inner circular region (top) and outer shell (bottom) of SNR J0456--6533. We plot the source spectrum (MOS1: black, MOS2: red, pn: blue) and the best-fit models. The thick solid lines show the contribution of the VNEI source emission component, which represents the emission of the SNR. All additional lines represent various background components as described in Section \ref{section:J0456-6533}.} 
    \label{fig:J0456-6533_spectra}
\end{figure}

For the fit of the source spectrum, the local X-ray background was not subtracted, but all background spectrum components were included in the model. The parameters of the background spectrum were fixed to the best-fit values, and the background spectrum was scaled by the ratio of the areas in the source and background regions.
For the source emission, we used one absorbed NEI plasma model
with variable element abundances VNEI. The Galactic column density towards the source was frozen to the value measured by \citet{Dickey-Lackman90}, while we included an additional absorption component for the LMC with the average LMC abundances of 0.5 times the solar value \citep{Westerlund97}. This absorption column density was a free parameter during the fit. 
We used the same model for all source extraction regions. 
First, we froze all the abundances in the emission component to $0.5$.
In the inner, outer, and entire regions, the fitted absorption in the LMC is very low ($N_\text{H} < 0.07 \times 10^{22}$ \SI{}{\cm^{-2}}) and is not constrained as shown in Fig.\,\ref{J0456-6533XMM_inner_contours}. We get an upper limit of $0.07 \times 10^{22} \,\mathrm{cm}^{-2}$ at the $90\%$ confidence level. 
Additionally, no difference in the LMC absorption between the outer and the inner regions is observed. We conclude that the source is not significantly absorbed by material in the LMC. Spectra are shown in Fig. \ref{fig:J0456-6533_spectra}. \par 
The ionisation time scale $\tau$ for the inner region is ${>} 10^{13}\,\mathrm{s}\,\mathrm{cm}^{-3}$, which indicates that the plasma is in ionisation equilibrium. By contrast, for the outer part, we have two statistically equivalent models: the first one presents a low temperature (${\sim} \SI{0.12}{\keV}$) and a large $\tau$ (${\sim} 10^{13}\,\mathrm{s}\,\mathrm{cm}^{-3}$), whereas the second model suggests a low $\tau$ (${\sim} 10^{10}\,\mathrm{s}\,\mathrm{cm}^{-3}$) and a higher temperature (${\sim} \SI{0.33}{\keV}$). The low counts of the outer region prevent us from constraining the best-fit model with enough statistical significance. With a measured radius of $\sim$37 kpc (assuming a distance of 50 kpc), the SNR is rather evolved and the forward shock has decelerated significantly. Therefore, it is likely that the shock temperature is low and the shocked ISM is close to ionization equilibrium (high $\tau$).
We point out that in the outer region, we observe most of the emission in the East, while in the West the outer emission is particularly faint. This could be due to a gradient in the density of the ISM surrounding the SNR (the X-ray emission depends on the density squared). Assuming that the outer ring emission is from the shocked ISM we fit the VNEI component for the source emission with all the abundances frozen to $0.5 \,\mathrm{Z}_{\odot}$. The best-fit temperature is lower than the temperature measured in the inner region. We conclude that the softer emission in the outer region is due to the lower temperature of the shocked ISM.   
In general, the temperature in the inner part is higher than the temperature measured in the outer part. The fitted temperatures are reported in Table \ref{tab:J0456-6433_fitvalues}. The uncertainties of all parameters listed were calculated using the XSPEC command \texttt{steppar} within a confidence interval of $90 \%$. \par
The green colour of the inner part is probably due to a combination of a higher temperature plasma with respect to the outer part, and a difference in elemental abundances.  
To investigate this possibility we let the Fe abundances free to vary for all regions. 
The fit results in a high Fe abundance for the inner region (see Table \ref{tab:J0456-6433_fitvalues}). We also tried to fit the Fe abundance in the outer region but it was not constrained, probably due to the lack of photon statistics.
\par
In the inner part, we observe an enhancement of Fe abundance. Despite this fact, the Fe abundance and the temperature tend to be slightly degenerate (see Fig.\,\ref{J0456-6533XMM_inner_contours}). We observe that the Fe value is always higher than $0.5 \,\mathrm{Z}_{\odot}$, indicating that the green colour in the inner region is most likely due to Fe emission of the ejecta. This suggests that the ejecta have been heated by a reverse shock that occurred in the past. In addition, the high abundance of Fe suggests that the SNR has a type Ia SN origin. Figure \ref{fig:J0456-6533_spectra} shows the spectra of the inner and outer regions.
In order to compare the \xmm\ and eROSITA observations we calculated the luminosity using the best-fit model for the entire region of the remnant obtained by fitting the same model to the \xmm \ and the eROSITA data.
Also for the entire remnant we obtain a high $\tau$ value (${\sim} 10^{13}\,\mathrm{s}\,\mathrm{cm}^{-3}$) and low foreground absorption. The fitted temperature is (${\sim} \SI{0.22}{\keV}$) which is between the hot temperature of the inner part the the cold temperature of the outer ring (see Table \ref{tab:J0456-6433_fitvalues}).
Because the eROSITA spectrum has fewer counts, we are not able to model the emission of the entire remnant with two separate VNEI components for the outer shell and the inner part. For this reason, we compared the eROSITA and spacially integrated \xmm\ data by fitting each with a single VNEI component where all abundances 
were set to $0.5 \,\mathrm{Z}_{\odot}$. This is justified as the emission of the outer shell dominates the emission of the remnant. Due to the lack of statistics we used the \texttt{cstat} implemented in \texttt{Xspec} and we used the ungrouped spectra. Also with eROSITA the best-fit temperature for the entire remnant is (${\sim} \SI{0.22}{\keV}$) and the foreground absorption is low (see Table \ref{tab:J0456-6433_fitvalues}). The ionization time scale is not constrained and indicates the ionization equilibrium of the plasma ($\tau {\sim} 10^{13}\,\mathrm{s}\,\mathrm{cm}^{-3}$). 
For consistency, we compare the luminosity determined with \xmm\ and with eROSITA. 
We calculated the absorbed flux using the \texttt{flux} command in XSPEC removing the contribution from the background. 
We assumed a distance of $50 \, \mathrm{kpc}$ \citep[][]{2014AJ....147..122D} and convert the resulting flux to luminosity. 
The value of the luminosity for the \xmm\ data is averaged over all three detectors EPIC-MOS1, MOS2, and pn and over TM 12346 for eROSITA. The uncertainties were calculated using Markov Chain Monte Carlo (MCMC) for the best-fit model.
For \xmm\ we find the luminosity to be
$L_\text{X, XMM} = 4.1^{+0.2}_{-0.2} \times 10^{34}\, \mathrm{erg} \, \mathrm{s}^{-1}$ while the eROSITA luminosity is $L_\text{X, eROSITA} = 4.9^{+0.3}_{-0.5} \times 10^{34}\, \mathrm{erg} \, \mathrm{s}^{-1}$ both in the energy band $0.3\mbox{--}8.0\,\mathrm{keV}$. Both uncertainties correspond to 90\% confidence intervals. The possible inconsistency might be caused by the lower spatial resolution of eROSITA combined with the low photon statistics for the SNR, which makes it difficult to define the source region and the difficult estimation of the local X-ray background in particular with \xmm.

    \begin{table}[h]
       \caption{Fit values for the \xmm \ observation of MCSNR J0456--6533.}
        \renewcommand\arraystretch{1.2}
        \begin{tabular}{@{} p{.07\textwidth}@{} p{.095\textwidth}@{}
        p{.075\textwidth}@{} 
         p{.075\textwidth}@{} p{.085\textwidth}@{} p{.053\textwidth}@{} p{.04\textwidth}@{} @{}}\toprule\midrule
            Region 
            & $\text{N}_\text{H}$ [$10^{22} \,\mathrm{cm}^{-2}$] 
            & $kT$ \newline [keV] 
            & Fe(=Ni) [Z$_{\odot}$] &  EM [$10^{57} \,\mathrm{cm}^{-3}$] & d.o.f.\tablefootmark{a} & $\chi^2_\text{red.}$ \\\midrule
             inner 
             & \multirow{3.5}{1cm}{\centering $< 0.07$}
             & $0.30^{+0.02}_{-0.04}$ 
            & $1.16^{+0.68}_{-0.30}$ & $1.42^{+0.18}_{-0.23}$ & 552 & 1.38 \\ [0.1cm]
             outer 
             &  
             & $0.12^{+0.02}_{-0.01}$  
            & $0.5$ & $6.76^{+2.87}_{-2.10}$ & 663 & 1.47\\ [0.1cm]
             entire 
             & 
             & $0.22^{+0.01}_{-0.01}$ 
             & $0.5$
             & $8.04^{+0.70}_{-0.60}$ & 707 & 1.67 \\ [0.1cm]
            eR\tablefootmark{c} 
             & $<0.01$
             & $0.22^{+0.01}_{-0.01}$ 
             & $0.5$
             & $8.70^{+0.10}_{-0.10}$ & 4905 & 0.72\tablefootmark{b} \\
             \bottomrule
        \end{tabular}
        \tablefoot{The uncertainties correspond to the $90\%$ confidence intervals of the fit parameters calculated with \texttt{steppar} in XSPEC. The last row shows the fit values of eROSITA (see discussion in Section \ref{section:J0456-6533}).
        \tablefoottext{a}{degrees of freedom.} 
        \tablefoottext{b}{reduced C-statistic}
        \tablefoottext{c}{eROSITA}
        }
        \label{tab:J0456-6433_fitvalues}
    \end{table}


%

\section{Conclusions}
We investigated the SNR population in the LMC using the \SRG/eROSITA data, which provides a complete look at the LMC in the soft to medium X-ray band. The large field of view of eROSITA and the full coverage in eRASS:4 allowed us to study known SNRs and candidates and find new SNRs and candidates. 
In general, we used X-ray 1-$\sigma$ detection to identify new SNR candidates. We compared the [\ion{S}{II}]/H${\alpha}$ line ratios and the non-thermal radio emission to investigate the true nature of the source. The confirmation of a new SNR required at least two of the following criteria: 3$\sigma$ detection in X-ray, [\ion{S}{II}]/H${\alpha}>0.67$, and a non-thermal radio diffuse emission.
   \begin{enumerate}
      \item We use the above-described multi-wavelength analysis to confirm three of the eROSITA candidates as SNRs (see Sect.\,\ref{erositasnrs}).
      \item Using a $3\sigma$ threshold for the X-ray emission we were able to confirm 1 previously known candidate as an SNR: MCSNR J0454--7003 (see Sect.\,\ref{text:J0454-7003}).
      \item We performed a Gaussian gradient magnitude filter analysis on the eRASS:4 images of the LMC to identify possible new SNR candidates. By combining the X-ray data with optical and radio data, we propose 13 sources as new X-ray SNR candidates (see Sect.\,\ref{eROSITACandidate}). Using the HRs and the SFH around each source we investigated the origin of the SNR candidates. 
      \item We propose J0614--7251 as the first X-ray SNR candidate in the outskirts of the LMC (see Sect.\,\ref{text:J0614-7251}). The source presents X-ray emission above the $3\sigma$ level with a net count rate of $(2.76 \pm 0.12)\times 10^{-1}$ cts $\mathrm{s}^{-1}$, resulting in a prominent isolated source. From the SFH  we can not rule out a CC origin for this SNR candidate.
      \item  The results summarised in points 1--4 bring the total number of SNRs in the LMC to 77 and the number of SNR candidates to 47 (see Table \ref{tab:Summary}). 
      \item We have performed a spectral analysis of the newly detected MCSNR J0456--6533 using \xmm\ data from a follow-up observation of the eROSITA detection (see Sect.\,\ref{J0456-6533XMM}). We modelled the source emission with a single absorbed VNEI model component and investigated different regions of the source. The column density in the LMC is very low $(N_\text{H} < 0.07 \times 10^{22}\,\mathrm{cm^{-2}})$ which suggests an un-absorbed source. We do not observe a significant difference in the fitted temperature between the inner part and the outer shell of the remnant. We found an enhancement in Fe abundance in the inner part, which is most likely dominated by ejecta emission and suggests a type Ia SN as the progenitor of MCSNR J0456--6533.     
      \end{enumerate}

\begin{acknowledgements}
This work was supported by the Deutsche Forschungsgemeinschaft through the project
SA 2131/14-1.
M.S. and J.K. acknowledge support from the Deutsche Forschungsgemeinschaft through the grants SA 2131/13-1 and SA 2131/15-1. P.J.K. acknowledges support from the Science Foundation Ireland/Irish Research Council Pathway programme under Grant Number 21/PATH-S/9360. The paper is based on data from eROSITA, the soft X-ray instrument aboard \SRG, a joint Russian-German science mission supported by the Russian Space Agency (Roskosmos), in the interests of the Russian Academy of Sciences represented by its Space Research Institute (IKI), and the Deutsches Zentrum für Luft- und Raumfahrt (DLR). The \SRG\ spacecraft was built by Lavochkin Association (NPOL) and its subcontractors and is operated by NPOL with support from the Max-Planck Institute for Extraterrestrial Physics (MPE). The development and construction of the eROSITA X-ray instrument were led by MPE, with contributions from the Dr. Karl Remeis Observatory Bamberg, the Erlangen Center for Astroparticle Physics (ECAP), the University of Hamburg Observatory, the Leibniz Institute for Astrophysics Potsdam (AIP), and the Institute for Astronomy and Astrophysics of the University of Tübingen, with the support of DLR and the Max Planck Society. The Argelander Institute for Astronomy of the University of Bonn and the Ludwig Maximilians Universität Munich also participated in the science preparation for eROSITA.
The eROSITA data shown here were processed using the eSASS/NRTA software system developed by the German eROSITA consortium. Part of this work was supported by the German Deutsche Forschungsgemeinschaft, DFG project number Ts 17/2–1. This work uses observations obtained with \xmm\ -- an ESA science mission with instruments and contributions directly funded by ESA Member States and NASA.
In this paper we use data from ASKAP, The Australian SKA Pathfinder is part of the Australia Telescope National Facility, which is managed by CSIRO. Operation of ASKAP is funded by the  Australian Government with support from the National Collaborative Research Infrastructure Strategy. ASKAP uses the resources of the Pawsey Supercomputing Centre. The establishment of ASKAP, the Murchison Radio-astronomy Observatory and the Pawsey Supercomputing Centre are initiatives of the Australian Government, with support from the Government of Western Australia and the Science and Industry Endowment Fund. We acknowledge the Wajarri Yamatji people as the traditional owners of the Observatory site. This work makes use of MCELS. The MCELS was funded through the support of the Dean B. McLaughlin fund at the University of Michigan and through NSF grant 9540747. 
This research has made use of Aladin, SIMBAD and VizieR, operated at the CDS, Strasbourg, France.
\end{acknowledgements}

%
%
\bibliographystyle{aa} 
\bibliography{reference} 





   
  




\FloatBarrier
\begin{appendix}
\section{SNR candidates detected with eROSITA}
\label{eROSITACandidate}
In the eROSITA data, we detected 16 new diffuse sources in the X-ray images. Among them, we are able to confirm three as SNRs as presented in Sect.\,\ref{erositasnrs}. In the following section, we present the 13 new SNR candidates detected with eROSITA for the first time. As these sources are not confirmed as SNRs, we first report the name following the eROSITA convention and in bracket the name as usually used in the SNRs catalogues. The list of the eROSITA candidates is reported in Table \ref{Tab:eROSITACandidate}

\paragraph{\textbf{{4eRASSU J045145.7--671724 (J0451--6717)}:}\label{text:J0451-6717}} 
The source is shown in Fig.\,\ref{J0451-6717}. This candidate SNR might be associated with a radio pulsar detected by \citet{Manchester06} using the Parkes radio telescope. The properties of the radio pulsar, including its position, are still highly uncertain. In our radio images, there is a point source inside J0451--6717, which can be associated with the pulsar, but further observations to measure the pulse period are required for confirmation. In the optical, there is an elongated shell structure particularly visible in [\ion{S}{II}], where the [\ion{S}{II}]/H${\alpha}$ ratio is partially higher than $0.67$. In the X-ray images, there is faint diffuse emission, which correlates with the optical emission, however, only with a $1\sigma$ detection. Only very small portions of the remnant are detected with $3\sigma$ confidence. In the GGM image, we can clearly see an edge with an elliptical shape associated with the source. The region was observed with \xmm\ but the candidate is located at the rim of the field of view, which prevented a previous detection.
Since the $3\sigma$ detection in the eROSITA data is in a very small portion we conservatively keep the source as an SNR candidate. Recently, this source was accepted for an \xmm\ follow-up observation. With a deeper observation, we will be able to further constrain the true nature of this source.

\paragraph{\textbf{{4eRASSU J045625.5--683052 (J0456--6830)}:}\label{text:J0456-6830}}
The source is shown in Fig.\,\ref{J0456-6830}.
This source was identified for the first time in the eROSITA survey. In the X-ray three-colour image it is seen as a diffuse green emission, which means that the emission peaks at around $\SI{1.0}{\keV}$. Even if the source is relatively faint with a net count rate of  $(2.90 \pm 0.78)\times 10^{-2}$ cts $\mathrm{s}^{-1}$, it is clearly visible in X-rays. The brightest part is on the southeast, as the $1\sigma$ contour shows. The source is relatively compact with a diameter of about 160\arcsec\ and has a circular shape. Also, the GGM image shows the presence of a diffuse emission. In the optical band, there is no compact emission clearly related to the candidate, but it is surrounded by a larger structure dominated by [\ion{S}{II}]. The optical line ratio [\ion{S}{II}]/H$\alpha < 0.67$ does not indicate shock emission. Neither in the radio continuum nor in the non-thermal component any emission is detectable. Therefore, the source remains a candidate. The SFH  within 100 pc around the centre of the candidate indicates a relatively high and constant star formation activity from $10^9$ to $10^8 \, \mathrm{yr}$ ago. The HRs and the colour of the image suggest a possible thermonuclear origin.

\paragraph{\textbf{{4eRASSU J050750.8$-$714241 (J0507--7143)}:}\label{text:J0507-7143}} The source is shown in Fig.\,\ref{J0507-7143}. From its green colour in the three-colour image, we can conclude that the peak of the emission is at around $\SI{1.0}{\keV}$. The presence of a diffuse source is confirmed by the GGM image. The source is relatively faint in the X-ray band with a count rate of $(5.60 \pm 1.41)\times 10^{-2}$ $\mathrm{cts} \, \mathrm{s}^{-1}$. We do not confirm any emission in the optical or radio. In the radio image, there is a non-thermal source in the north which is an AGN in the background. The same AGN is visible as a point source in the X-ray image. We propose this X-ray source as an SNR candidate. The SFH around the source position has a peak between $10^9$ and $10^{10} \, \mathrm{yr}$ ago. The region around the candidate did not face a recent SF activity. From the X-ray colour of the source and the SFH, we can argue for a type Ia origin. 

\paragraph{\textbf{4eRASSU J051028.3--685329 (J0510--6853)}:\label{text:J0510-6853}} The source is shown in Fig.\,\ref{J0510-6853}. This source is located next to a molecular cloud [WHO2011] C173 \citep{2011ApJS..197...16W}. There is a faint X-ray source, which is also clearly seen in the GGM image. In the optical band, there is a shell structure in the east.
However, the radio [\ion{S}{II}]/H${\alpha}$ is below the limit of 0.67, characteristic of an SNR. The radio image shows a shell structure on the east with a $3\sigma$ non-thermal emission, with a possible shell in the west. The SFH shows no recent star formation. The hardness ratios $\mathrm{HR}_1 = -0.29 \pm 0.15$ and $\mathrm{HR}_2 = -0.70 \pm 0.33$ correspond to values characteristic of CC SNR (see Fig.\,\ref{FigureHR}). 
Therefore, despite the SFH, we can not rule out the CC scenario. Since there is a non-thermal shell-like source in radio but no X-ray is detected at the $3\sigma$ level, the source is classified as a candidate.

\paragraph{\textbf{{4eRASSU J052136.6--670741 (J0521--6707)}:}\label{text:J0521-6707}}
The source is shown in Fig.\,\ref{J0521-6707}. In the eROSITA image, there is a very faint diffuse source with a net count rate of $(3.10 \pm 1.37)\times 10^{-2}$ cts s$^{-1}$, confirmed by the GGM image. In the optical image, these is an enhancement of [\ion{S}{II}]/H${\alpha} > 0.67$ inside the shell.  In the radio image, there is a point source detected by \citet{1995A&AS..111..311F} with the catalogue name B0521--6710 at 4.75 GHz, and by \citet{1997A&AS..126..325M} with the name MDM 15 and detected at 2.4 GHz. In the latter paper, they measure a spectral index of $-0.3$. No diffuse emission can be observed in the radio images.
The SFH around the source indicates two relative peaks between $10^7\mbox{--}10^8$ years ago. The absence of the $3\sigma$ detection in X-rays prevents us from confirming the source as an SNR, and it is classified as a candidate.

\paragraph{\textbf{{4eRASSU J052126.5--685245 (J0521--6853)}:}\label{text:J0521-6853}}
The source is shown in Fig.\,\ref{J0521-6853}.
The source was detected in the ROSAT survey \citep{1999A&AS..139..277H}. 
In the eROSITA image, it is visible at $1\sigma$ level despite the source being located in a relatively bright environment. 
The GGM image shows an enhancement of the emission relative to the surroundings. At the same location, there is a diffuse emission in the optical image with H$\alpha$, [\ion{S}{II}], and [\ion{O}{III}] emission. There is a small bright region in the centre with a diameter of about 21\arcsec, embedded in a larger structure. There is no optical emission with [\ion{S}{II}]/H$\alpha$ $>0.67$ at the position of the X-ray emission. However, it is [\ion{S}{II}]/H$\alpha$ $>0.67$ in the surroundings. In the radio band, there is non-thermal emission north of the X-ray emission, wherein the optical ratio [\ion{S}{II}]/H$\alpha$ $>0.67$.
The SFH around the source shows no recent star formation activity. 
The absence of the $3\sigma$ detection prevents us from confirming the source as an SNR, and it is classified as a candidate.

\paragraph{\textbf{{4eRASSUJ052148.7--693649 (J0521--6936)}:}\label{text:J0521-6936}}
The source is shown in Fig.\,\ref{J0521-6936}.
The source was identified for the first time with eROSITA probably because it is located in a region where the surrounding is bright in X-rays. There is a diffuse source in X-rays, which is clearly visible in particular in the GGM image. The emission is brighter in the north. It peaks at around 1\,keV as the green colour suggests. There is a very small region with $3\sigma$ emission in the northern part.
In the optical image, the source is surrounded by diffuse emission in the east and the south with only small regions with  [\ion{S}{II}]/H$\alpha > 0.67$. The X-ray peak correlates with a bubble-like structure in optical.
In the radio image, we can not see any structure that can be related to the X-ray source.  
The SFH indicates two relatively recent peaks around $4 \times 10^7 \, \mathrm{yr}$ and $10^7 \, \mathrm{yr}$ ago. The low significance of the detection in the X-ray band prevents us from confirming the source as an SNR and classifying it as a candidate.

\paragraph{\textbf{4eRASSUJ052330.7--680400 (J0523--6804)}:\label{text:J0523-6804}} 
The source is shown in Fig. \ref{J0523-6804}.
The X-ray images show a circular object with a radius of 116\arcsec. The GGM also shows the circular edge of the source. We do not observe any emission in the optical band. In the radio image, there is non-thermal emission, especially in the south. Unfortunately, there is no $3\sigma$ detection in X-rays. 
The hardness ratios $\mathrm{HR}_1 = 0.04 \pm 0.21$ and $\mathrm{HR}_2=-0.89 \pm 0.32$ locate the source in the region in the HR$_1$--HR$_2$ diagram (Fig.\,\ref{FigureHR}) populated by thermonuclear SNRs. The green colour in the X-ray image shows a peak in the emission around 1\,keV typical of thermonuclear origin. The SFH around the source has a peak in the recent past around $10^6 \, \mathrm{yr}$. This star formation activity would be compatible with a possible CC origin but from the position of the source in the HR diagram \ref{FigureHR} and from the colour we propose the candidate to originate from an SN Ia. The absence of a $3\sigma$ detection in the X-ray image prevents us from confirming the source as an SNR. For this reason, we propose this source as an SNR candidate.

\paragraph{\textbf{4eRASSUJ052502.7--662125 (J0525--6621)}:\label{text:J0525-6621}} 
The source is shown in Fig. \ref{J0525-6621}.
The source has an irregular shape in the X-ray images, with a softer emission in the northeast. There is some $3\sigma$ emission inside the source, but its distribution is not sufficient to confirm the source as an SNR.  In the GGM image, we can observe a clearly enhanced edge. In the optical image, there is a strong enhancement of [\ion{S}{II}]/H$_{\alpha} > 0.67$ while in the radio image, there is a complex environment where the emission appears to be non-thermal. There is no structure in the radio emission that correlates with the X-ray or the optical emission. The HRs of the source indicate a type Ia SNR (see Fig.\,\ref{FigureHR}). There was quite constant star formation activity from $10^9$ until today with a relative peak in SFH between $10^8\mbox{--}10^7 \, \mathrm{yr}$ ago. We propose this source as a candidate.

\paragraph{\textbf{{4eRASSU J052849.7--671913 (J0528--6719)}:}\label{text:J0528-6719}}
The source is shown in Fig.\,\ref{J0528-6719}.
The source is detected at 1$\sigma$ level with a soft diffuse emission between $0.2 \mbox{--} 0.7 \, \mathrm{keV}$ and thus appears red in the eROSITA images. In X-rays, it has a half-shell structure in the south, which is also visible in the GGM image. There is no emission that can be related to the X-ray source, neither in the optical nor in the radio. The SFH has a recent peak, which might suggest a CC origin. 

\paragraph{\textbf{{4eRASSU J053224.5$-$655411 (J0532--6554)}:}\label{text:J0532-6554}}
The source is shown in Fig.\,\ref{J0532-6554}.
The source was detected in the ROSAT data by \cite{1999A&AS..139..277H}. Its ROSAT HRs suggested that it could be an AGN, SNR, or an X-ray binary. In the eROSITA images the source shows a diffuse emission with a count rate of $(3.80 \pm 0.50)\times 10^{-2}$  counts s$^{-1}$ and $\mathrm{HR}_1 = 0.37 \pm 0.12$, $\mathrm{HR}_2 = -0.65 \pm 0.14$. The contours in the X-ray image show a $3\sigma$ emission.
No particular structure can be observed in the optical or radio image. For this reason, we propose the source to be a candidate.
The SFH  shows a recent increase in the star formation activity, which is nevertheless not significant enough to suggest a CC origin.

\paragraph{\textbf{{4eRASSU J054949.7--700145 (J0549--7001)}:}\label{text:J0549-7001}} The source is shown in Fig.\,\ref{J0549-7001}.
This source has an elongated shape in the northeast-southwest direction with a brighter emission in the southwest. The source is relatively bright with count rates of $(5.80 \pm 0.82)\times 10^{-2}$ counts s$^{-1}$. The $1\sigma$ contours follow the diffuse structure and small portions inside have detection with significance above $3\sigma$. 
In the optical band, the region presents a complex filamentary emission, where it is not clear if any structure can be related to the candidate. No enhanced ratio of [\ion{S}{II}]/H$\alpha > 0.67$  is observed. In radio, there is no emission associated with the source either. Therefore, this source is an SNR candidate. The SFH  shows a small peak at around $5 \times 10^6 \,\mathrm{yr}$ ago. However, its X-ray colour is indicative of a thermonuclear origin.

\section{Known SNRs not included in \citet{Maggi16}, but observed with eROSITA}
\label{KnownSNRnotMaggi}
The catalogue of \citet{Maggi16} contained 51 confirmed SNRs. Here we summarise the SNRs, which were confirmed in works performed later. The eROSITA X-ray properties of the following SNRs are reported in Table \ref{Tab:SNRNotInMaggi}.
\paragraph{\textbf{{MCSNR J0447--6918}:}\label{text:J0447-6918}} 
The source is shown in Fig.\,\ref{J0447-6918}.
The source was recently confirmed by \cite{Kavanagh22} using \xmm\ data. The source was a radio candidate proposed by \cite{Bozzetto_2017}. The remnant presents a diffuse emission in the radio continuum. We detect non-thermal emissions, especially coming from the centre of the source. In the optical band, we can observe a half-ring structure in the south with high ${\mathrm{[\ion{S}{II}]/H}\alpha}$. There is some additional filamentary emission which extends from the inner part of the shell to the northeast. The optical emission correlates with the diffuse radio emission. In X-rays, we also detect an elongated faint emission, which is more visible using the GGM filter and it agrees with the shell seen in the optical and radio images. 

\paragraph{\textbf{{MCSNR J0449--6903}:}\label{text:J0449-6903}}
The source is shown in Fig.\,\ref{J0449-6903}.
This source was another radio candidate proposed by \citet{Bozzetto_2017} and confirmed by \citet{Kavanagh22}. In radio, the source presents a prominent diffuse emission. The remnant shows that there is a strong non-thermal emission which correlates with the entire diffuse emission. In the optical, there is a faint shell of [\ion{O}{III}], where the interior has an enhancement of ${\mathrm{[\ion{S}{II}]/H}\alpha}$. Neither in the eROSITA images nor in the GGM image we can detect any significant emission.

\paragraph{\textbf{{MCSNR J0456--6950}:}\label{text:J0456-6950}} 
The source is shown in Fig.\,\ref{J0456-6950}.
This is another radio candidate \citep{Bozzetto_2017} confirmed as an SNR by \cite{Kavanagh22}. The radio image presents a diffuse spherical emission with the presence of non-thermal emission in the entire remnant. Neither in the eROSITA images nor in the optical images, any clear emission can be associated with the remnant. 

\paragraph{\textbf{{MCSNR J0504--6723}:}\label{text:J0504-6723}} The source is shown in Fig.\,\ref{J0504-6723}.
This remnant was proposed as an X-ray candidate by \cite{Haberl99} and confirmed by \cite{Kavanagh22}. Also in the eROSITA images the source appears as a bright diffuse emission with a peak at around $1.0$\,keV. In radio we do not detect any non-thermal emission and the continuum does not show a structure which can be connected to the remnant. In the optical image, we can clearly see an enhancement of the ${\mathrm{[\ion{S}{II}]/H}\alpha}$  ratio, especially in the north. In the rest of the region, we can see filaments of H$\alpha$ and [\ion{S}{II}]. 

\paragraph{\textbf{{MCSNR J0506--6815}:}\label{text:J0506-6815}} 
The source is shown in Fig.\,\ref{J0506-6815}.
This source was found by \citet{Filipovich23} and confirmed as an SNR in the same study. They measured a spectral index of $\alpha \sim -0.47$ and used an \xmm\ observation, in which soft X-ray emission was detected. There is a clear circular emission in the radio with a non-thermal component. Also in the optical band, a clear shell is visible, which correlates with the radio emission. The X-ray emission is also clearly visible in the eROSITA data. 

\paragraph{\textbf{{MCSNR J0507--6847}:}\label{text:J0507-6847}} 
The source is shown in Fig.\,\ref{J0507-6847}.
The source was studied by \cite{Chu_2000} using ROSAT data. 
We can see that in  X-rays there is a large elliptical shell with a point source at its centre as can be seen from the contours. The SNR is associated with the high-mass X-ray binary pulsar XMMU\,J050722.1--684758 \citep{Maitra21}. The SNR presents an enhancement of ${\mathrm{[\ion{S}{II}]/H}\alpha}$ in the north. Also, the radio continuum shows a faint shell structure associated with the SNR. 

\paragraph{\textbf{{MCSNR J0510--6708}:}\label{text:J0510-6708}} 
The source is shown in Fig.\,\ref{J0510-6708}.
\cite{Bozzetto_2017} proposed the source as a radio candidate in particular because of the presence of a compact radio source in the centre of the emission. The source was then confirmed as an SNR by \cite{Kavanagh22}. In the radio image, only a point source can be seen. 
In the optical images, there is a faint circular structure where the emission is mainly due to [\ion{S}{II}]. Inside the diffuse emission, the ratio is [\ion{S}{II}]/H$\alpha > 0.67$ which suggests that the gas has been heated by a shock. In the eROSITA images, we can see a very faint diffuse emission with a count rate of $(2.10 \pm 0.88)\times 10^{-2}$ cts s$^{-1}$ but no significant detection. 

\paragraph{\textbf{{MCSNR J0512--6716}:}\label{text:J0512-6716}} 
The source is shown in Fig.\,\ref{J0512-6716}.
The source was proposed as a radio candidate \citep{Bozzetto_2017} and was recently confirmed as an SNR by \cite{Kavanagh22}. In the radio image, there is a diffuse elliptical emission with a brighter half-shell in the southeast. The contours in the image reveal that the emission from the bright shell is strongly non-thermal. In the optical images, there are filaments and a half shell of [\ion{O}{III}], which correlates with the radio shell. There is no enhancement of [\ion{S}{II}]/H$\alpha$ in the MCELS images. There is faint diffuse X-ray emission in the eROSITA images.

\paragraph{\textbf{{MCSNR J0513--6724}:}\label{text:J0513-6724}}
The source is shown in Fig.\,\ref{J0513-6724}.
This source was proposed as a radio candidate and was confirmed as an SNR by \citet{Maitra19} using \xmm\ data.
The source is clearly visible in the radio band as a diffuse emission with significant non-thermal emission. In the optical band, there is a faint shell structure which overlaps with the radio emission. In the eROSITA image, there is a $3\sigma$ detection at the same position of the radio non-thermal emission.  

\paragraph{\textbf{{MCSNR J0522--6543}:}\label{text:J0522-6543}} 
The source is shown in Fig.\,\ref{J0522-6543}.
The source was proposed and confirmed as a bona-fide SNR by \citet{Filipovich23} with radio observation. In radio, there is a diffuse emission with a bright non-thermal point source roughly at the centre of the diffuse emission. The optical emission correlates with the radio emission with high [\ion{O}{III}] emission in the interior. \citet{Filipovich23} measured a spectral index in the radio of $\alpha=-0.51 \pm 0.05$ for the entire structure while for the point source, they measured a steeper photon index of $\alpha=-1.00 \pm 0.04$. They argued that most likely the point source is a background active galactic nucleus (AGN). In the eROSITA images, neither diffuse emission related to the SNR nor any counterpart of the AGN is detected. 

\paragraph{\textbf{{MCSNR J0522--6740}:}\label{text:J0522-6740}} 
The source is shown in Fig.\,\ref{J0522-6740}.
The SNR was proposed by \citet{Yew21} and confirmed in the same paper using \xmm\ data. In the optical band, we can see an almost complete shell structure especially from [\ion{S}{II}]. In the eROSITA image, we can detect a $1\sigma$ emission which defines an extended source which correlates with the optical emission. Small regions of $3\sigma$ can be observed with eROSITA. In the radio continuum, we can not observe any particular emission. 

\paragraph{\textbf{{MCSNR J0527--7134}:}\label{text:J0527-7134}} 
The source is shown in Fig.\,\ref{J0527-7134}.
It is another radio candidate \citep{Bozzetto_2017}, which was confirmed by \cite{Kavanagh22}. We can clearly see a diffuse non-thermal emission in the radio band, while in the optical images, we see a compact circular structure where [\ion{S}{II}]/H$\alpha > 0.67$. In the eROSITA images, there is diffuse emission with a clump of soft emission roughly at the centre.

\paragraph{\textbf{{MCSNR J0529--7004}:}\label{text:J0529-7004}} 
The source is shown in Fig.\,\ref{J0529-7004}.
This optical candidate suggested by \citet{Yew21}  was confirmed as an SNR by \citet{2022A&A...661A..37S} using eROSITA data of the calibration and performance verification phase of the telescopes. In the MCELS images, we can see a shell structure that correlates with a faint radio emission. In our eROSITA images, we can see a bright extended structure in the interiors with a net count rate of $(1.50 \pm 0.26)\times 10^{-1}$ cts $\mathrm{s}^{-1}$.

\paragraph{\textbf{{MCSNR J0542--7104}:}\label{text:J0542-7104}} 
The source is shown in Fig.\,\ref{J0542-7104}.
This optical candidate suggested by \citet{Yew21}  was confirmed in the same work using also \xmm\ data. In the MCELS images, we can see a half-shell structure with an enhancement of [\ion{S}{II}]/\ion{H}{$\alpha$}  to the east. In our eROSITA images, there is a bright extended structure with a peak around $\SI{1}{keV}$, partly with $3\sigma$ significance.

\section{Previous candidates which remain candidates}
\label{canNotConf}
\setlength{\parskip}{0pt}
The eROSITA X-ray properties of the following SNR candidates are reported in Table \ref{Tab:PreviousCandRemainCand}.
\paragraph{\textbf{{J0444--6758}:}\label{text:J0444-6758}} 
The source is shown in Fig.\,\ref{J0444-6758}.
This source was proposed as a candidate in \citet{Yew21}. 
In the optical images, there is a relatively compact emission from [\ion{O}{III}] surrounded by a H${\alpha}$ shell. In the centre the [\ion{S}{II}]/H${\alpha}$ ratio is higher than $0.67$. In our radio image, we cannot identify any diffuse structure related to the candidate. In the eROSITA images, there is no significant X-ray emission to confirm this candidate as an SNR. Therefore, the source stays a candidate. 

\paragraph{\textbf{{J0450--6818}:}\label{text:J0450-6818}} The source is shown in Fig.\,\ref{J0450-6818}. This source was also proposed as a candidate in \citet{Yew21} in the optical band. No clear structure is visible in the X-ray images, neither in the count rate image nor in the GGM. For this reason, the source remains a candidate.

\paragraph{\textbf{{J0451--6906}:}\label{text:J0451-6906}} The source is shown in Fig.\,\ref{J0451-6906}. This source was detected with ASKAP and proposed as an SNR candidate by \citet{Filipovich23}. 
In the optical band, there is emission with [\ion{S}{II}]/H${\alpha} > 0.67$. No X-ray emission can be confirmed, and therefore, it remains a candidate. 

\paragraph{\textbf{{J0451--6951}:}\label{text:J0451-6951}} The source is shown in Fig.\,\ref{J0451-6951}. The radio source was proposed by \cite{Filipovich23} as an SNR candidate. There is a faint diffuse emission in the radio images. The optical images show a diffuse emission in [\ion{S}{II}]  but the shape of this emission suggest that is related to a larger environmental structure instead of the candidate. Also in this case no significant X-ray emission is detected, neither in the count rate image nor in the GGM image. Therefore, the source remains a candidate.

\paragraph{\textbf{{J0452--6638}:}\label{text:J0452-6638}} The source is shown in Fig.\,\ref{J0452-6638}. This source was also detected with ASKAP and proposed as an SNR candidate by \citet{Filipovich23}. The radio image shows a shell, which is brighter on the southwest and northwest. A clumpy non-thermal emission is found. In the optical image a shell of [\ion{O}{III}] surrounds a ring-like structure seen in [\ion{S}{II}]. There is another clump emission in the east with [\ion{S}{II}]/H${\alpha} > 0.67$ indicative of shocked material. Very faint X-ray emission is seen in the eROSITA data. Also, the GGM image suggests the presence of an edge in correspondence with the diffuse emission. 
We detect a $1\sigma$ emission in the southwest and a bright clump in the east with a $3\sigma$ detection. The eastern clump correlates with the clump in the optical image. 
Despite these features, we need deeper observations to confirm this source as an SNR and hence the source remains a candidate. 

\paragraph{\textbf{{J0455--6830}:}\label{text:J0455-6830}} The source is shown in Fig.\,\ref{J0455-6830}. It is an optical candidate \citep{Yew21}, which is visible in MCELS images as a small circular structure with relatively strong emission of [\ion{O}{III}] in the centre, while the [\ion{S}{II}] and H${\alpha}$ emission is a little more extended. Neither in X-ray nor in radio we can detect any emission which is correlated with this candidate. The radio image presents a point-like source that is most likely not related to the candidate. Therefore, it remains a candidate. 

\paragraph{\textbf{{J0457--6739}:}\label{text:J0457-6739}} The source is shown in Fig.\,\ref{J0457-6739}. The source is a radio candidate proposed by \citet{Bozzetto_2017}. This source is embedded in the \ion{H}{II} region DEM L40. 
As argued by \citet{Bozzetto_2017} at the inner side of the \ion{H}{II} region there is an enhanced [\ion{S}{II}]/H${\alpha}$ $\sim$ $0.4$, but less than $0.67$. No X-ray emission is detected. Therefore, it remains a candidate. 

\paragraph{\textbf{{J0457--6823}:}\label{text:J0457-6823}} The source is shown in Fig.\,\ref{J0457-6823}. This source has been proposed as a radio candidate by \citet{Filipovich23}. In the radio band, the source has an elongated structure, with a relatively strong shell emission in the southwest. The source was detected in ROSAT data by \citet{Haberl99} as a faint extended source ([HP99] 655). The source was located largely off-axis during the ROSAT observation. In the optical band, there is the same elongated structure, emitting in H${\alpha}$, [\ion{S}{II}], and [\ion{O}{III}]. In the eROSITA data, there is no significant detection of X-ray emission. Therefore, it remains a candidate. 

\paragraph{\textbf{{J0457--6923}:}\label{text:J0457-6923}} The source is shown in Fig.\,\ref{J0457-6923}. This source was first suggested as a potential optical candidate by \citet{Bozzetto_2017}, with [\ion{S}{II}]/H${\alpha}$ $>$ $0.4$. Indeed in the optical band, there is emission with a ratio of [\ion{S}{II}]/H${\alpha}$ $>$ $0.67$, which has a circular shape. In the radio continuum image, there is diffuse non-thermal emission which seems to correlate with the optical emission. In the X-ray image, we detect a significant emission at the position of the candidate. However, we also point out that the candidate is embedded in a complex region and it is difficult to distinguish it from the near \ion{H}{II} region LHA 120-N 94B. Therefore we can not confirm the source as an SNR and the source remains a candidate. 

\paragraph{\textbf{{J0459--6757}:}\label{text:J0459-6757}} The source is shown in Fig.\,\ref{J0459-6757}. This source was proposed as a candidate by \citet{Filipovich23} using the ASKAP data. In the radio continuum, the source has a shell emission, especially in the south. This emission correlates with the emission in the optical band where an elongated structure is visible with an enhancement in the south, which is embedded in a larger \ion{H}{II} region (LHA 120-N 16A). There is an AGN in the background (MACHO 24.3321.1348) at the position of the source, which affects the X-ray observation. The AGN emission was removed for further analysis. We are not able to confirm the source as an SNR and it remains a candidate.

\paragraph{\textbf{{J0459--7008b}:}\label{text:J0459-7008b}} The source is shown in Fig.\,\ref{J0459-7008b}. This source was proposed as a radio candidate using the ASKAP survey data. It is near the known MCSNR J0459--7008, and both are embedded in the \ion{H}{II} region forming a superbubble (LHA 120-N 186). The candidate is detected southwest of MCSNR J0459--7008, which is located on the northern edge of the superbubble. Even though the environment is complex and crowded, the [\ion{S}{II}]/H${\alpha}$ $>$ $0.67$ contours clearly show emission correlated with the radio emission. Due to the low statistics in the X-ray images, it is difficult to separate the emission from the candidate from that of MCSNR J0459--7008  or from the superbubble N186. Radio, optical, and GGM images suggest an SNR inside the superbubble, most likely caused by an explosion inside the superbubble. The lack of a $3\sigma$ detection in X-rays prevents us from confirming the source as an SNR, and it remains a candidate.

\paragraph{\textbf{{J0500--6512}:}\label{text:J0500-6512}} The source is shown in Fig.\,\ref{J0500-6512}. This source is an optical candidate proposed by  \citet{Yew21}. In the optical band, \citet{Yew21} detected a large shell-like structure where the emission is mainly due to H${\alpha}$ and [\ion{S}{II}]. In our radio data, there is no related structure neither in the continuum image nor in the non-thermal contours. Faint X-ray emission is instead visible in the eROSITA data, but only with $1\sigma$ significance. Since we do not detect emission at a $3\sigma$ level, we can not confirm the source as an SNR and it remains a candidate. Also, the GGM image highlights an edge structure compatible with the same elongated structure as in H${\alpha}$. We have proposed an observation with \xmm, which has been carried out recently. The analysis is ongoing. 

\paragraph{\textbf{{J0502--6739}:}\label{text:J0502-6739}} The source is shown in Fig.\,\ref{J0502-6739}. It was proposed as an optical candidate \citep{Yew21} as it shows a clear shell structure in the optical with brighter emission in the west. We can also observe a strong enhancement of [\ion{S}{II}]/H${\alpha}$ $>$ $0.67$ in most of the source. Neither in radio nor in the X-ray we can see any structure that can be related with that in the optical. Therefore, the source remains a candidate. 

\paragraph{\textbf{{J0504--6901}:}\label{text:J0504-6901}} The source is shown in Fig.\,\ref{J0504-6901}. The source is located in a larger emission region known as DEM L64 and was already observed in \citet{Filipovic95}, \citet{Filipovic96}, and \citet{Filipovic98} using the Parkes telescopes. The source has a complex shape in the radio image with a shell structure in the southwest and a more diffuse but compact emission in the northeast. In the whole emission region, we can detect non-thermal radio emission. In the optical band, due to the fact that the source is embedded in a \ion{H}{II} region, it is hard to disentangle the contribution of the candidate from the \ion{H}{II} region. In the X-ray count rate image, no emission can be observed. Therefore, the source remains a candidate. 

\paragraph{\textbf{{J0506--6509}:}\label{text:J0506-6509}} The source is shown in Fig.\,\ref{J0506-6509}. This source is an optical candidate \citep{Yew21}. In the MCELS image, there is a circular shell structure that is relatively bright, especially in the west. In the radio images, there is no source that can be clearly associated with the candidate. Also in the X-ray images, the emission is very faint with a net count rate of ${<} 10^{-2}$ cts $\mathrm{s}^{-1}$, which is not enough to confirm this candidate as an SNR. 

\paragraph{\textbf{{J0507--7110}:}\label{text:J0507-7110}} The source is shown in Fig.\,\ref{J0507-7110}. The source was proposed as a candidate in the radio band with a shell structure particularly enhanced in the south \citep{Bozzetto_2017}. The structure correlates with the optical emission which has the same shell structure and enhancement of [\ion{S}{II}]/H${\alpha}$ $>$ $0.67$. There are two X-ray point sources in the region of the candidate, which prevents us from seeing possible diffuse emission correlated with the candidate. The point sources are significantly detected and are listed in the eRASS:4 point source catalogue with an extent likelihood $\texttt{EXT\_LIKE}=0$.
The point sources were removed from the analysis.
We cannot confirm the source as an SNR and it remains a candidate. 

\paragraph{\textbf{{J0508--6928}:}\label{text:J0508-6928}} The source is shown in Fig.\,\ref{J0508-6928}. This is an optical candidate proposed by \citet{Yew21}. In the optical, we have a half shell in the northeast with a small increase of [\ion{S}{II}]/H${\alpha}$ in part of the shell. No emission is detected in the other bands, neither in radio nor in X-rays, therefore the source stays a candidate. 

\paragraph{\textbf{{J0509--6402}:}\label{text:J0509-6402}} The source is shown in Fig.\,\ref{J0509-6402}. The source was proposed as an optical candidate \citep{Yew21}. Even if the source is almost on the edge of the MCELS survey it was detected as an elliptical shell in H${\alpha}$ and [\ion{S}{II}]. In the radio images, we do not see any diffuse emission but a point source which lies roughly in the centre of the optical shell. The point source has a strong non-thermal emission.
In the X-ray image, there is faint diffuse emission, which correlates with the optical emission, with contours at $1\sigma$ in the northeast and southwest. We can also detect a small portion at $3\sigma$. This detection is confirmed also by the GGM image which highlights an edge structure compatible with the optical diffuse emission. Despite this information, the detection is not significant enough to safely confirm the source as an SNR.

\paragraph{\textbf{{J0513--6731}:}\label{text:J0513-6731}} The source is shown in Fig.\,\ref{J0513-6731}. This source was proposed as a radio candidate. The source is clearly visible in the radio band as a diffuse emission with significant non-thermal emission. In the optical band, the source has a circular emission mainly of [\ion{O}{III}]. In the eROSITA image, the emission is mainly observed in $0.2\mbox{--}\SI{0.7}{\keV}$ as a diffuse emission from the inner part of the remnant. Despite the fact we can see some diffuse emission, we did not obtain a $3\sigma$ detection. Therefore the source remains a candidate.

\paragraph{\textbf{{J0517--6757}:}\label{text:J0517-6757}} The source is shown in Fig.\,\ref{J0517-6757}. It was proposed as an optical candidate and is located northeast of the known SNR MCSNR J0517--6759 \citep{Yew21}. In the optical band, there is a relatively faint diffuse emission with [\ion{S}{II}]/H${\alpha} > 0.67$ in the north. In radio, there is a bright radio source which has an X-ray counterpart and appears to be a point source. As this source is probably associated with a blazar candidate proposed in \citet{2018ApJ...867..131Z}, we exclude that this emission is related to the SNR candidate.
There is no other emission compatible with the optical candidate. Therefore, we can not confirm the source as an SNR and it remains a candidate. 

\paragraph{\textbf{{J0528--7018}:}\label{text:J0528-7018}} The source is shown in Fig.\,\ref{J0528-7018}. This optical candidate proposed by \citet{Yew21} presents a shell structure in which most of the emission is found in the southwest. Also  [\ion{S}{II}]/H${\alpha}$ is higher than $0.67$ in most part of the source. In radio, we can barely see an emission that correlates with the optical image. Due to the relatively bright surroundings, no clear X-ray structure which can be associated with an SNR is detected, even if the net count inside the candidate region is $\sim 400$. 
For these reasons, the source remains a candidate.

\paragraph{\textbf{{J0534--6700}:}\label{text:J0534-6700}} The source is shown in Fig.\,\ref{J0534-6700}. It is a radio candidate from the ASKAP survey proposed by \citet{Filipovich23} and it is located near to the LMC super-giant shell 4 \citep{Meaburn80}. The optical image shows the presence of nearby filaments which do not correlate with the source. \cite{Filipovich23} argued that most likely the candidate is an old SNR so we do not expect any X-ray emission. In the X-ray images instead, we can observe a soft "horn-shaped" structure in the southern part of the candidate which correlates with a similar "horn-shaped" structure described in \cite{Filipovich23}. We point out also that inside the candidate region, there are two point sources detected in the eRASS:4 point source catalogue and detected with $\texttt{DETLIKE} > 20$ and $\texttt{EXTLIKE} = 0$. We excluded the point source from further analysis.
For these reasons, the source remains a candidate.

\paragraph{\textbf{{J0534--6720}:}\label{text:J0534-6720}} The source is shown in Fig.\,\ref{J0534-6720}. The radio candidate proposed based on ASKAP interferometry has a ring structure with an enhanced emission in the south. The region presents a high non-thermal emission. \cite{Filipovich23} detected an enhancement in the [\ion{S}{II}]/H${\alpha}$ ratio, which, however, is $< 0.67$ in the whole candidate region. In the X-ray data, we see a diffuse soft emission. Nevertheless, we notice that in the X-ray images, most of the emission is coming from the southern part. Despite this emission, we can not confirm the source as an SNR and it remains a candidate.

\paragraph{\textbf{{J0538--6921}:}\label{text:J0538-6921}} The source is shown in Fig.\,\ref{J0538-6921}. The radio candidate proposed by \citet{Bozzetto_2017} shows a bright circular non-thermal emission. In the optical band, the region is covered by filaments. In the X-ray image, there is a diffuse emission but it is difficult to separate the contribution of the candidate from the surroundings since the candidate is located in a crowded region near 30 Doradus. The candidate was analysed by \citet{2022A&A...661A..37S} using eROSITA observations during the calibration and performance verification phase. In spite of the deeper observations, it is not possible to distinguish a clear structure which is associated with an SNR. The source remains a candidate.

\paragraph{\textbf{{J0538--7004}:}\label{text:J0538-7004}} The source is shown in Fig.\,\ref{J0538-7004}. The optical candidate proposed by \cite{Yew21} appears as a filled shell. The candidate is relatively small and has an estimated diameter of $D=19.4$\,pc \citep{Yew21}. J0538--7004 shows an enhancement of [\ion{S}{II}]/H${\alpha}$ $\sim 0.8$. The compactness of this candidate indicates a young SNR which would suggest the presence of X-ray and radio emission. However, we do not detect any significant emission, neither in radio nor in X-rays. As argued in \cite{Yew21} the enhancement of [\ion{O}{III}] without X-ray and radio emission might suggest that the candidate is most likely an \ion{H}{II} region instead of an SNR. For this reason, the source remains a candidate. 

\paragraph{\textbf{{J0539--7001}:}\label{text:J0539-7001}} The source is shown in Fig.\,\ref{J0539-7001}. This X-ray source was detected in the ROSAT survey by \citet{Haberl99} and classified as an SNR candidate. In the radio band, there is a point source almost at the centre of the SNR candidate, which can be seen in the ASKAP images. In \citet{Bozzetto_2017} a spectral index of $\alpha = -0.47$ was measured. In the eROSITA three-colour image, the SNR appears as a bright elongated source in green indicating a peak in the emission in the medium energy band, probably due to  Fe-L emission. This feature suggests a possible thermonuclear origin of the remnant, which is also indicated by its position in the HR diagrams shown in Fig.\,\ref{FigureHR} with $\mathrm{HR}_1 = 0.71 \pm 0.04$, $\mathrm{HR}_2 = -0.88 \pm 0.04$, and $\mathrm{HR}_3 = -0.82 \pm 0.47$. The source is clearly detected with $3\sigma$ confidence in X-rays. Despite the prominent eROSITA emission, we can just confirm the X-ray nature of this source. The lack of multi-wavelength evidence does not allow us to confirm the candidate as SNR.

\paragraph{\textbf{{J0542--6852}:}\label{text:J0542-6852}} The source is shown in Fig.\,\ref{J0542-6852}. This source was detected for the first time by \citet{Filipovich23} using ASKAP data. 
In the optical band, it is possible to see a shell-like structure in [\ion{O}{III}] and an enhancement of [\ion{S}{II}]/H${\alpha} > 0.67$. The source has a very faint X-ray emission in \xmm\ data in the medium band \citep{Filipovich23}. In the eROSITA data, there is diffuse emission which overlaps with the radio emission. This diffuse emission is detected with a confidence of $1\sigma$, and only a small portion of the diffuse emission exceeds the $3\sigma$ detection, although it is not sufficient to confirm the source as an SNR. The source remains a candidate.

\paragraph{\textbf{{J0543--6906}:}\label{text:J0543-6906}} The source is shown in Fig.\,\ref{J0543-6906}. The source was detected by \cite{Filipovich23} for the first time as a ring structure in radio. In optical the region is covered by different filaments but there is no clear emission from the candidate. In the eROSITA image, the candidate is embedded in a large diffuse structure and it is hard to distinguish the candidate. The contours at $1\sigma$ indicate a possible bubble structure that is correlated with the radio emission. Also, the GGM image suggests a ring structure which agrees with the radio emission. Despite the correlation, we do not detect a $3\sigma$ emission. Therefore the source remains a candidate.

\paragraph{\textbf{{J0543--6923}:}\label{text:J0543-6923}} The source is shown in Fig.\,\ref{J0543-6923}. For this radio candidate proposed by \cite{Filipovich23} there is a circular non-thermal shell in the radio image. The ring is connected in the east with an elongated structure, which made \cite{Filipovich23} argue that is likely related to a superbubble. In the MCELS data, the candidate region is filled with filaments and it is difficult to distinguish any structure related to the candidate. Also in the X-ray images, there is diffuse emission in the surroundings and no particular structure related to the candidate is visible. The source thus stays a candidate. 

\paragraph{\textbf{{J0543--6928}:}\label{text:J0543-6928}} The source is shown in Fig.\,\ref{J0543-6928}. This is another radio candidate from \cite{Filipovich23}. In the radio band, the source has a slightly elongated shape. There is non-thermal radio emission in the centre. \cite{Filipovich23} reported that no counterpart is detected, neither in the optical nor with \xmm. We confirm that no evident structure is visible in the eROSITA images, for this reason, it remains a candidate. 

\paragraph{\textbf{{J0548--6941}:}\label{text:J0548-6941}} The source is shown in Fig.\,\ref{J0548-6941}. The source is an optical candidate from \cite{Yew21}, which has a half-shell structure with a small part with high [\ion{S}{II}]/H${\alpha}$. Unfortunately, we can not detect any structure in the X-ray or in the radio image. 
For this reason, it remains a candidate.

\paragraph{\textbf{{J0549--6618}:}\label{text:J0549-6618}} The source is shown in Fig.\,\ref{J0549-6618}. Proposed as an optical candidate, the source has a complex shell structure filled in the centre, where \cite{Yew21} measured [\ion{S}{II}]/H${\alpha} \sim 1.0$. In the radio band, there is no significant emission in the region. In the X-ray images, there could be a very faint soft emission but it is not significant and for these reasons, we can not confirm the source as an SNR, and it remains a candidate. 

\paragraph{\textbf{{J0549--6633}:}\label{text:J0549-6633}} The source is shown in Fig.\,\ref{J0549-6633}. The source is an optical candidate proposed in \cite{Yew21}.
In the MCELS data, the emission is mainly due to [\ion{S}{II}]. \cite{Yew21} argued that the source is outside of the spectral range to be an SNR and is more likely a superbubble. However, they also point out that there are only a few OB stars in the surroundings, which does not support the superbubble nature. No radio emission is detected. In the X-ray images, there is a diffuse elongated structure coinciding with the optical source. Most of the emission is detected in the energy band $0.7\mbox{--}\SI{1.1}{\keV}$. The GGM image also shows the edges of this source. However, since there is no $3\sigma$ detection in X-ray we can not confirm the nature of the diffuse emission. The source thus remains an SNR candidate. 

\paragraph{\textbf{{J0624--6948}:}\label{text:J0624-6948}} The source is shown in Fig.\,\ref{J0624-6948}. This source, also known as the LMC Odd Radio Circle, was detected by \cite{2022MNRAS.512..265F} in radio data using the ASKAP survey of the LMC. Similar to J0614--7251 described in Sect.\,\ref{text:J0614-7251},  J0624--6948 is also located far from the LMC main body and lies between the LMC and the Galactic plane. The positions of these sources make them extremely interesting. The positions can be the consequence of tidal interaction between the Small Magellanic Cloud, LMC, and the Milky Way. Unfortunately, MCELS does not cover the position of the source. In Fig.\,\ref{J0624-6948} we show the comparison between the eROSITA data, the GGM filter applied on the X-ray data, and the ASKAP image. In radio there is clearly a circular structure as studied in \cite{2022MNRAS.512..265F}. We detect diffuse X-ray emission up to $3\sigma$ level, which is represented by the red contours in Fig.\,\ref{J0624-6948}. However, inside the source region, two point sources are reported in the eRASS:4 point source catalogue, which makes the calculation of the significance contours difficult. Recently, we observed the source with \xmm \ and the analysis is ongoing. 
\onecolumn

\section{Tables}
\label{appendixTables}

\begin{longtable}
    {@{} p{.15\textwidth}@{} p{.15\textwidth}@{} p{.15\textwidth}@{} p{.2\textwidth}@{} p{.23\textwidth}@{} p{.12\textwidth}@{} @{}}
    \caption{ \label{Tab:TotalMCSNR}  Complete catalogue of the known X-ray SNRs in the LMC (for more details on the sources discussed in this work see Sect. \ref{erositasnrs}, \ref{KnownSNRnotMaggi}, and Tables \ref{Tab:eROSITASNR}, \ref{Tab:SNRNotInMaggi}, and Fig. \ref{CommonMainText}, \ref{appendix:SNRNotInMaggi}).} \\[-0.2cm]\toprule\midrule
     MCSNR & RA [J2000] & DEC [J2000] & Size [\arcmin] (PA [$\deg$]) & Rate [cts $\mathrm{s}^{-1}$] & Ref  \\\midrule
     \endfirsthead
     \caption{\textbf{continued.} Complete catalogue of the known X-ray SNRs in the LMC.}  \\[-0.2cm]\toprule\midrule
     MCSNR & RA [J2000] & DEC [J2000] & Size [\arcmin] (PA [$\deg$]) & Rate [counts $\mathrm{s}^{-1}$] & Ref \\\midrule
     \endhead
     \midrule
     \endfoot
     \parbox{\textwidth}{\tablebib{B17: \citet{Bozzetto_2017}; B23: \citet{Filipovich23}; BFC12a: \citet{BFC12a}; BFC12b: \citet{BFC12b}; BFC13: \citet{BFC13}; BGS06: \citet{BGS06}; BKM14: \citet{BKM14}; CDS95: \citet{CDS95}; CKS97: \citet{CKS97}; CMG93: \citet{CMG93}; DFB12: \citet{DFB12}; GSH12: \citet{GSH12}; HP99: \citet{Haberl99}; KPS10: \citet{KPS10}; KSB15a: \citet{KSB15a}; KSB15b: \citet{KSB15b}; KSP13: \citet{KSP13}; Lea17: \citet{Lea17}; LHG81: \citet{LHG81}; Ma19: \citet{Maitra19}; MC73: \citet{MC73}; MFD83: \citet{MFD83}; MFD84: \citet{MFD84}; MFT85: \citet{MFT85}; MHB12: \citet{MHB12}; MHK14: \citet{MHK14}; SCM94: \citet{SCM94}; TM84: \citet{TM84}; TW: this work; WM66: \citet{WM66}; Y21: \citet{Yew21}.}}
     \parbox{\textwidth}{
     \tablefoot{The first column reports the name of the source. The second, third, and fourth columns report the position and the size of the source in the eROSITA images. The fifth column reports the net count in the eROSITA data in the energy band $0.2\mbox{--}5.0\,\mathrm{keV}$ collected by TM 12346. The last column gives the reference to the paper where the source was first  identified as an SNR.
     {\tablefoottext{a}{In eROSITA data, there is diffuse emission, but there is $1\sigma$ emission only in the central region.}
     \tablefoottext{b}{No $1\sigma$ contours are detected in eROSITA data. Despite this around 48 net counts are detected}
     \tablefoottext{c}{An extended emission is visible in eROSITA data, but there is no $1\sigma$ detection.}
      \tablefoottext{d}{Small portion of $1\sigma$ contours are detected in eROSITA data. The contours are not clearly associable with an extended source.}
      \tablefoottext{e}{This source was confirmed as bona-fide SNR in \citet{Filipovich23}. However, no emission is detectable in eROSITA data. For all sources, count rates are derived from the eROSITA data in the entire SNR region. See Section \ref{section:luminosity} .}}
     }}
    \endlastfoot
J0447$-$6918\tablefootmark{a} & 04:47:12.2 & $-$69:19:16 & $6.0\times3.8 \ (320)$ & $(2.40  \pm 1.10) \times 10^{-2}$ & B17 \\
J0448$-$6700 & 04:48:25.2 & $-$67:00:25 & 5.4 & $(2.22 \pm 0.15)\times 10^{-1}$ & BGS06 \\
J0449$-$6903\tablefootmark{b} & 04:49:34.0 & $-$69:03:34 & $3.6\times3.4 \ (0)$ & $(3.00 \pm 0.87)\times 10^{-2}$ & B17 \\
J0449$-$6920 & 04:49:20.0 & $-$69:20:20 & 3.3 & $(6.30 \pm 0.96)\times 10^{-2}$ & KPS10 \\
J0450$-$7050 & 04:50:27.0 & $-$70:50:15 & 6.4 & $(2.53 \pm 0.19)\times 10^{-1}$ & MFT85 \\
J0453$-$6655 & 04:53:10.2 & $-$66:54:52 & 4.3 & $(5.14 \pm 0.18)\times 10^{-1}$ & SCM94 \\
J0453$-$6829 & 04:53:37.7 & $-$68:29:38 & 3.3 & $6.23 \pm 0.06$ & LHG81 \\
J0454$-$6626 & 04:54:49.0 & $-$66:25:32 & 2.5 & $(3.96 \pm 0.15)\times 10^{-1}$ & MC73 \\
J0454$-$6713 & 04:54:27.2 & $-$67:13:20 & 2.4 & $(7.58 \pm 0.22)\times 10^{-1}$ & SCM94 \\
J0454$-$7003 & 04:54:19.8 & $-$70:03:27 & $2.6\times2.4 \ (80)$ & $(4.80 \pm 0.52)\times 10^{-2}$ & Y21 \\
J0455$-$6839 & 04:55:29.2 & $-$68:39:01 & $6.4\times5.2$ ($90$) & $(7.57 \pm 0.30)\times 10^{-1}$ & MC73 \\
J0456$-$6533 & 04:56:50.7 & $-$65:32:44 & 5.7 & $(1.80 \pm 0.14)\times 10^{-1}$ & TW \\
J0456$-$6950\tablefootmark{c} & 04:56:38.0 & $-$69:50:55 & $5.4\times4.2 \ (0)$ & $(1.00   \pm 1.33)\times 10^{-2}$ & B17 \\
J0459$-$7008 & 04:59:51.9 & $-$70:07:50 & 2.9   & $(2.11   \pm 0.13)\times 10^{-1}$   & MC73       \\
J0504$-$6723 & 05:04:46.1 & $-$67:23:59 & 5.5   & $(1.42   \pm 0.15)\times 10^{-1}$    & HP99       \\
J0505$-$6753 & 05:05:41.9 & $-$67:52:45 & 2.2   & $(1.80   \pm 0.01)\times 10^{1}$    & LHG81      \\
J0505$-$6802 & 05:05:54.7 & $-$68:01:50 & $2.2\times1.8 \ (45)$ & $(1.09   \pm 0.01)\times 10^{1}$    & MC73       \\
J0506$-$6541 & 05:05:59.8 & $-$65:42:37 & 8.6   & $(2.22 \pm 0.18)\times 10^{-1}$    & KPS10      \\
J0506$-$6815 & 05:06:07.1 & $-$68:15:43 & $4.4\times4.0 \ (30)$  & $(9.20   \pm 1.25)\times 10^{-2}$    & B17        \\
J0506$-$7009 & 05:06:15.8 & $-$70:09:20 & $3.2\times2.4 \ (140)$ & $(9.50   \pm 0.95)\times 10^{-2}$    & TW        \\
J0506$-$7026 & 05:06:50.0 & $-$70:25:53 & 5.5   & $(5.88 \pm 0.23)\times 10^{-1}$    & Lea17   \\
J0507$-$6847 & 05:07:33.6 & $-$68:47:27 & $11.0\times9.2 \ (120)$ & $(8.67  \pm 0.55)\times 10^{-1}$    & B17        \\
J0508$-$6830 & 05:08:50.0 & $-$68:30:50 & 3.4   & $(6.20 \pm 0.91)\times 10^{-2}$    & MHK14      \\
J0508$-$6902 & 05:08:37.0 & $-$69:02:54 & 3.6   & $(1.94 \pm 0.13)\times 10^{-1}$    & BKM14      \\
J0509$-$6731 & 05:09:30.6 & $-$67:31:20 & 1.5   & $5.61 \pm 0.05$    & LHG81      \\
J0509$-$6844 & 05:08:59.0 & $-$68:43:35 & 2.4   & $(2.06 \pm 0.01)\times 10^{1}$     & MC73       \\
J0510$-$6708\tablefootmark{d} & 05:10:11.4 & $-$67:08:04 & 2.0   & $(2.10 \pm 0.88)\times 10^{-2}$    & B17        \\
J0511$-$6759 & 05:11:17.4 & $-$67:59:10 & 4.1   & $(1.19 \pm 0.11)\times 10^{-1}$    & MHK14      \\
J0512$-$6707 & 05:12:27.0 & $-$67:07:18 & 2.6   & $(5.80 \pm 0.66)\times 10^{-2}$    & KSB15b     \\
J0512$-$6716 & 05:12:24.7 & $-$67:16:55 & $7.8\times7.4 \ (45)$  & $(1.40  \pm 0.19)\times 10^{-1}$    & B17        \\
J0513$-$6724 & 05:13:43.0 & $-$67:24:10 & $2.6\times2.0 \ (0)$   & $(5.00  \pm 0.59)\times 10^{-2}$    & Ma19       \\
J0513$-$6912 & 05:13:11.6 & $-$67:12:20 & $4.0\times3.2 \ (55)$   & $(2.36  \pm 0.15)\times 10^{-1}$    & MFT85       \\
J0514$-$6840 & 05:14:12.9 & $-$68:40:15 & 4.8   & $(1.60  \pm 0.14)\times 10^{-1}$    & MHK14      \\
J0517$-$6759 & 05:17:11.7 & $-$67:58:50 & 5.3   & $(1.16  \pm 0.14)\times 10^{-1}$    & MHK14      \\
J0518$-$6939 & 05:18:41.7 & $-$69:39:20 & 3.4   & $(4.42  \pm 0.18)\times 10^{-1}$    & MC73       \\
J0519$-$6902 & 05:19:33.3 & $-$69:02:21 & 2.6   & $(1.51  \pm 0.01)\times 10^{1}$    & LHG81      \\
J0519$-$6926 & 05:19:44.0 & $-$69:26:08 & 4.6   & $1.28   \pm 0.03$    & MFD83      \\
J0521$-$6543 & 05:21:39.0 & $-$65:43:07 & 6.6   & $(5.80  \pm 1.09)\times 10^{-2}$     & BGS06      \\
J0522$-$6543\tablefootmark{e} & 05:22:53.5 & $-$65:43:09 & $5.8\times5.2 \ (26)$  & $< 10^{-2}$     & B23        \\
J0522$-$6740 & 05:22:33.7 & $-$67:41:04 & $3.2\times3.0 \ (90)$  & $(5.30  \pm 1.19)\times 10^{-2}$   & Y21        \\
J0523$-$6753 & 05:23:05.5 & $-$67:53:20 & 4.8   & $(4.14  \pm 0.19)\times 10^{-1}$    & CMG93      \\
J0524$-$6624 & 05:24:20.8 & $-$66:24:28 & 4.3   & $(3.80 \pm 0.60)\times 10^{-2}$    & MFT85      \\
J0525$-$6559 & 05:25:24.1 & $-$65:59:26 & 3.7   & $(1.61   \pm 0.01)\times 10^{1}$     & MC73       \\
J0525$-$6938 & 05:25:02.9 & $-$69:38:43 & 3.7   & $(1.30   \pm 0.01)\times 10^{2}$     & WM66       \\
J0526$-$6605 & 05:25:59.4 & $-$66:05:04 & 2.7   & $(2.50   \pm 0.01)\times 10^{1}$     & WM66       \\
J0527$-$6550 & 05:27:54.0 & $-$65:49:38 & 5.2   & $(4.75   \pm 0.13)\times 10^{-1}$     & LHG81      \\
J0527$-$6714 & 05:28:07.9 & $-$67:13:43 & 6.9   & $(3.70   \pm 1.53)\times 10^{-2}$     & TM84       \\
J0527$-$6912 & 05:27:39.7 & $-$69:12:20 & 3.9   & $(4.47   \pm 0.16)\times 10^{-1}$    & MFD84      \\
J0527$-$7104 & 05:28:01.2 & $-$71:04:23 & $5.6\times3.2 \ (65)$  & $(1.27   \pm 0.15)\times 10^{-1}$     & KSP13      \\
J0527$-$7134 & 05:27:49.9 & $-$71:34:08 & $3.8\times3.6 \ (45)$  & $(6.90   \pm 0.92)\times 10^{-2}$     & B17        \\
J0528$-$6727 & 05:28:05.0 & $-$67:27:20 & 5.7  & $(1.80 \pm 0.15)\times 10^{-1}$     & MHB12      \\
J0529$-$6653 & 05:29:49.2 & $-$66:53:34 & 3.5   & $(3.95 \pm 0.11)\times 10^{-1}$     & BFC12a     \\
 J0529$-$7004 & 05:29:11.4 & $-$70:04:40 & 1.6   & $(1.50 \pm 0.26)\times 10^{-1}$ & Y21        \\
J0530$-$7008 & 05:30:39.0 & $-$70:07:30 & 5.8   & $(3.48   \pm 0.30)\times 10^{-1}$ & DFB12      \\
J0531$-$7100 & 05:31:56.0 & $-$71:00:19 & 4.0   & $1.14   \pm 0.02$    & MC73       \\
J0532$-$6732 & 05:32:14.0 & $-$67:32:10 & 5.3   & $1.17   \pm 0.02$    & MFT85      \\
J0533$-$7202 & 05:33:53.3 & $-$72:02:57 & 4.4   & $(2.59   \pm 0.14)\times 10^{-1}$ & BFC13      \\
J0534$-$6955 & 05:34:00.4 & $-$69:55:03 & 3.3   & $3.36   \pm 0.04$    & LHG81      \\
J0534$-$7033 & 05:34:14.9 & $-$70:33:46 & 4.1   & $(7.12   \pm 0.20)\times 10^{-1}$ & LHG81      \\
J0535$-$6602 & 05:35:44.9 & $-$66:02:09 & 3.1   & $(8.33   \pm 0.01)\times 10^{1}$  & WM66       \\
J0535$-$6916 & 05:35:27.7 & $-$69:16:15 & 1.4   & $6.61   \pm 0.05$    & HISTORICAL \\
J0535$-$6918 & 05:35:47.2 & $-$69:18:14 & 2.2   & $(1.29   \pm 0.17)\times 10^{-1}$ & CDS95      \\
J0536$-$6735 & 05:35:56.2 & $-$67:34:07 & $6.8\times3.6 \ (66)$  & $1.37   \pm 0.02$  & MFT85      \\
J0536$-$6913 & 05:36:15.4 & $-$69:13:07 & 1.8   & $(1.22   \pm 0.10)\times 10^{-1}$ & KSB15a     \\
J0536$-$7039 & 05:36:01.3 & $-$70:38:26 & 4.5   & $(7.00   \pm 0.19)\times 10^{-1}$ & LHG81      \\
J0537$-$6628 & 05:37:30.9 & $-$66:27:52 & $4.4\times3.0 \ (315)$ & $(1.31   \pm 0.06)\times 10^{-1}$     & KPS10      \\
J0537$-$6910 & 05:37:47.4 & $-$69:10:17 & 2.6   & $2.86   \pm 0.03$    & MC73       \\
J0540$-$6920 & 05:40:10.3 & $-$69:19:59 & 1.4   & $(1.08 \pm 0.01)\times 10^{1}$ & MC73       \\
J0540$-$6944 & 05:40:06.1 & $-$69:44:00 & 1.4   & $(9.46 \pm 0.18)\times 10^{-1}$ & CKS97      \\
J0541$-$6659 & 05:41:49.5 & $-$66:58:44 & 5.6   & $(3.45 \pm 0.10)\times 10^{-1}$ & GSH12      \\
J0542$-$7104 & 05:42:42.0 & $-$71:04:29 & 1.7   & $(5.70 \pm 0.75)\times 10^{-2}$ & Y21        \\
J0543$-$6624 & 05:43:48.6 & $-$66:23:51 & 5.6   & $(1.05 \pm 0.07)\times 10^{-1}$     & TW        \\
J0543$-$6858 & 05:43:05.9 & $-$68:59:03 & 5.6   & $(7.09 \pm 0.20)\times 10^{-1}$     & LHG81      \\
J0547$-$6941 & 05:47:23.2 & $-$69:41:23 & 2.7   & $(6.30 \pm 0.13)\times 10^{-1}$     & MC73       \\
J0547$-$6943 & 05:46:59.2 & $-$69:43:05 & $4.4\times3.0 \ (35)$  & $(6.20   \pm 0.14)\times 10^{-1}$     & MC73       \\
J0547$-$7025 & 05:47:48.2 & $-$70:24:54 & 3.4 & $1.70   \pm 0.02$     & MFD83      \\
J0550$-$6823 & 05:50:30.9 & $-$68:23:43 & $7.0\times5.6 \ (25)$  & $(4.94   \pm 0.12)\times 10^{-1}$     & BFC12b  \\
   \bottomrule   
\end{longtable}
\begin{longtable}{@{} p{.12\textwidth}@{} p{.105\textwidth}@{} p{.11\textwidth}@{} p{.155\textwidth}@{} p{.145\textwidth}@{} wr{.12\textwidth}@{} wr{.12\textwidth}@{} wr{.12\textwidth}@{}  @{}}
\caption{SNRs discovered and confirmed with eROSITA (Sect. \ref{erositasnrs} and Fig. \ref{CommonMainText}). X-ray SNRs in the LMC detected for the first time by eROSITA and confirmed as SNRs in this work. The first column reports the name of the source. The second, third, and fourth columns report the position and the size of the source in the eROSITA images. The fifth column reports the net count in the eROSITA data in the energy band $0.2\mbox{--}5.0\,\mathrm{keV}$ collected by TM 12346. The last three columns report the HR values in the different bands as described in Sect.\,\ref{HR}.}\\[-0.2cm]\toprule\midrule
    MCSNR & RA [J2000] & DEC [J2000] & Size [\arcmin] (PA [$\deg$]) & Rate [counts $\mathrm{s}^{-1}$] & \multicolumn{1}{l}{HR1} & \multicolumn{1}{l}{HR2} & \multicolumn{1}{l}{HR3}   \\
    \midrule
     \endfirsthead
     \caption{\textbf{continued.} X-ray SNRs in the LMC were detected for the first time by eROSITA and confirmed as SNRs in this work. The first column reports the name of the source. The second, third, and fourth columns report the position and the size of the source in the eROSITA images. The fifth column reports the net count in the eROSITA data in the energy band $0.2-\SI{5.0}{\keV}$ collected by TM 12346. The last three columns report the HR values in the different bands as described in Sect.\,\ref{HR}.}\\\toprule\midrule
    MCSNR & RA [J2000] & DEC [J2000] & Size [\arcmin] (PA [$\deg$]) & Rate [counts $\mathrm{s}^{-1}$] & HR1 & HR2 & HR3   \\\midrule
     \endhead
     \midrule
     \endfoot
     \endlastfoot
J0456$-$6533 & 04:56:50.7 & $-$65:32:44 & 5.7   & $(1.80 \pm 0.14)\times 10^{-1}$  & $-0.20 \pm 0.06$ & - & - \\
J0506$-$7009 & 05:06:15.8 & $-$70:09:20 & $3.2\times2.4 \ (140)$ & $(9.50 \pm 0.95)\times 10^{-2}$ & $0.68 \pm 0.11$ & $-0.68 \pm 0.08$ & $-0.29 \pm 0.29 $ \\
J0543$-$6624 & 05:43:48.6 & $-$66:23:51 & 5.6   & $(1.05 \pm 0.07)\times 10^{-1}$  & $-0.60 \pm 0.05$ & - & - 
\\\bottomrule  
 \label{Tab:eROSITASNR}
\end{longtable}
\begin{longtable}
    {@{} p{.12\textwidth}@{} p{.105\textwidth}@{} p{.11\textwidth}@{} p{.155\textwidth}@{} p{.145\textwidth}@{} wr{.12\textwidth}@{} wr{.12\textwidth}@{} wr{.12\textwidth}@{}  @{}}
     \caption{Previous MCELS candidate confirmed with eROSITA (Sect. \ref{previousCandConf} and Fig.\ref{J0454-7003}). SNR candidate proposed in \citet{Yew21} work and confirmed as an SNR in this work with eROSITA. }\\[-0.2cm]\toprule\midrule
    MCSNR & RA [J2000] & DEC [J2000] & Size [\arcmin] (PA [$\deg$]) & Rate [counts $\mathrm{s}^{-1}$] & \multicolumn{1}{l}{HR1} & \multicolumn{1}{l}{HR2} & \multicolumn{1}{l}{HR3} \\
    \midrule
    \endfirsthead
J0454$-$7003 & 04:54:19.8 & $-$70:03:27 & $2.6\times2.4 \ (80)$  & $(4.80 \pm 0.52)\times 10^{-2}$ & $-0.66 \pm 0.13$ & $-0.44\pm 0.44$ & $-0.10 \pm 0.88$ 
\\\bottomrule  
 \label{Tab:PreviousCandConfi}
\end{longtable}
\begin{longtable}
    {@{} p{.16\textwidth}@{} p{.095\textwidth}@{} p{.105\textwidth}@{} p{.15\textwidth}@{} p{.16\textwidth}@{} wr{.11\textwidth}@{} wr{.115\textwidth}@{} wr{.115\textwidth}@{}  @{}}
     \caption{SNR candidates detected with eROSITA (Sect. \ref{eROSITACandidate} and Fig. \ref{eROSITACandi_images}). New X-ray SNR candidates in the LMC found with eROSITA for the first time.}
    \\[-0.2cm] \toprule\midrule
    4eRASSU & RA [J2000] & DEC [J2000] & Size [\arcmin] (PA [$\deg$]) & Rate [counts $\mathrm{s}^{-1}$] & \multicolumn{1}{l}{HR1} & \multicolumn{1}{l}{HR2} & \multicolumn{1}{l}{HR3}   \\
    \midrule
     \endfirsthead
\caption{\textbf{continued.} New X-ray SNR candidates in the LMC were found with eROSITA for the first time.} \\[-0.2cm]\toprule\midrule
4eRASSU & RA [J2000] & DEC [J2000] & Size [\arcmin] (PA [$\deg$]) & Rate [counts $\mathrm{s}^{-1}$] & \multicolumn{1}{l}{HR1} & \multicolumn{1}{l}{HR2} & \multicolumn{1}{l}{HR3}   \\ \midrule
    \endhead
     \midrule
     \endfoot
     \endlastfoot
J045145.7$-$671724 & 04:51:45.7 & $-$67:17:24 & $9.8\times5.6 \ (60)$  & $(2.06 \pm 0.19)\times 10^{-1}$ & $0.00 \pm 0.09$ & $-0.64 \pm 0.13$ & $-0.20 \pm 0.42$\\
J045625.5$-$683052 & 04:56:25.5 & $-$68:30:52 & 2.7   & $(2.90 \pm 0.78)\times 10^{-2}$  & $0.89 \pm 0.35$ & $-0.79\pm 0.23$ & -    \\
J050750.8$-$714241 & 05:07:50.8 & $-$71:42:41 & $4.8\times3.8 \ (359)$ & $(5.60 \pm 1.41)\times 10^{-2}$ & $0.58 \pm 0.22$ & - & -  \\
J051028.3$-$685329 & 05:10:28.3 & $-$68:53:29 & 3.7   & $(7.20 \pm 1.13)\times 10^{-2}$ & $0.29 \pm 0.15$ & $-0.70 \pm 0.33$ & $-0.09 \pm 0.97$ \\ 
J052136.6$-$670741 & 05:21:36.6 & $-$67:07:41 & 6.6 & $(3.10 \pm 1.37)\times 10^{-2}$ & $0.23 \pm 0.41$ & - & -  \\
J052126.5$-$685245 & 05:21:26.5 & $-$68:52:45 & 5.2 & $(8.40 \pm 1.41)\times 10^{-2}$ & $0.50 \pm 0.20$ & $-0.75 \pm 0.16$ & $-0.49 \pm 0.78$  \\
J052148.7$-$693649 & 05:21:48.7 & $-$69:36:49 & 5.2   & $(1.36 \pm 0.20)\times 10^{-1}$  & $0.38 \pm 0.14$ & - & -  \\
J052330.7$-$680400 & 05:23:30.7 & $-$68:04:00 & 3.9   & $(4.10 \pm 1.02)\times 10^{-2}$  & $0.04 \pm 0.21$ & $-0.89 \pm 0.32$ & -  \\
J052502.7$-$662125 & 05:25:02.7 & $-$66:21:25 & 5.1   & $(1.19 \pm 0.11)\times 10^{-1}$ & $-0.01 \pm 0.07$ & - & -  \\
J052849.7$-$671913 & 05:28:49.7 & $-$67:19:13 & 3.8   & $(3.70 \pm 0.82)\times 10^{-2}$ & $-0.03 \pm 0.22$  & $-0.45 \pm 0.25$ & - \\
J053224.5$-$655411 & 05:32:24.5 & $-$65:54:11 & 2.8   & $(3.80 \pm 0.50)\times 10^{-2}$  & $0.37 \pm 0.12$& $-0.65 \pm 0.14$ & - \\
J054949.7$-$700145 & 05:49:49.7 & $-$70:01:45 & $5.2\times3.4 \ (160)$ & $(5.80 \pm 0.82)\times 10^{-2}$ & $0.76 \pm 0.04$ & - & - \\
J061438.1$-$725112 & 06:14:38.1 & $-$72:51:12 & 3.6   & $(2.76 \pm 0.12)\times 10^{-1}$ & $-0.27 \pm 0.04$& $-0.75 \pm 0.06$ & $-0.73 \pm 0.40$ 
\\\bottomrule  
 \label{Tab:eROSITACandidate}
\end{longtable}
\begin{longtable}{@{} p{.105\textwidth}@{} p{.095\textwidth}@{} p{.105\textwidth}@{} p{.15\textwidth}@{} p{.16\textwidth}@{} wr{.11\textwidth}@{} wr{.11\textwidth}@{} wr{.11\textwidth}@{} wr{.06\textwidth}@{} @{}}
    \caption{Known SNRs not included in \citet{Maggi16}, but observed with eROSITA (Appendix \ref{KnownSNRnotMaggi} and Fig. \ref{appendix:SNRNotInMaggi}). X-ray SNRs confirmed in \cite{Maitra19}, \citet{Yew21}, \cite{Kavanagh22}, \citet{Maitra21}, \citet{2022A&A...661A..37S}, and \cite{Filipovich23}.}\\[-0.2cm]\toprule\midrule
    \label{Tab:SNRNotInMaggi}
    MCSNR & RA [J2000] & DEC [J2000] & Size [\arcmin] (PA [$\deg$]) & Rate [counts $\mathrm{s}^{-1}$] & \multicolumn{1}{l}{HR1} & \multicolumn{1}{l}{HR2} & \multicolumn{1}{l}{HR3} & Conf  \\
    \midrule
    \endfirsthead
    \caption{\textbf{continued.} X-rays SNRs confirmed in \cite{Maitra19}, \citet{Yew21}, \cite{Kavanagh22}, \citet{Maitra21}, \citet{2022A&A...661A..37S}, and \cite{Filipovich23}.}\\[-0.2cm]\toprule\midrule
    \label{Tab:SNRNotInMaggi}
    MCSNR & RA [J2000] & DEC [J2000] & Size [\arcmin] (PA [$\deg$]) & Rate [counts $\mathrm{s}^{-1}$] & HR1 & HR2 & HR3 & Conf  \\
    \endhead
    \midrule
    \endhead
    \midrule
    \endfoot
    \parbox{\textwidth}{\tablebib{Ma19: \citet{Maitra19}; Y21: \citet{Yew21}; Ma21: \citet{Maitra21}; K22: \citet{Kavanagh22}; Sa22: \citet{2022A&A...661A..37S}; B23: \citet{Filipovich23} .}}
    \endlastfoot
J0447$-$6918 & 04:47:12.2 & $-$69:19:16 & $6.0\times3.8 \ (320)$ & $(2.40  \pm 1.10) \times 10^{-2}$ & $0.17 \pm 0.31$ & $-0.87 \pm 0.55$ & - & K22 \\
J0449$-$6903 & 04:49:34.0 & $-$69:03:34 & $3.6\times3.4 \ (0)$   & $(3.00 \pm 0.87)\times 10^{-2}$  & $-0.13 \pm 0.46$ & $0.42 \pm 0.31$ & $-0.66 \pm 0.5$ &  K22  \\
J0456$-$6950 & 04:56:38.0 & $-$69:50:55 & $5.4\times4.2 \ (0)$   & $(1.00   \pm 1.33)\times 10^{-2}$  & - & - & - & K22  \\
J0504$-$6723 & 05:04:46.1 & $-$67:23:59 & 5.5    & $(1.42   \pm 0.15)\times 10^{-1}$  & $0.65 \pm 0.10$ & $-0.88 \pm 0.08$ & - & K22 \\
J0506$-$6815 & 05:06:07.1 & $-$68:15:43 & $4.4\times4.0 \ (30)$  & $(9.20   \pm 1.25)\times 10^{-2}$  & $-0.03 \pm 0.16$ & $-0.04 \pm 0.15$ & $-0.85 \pm 0.30$ & B23 \\
J0507$-$6847 & 05:07:33.6 & $-$68:47:27 & $11.0\times9.2 \ (120)$  & $(8.67  \pm 0.55)\times 10^{-1}$  & - & - & - & Ma21 \\
J0510$-$6708 & 05:10:11.4 & $-$67:08:04 & 2.0    & $(2.10 \pm 0.88)\times 10^{-2}$  & $0.23 \pm 0.43$ & - & - &  K22  \\
J0512$-$6716 & 05:12:24.7 & $-$67:16:55 & $7.8\times7.4 \ (45)$  & $(1.40  \pm 0.19)\times 10^{-1}$ & $-0.41 \pm 0.10$ & $-0.80 \pm 0.30$ & - &   K22 \\
J0513$-$6724 & 05:13:43.0 & $-$67:24:10 & $2.6\times2.0 \ (0)$  & $(5.00  \pm 0.59)\times 10^{-2}$  & $-0.05 \pm 0.13$ & $-0.23 \pm 0.14$ & $-0.67 \pm  0.29$ &  Ma19  \\
J0522$-$6543 & 05:22:53.5 & $-$65:43:09 & $5.8\times5.2$ (26)  &  $< 10^{-2}$  & - & - & - &   B23  \\
J0522$-$6740 & 05:22:33.7 & $-$67:41:04 & $3.2\times3.0$ (90) & $(5.30  \pm 1.19)\times 10^{-2}$ & $0.38 \pm 0.21$ & - & - &   Y21  \\
J0527$-$7134 & 05:27:49.9 & $-$71:34:08 & $3.8\times3.6 \ (45)$  & $(6.90   \pm 0.92)\times 10^{-2}$ & $-0.27 \pm 0.11$ & $-0.74 \pm 0.25$ & - &   K22  \\
J0529$-$7004 & 05:29:11.4 & $-$70:04:40 & 1.6   & $(2.38 \pm 0.26)\times 10^{-1}$ & $0.00 \pm 0.17$ & $-0.76 \pm 0.21$ & $-0.23 \pm 0.87$ &   Sa22 \\
J0542$-$7104 & 05:42:42.0 & $-$71:04:29 & 1.7   & $(5.70 \pm 0.75)\times 10^{-2}$ & $0.74 \pm 0.13$ & $-0.71 \pm 0.11$ & - &   Y21 
\\\bottomrule 
\end{longtable}

\begin{longtable}{@{}  p{.11\textwidth}@{} p{.095\textwidth}@{} p{.105\textwidth}@{} p{.15\textwidth}@{} p{.15\textwidth}@{} wr{.11\textwidth}@{} wr{.11\textwidth}@{} wr{.11\textwidth}@{} wr{.06\textwidth}@{} @{}}
    \caption{Previous candidates which remain candidates (Appendix \ref{canNotConf} and Fig. \ref{CandiRemainCandi_images}). X-ray SNR candidates in the LMC proposed in previous works which do not have a 3$\sigma$ eROSITA counterpart.}\\[-0.2cm]\toprule\midrule
    ID & RA [J2000] & DEC [J2000] & Size [\arcmin] (PA [$\deg$]) & Rate [counts $\mathrm{s}^{-1}$] & HR1 & HR2 & HR3 &  Ref  \\\midrule
     \endfirsthead
    \caption{\textbf{continued.} X-ray SNR candidates in the LMC proposed in previous works do not have a 3$\sigma$ eROSITA counterpart.}\\[-0.2cm]\toprule\midrule
    ID & RA [J2000] & DEC [J2000] & Size [\arcmin] (PA [$\deg$]) & Rate [counts $\mathrm{s}^{-1}$] & \multicolumn{1}{l}{HR1} & \multicolumn{1}{l}{HR2} & \multicolumn{1}{l}{HR3} &  Ref  \\\midrule
    \endhead
    \midrule
    \endfoot
    \parbox{\textwidth}{\tablebib{HP99:  \citet{Haberl99};
    B17: \citet{Bozzetto_2017}; B23: \citet{Filipovich23}; F22: \citet{2022MNRAS.512..265F} Y21: \citet{Yew21}.}}
    \parbox{\textwidth}{\tablefoot{\tablefoottext{a}{This source is an X-ray candidate proposed by \citet{Haberl99} with no confirmation at other wavelengths, therefore the eROSITA 3$\sigma$ detection cannot be used to confirm the SNR nature of this source.}}}
    \endlastfoot
J0444$-$6758 & 04:44:27.8 & $-$67:58:13 & $2.6\times1.8 \ (80)$ & $< 10^{-2}$  & - & - & - & Y21 \\
J0450$-$6818 & 04:50:12.4 & $-$68:18:05 & $7.4\times6.6 \ (105)$ & $< 10^{-2}$  &- & - & - & Y21 \\
J0451$-$6906 & 04:51:38.9 & $-$69:06:26 & $10.0\times6.4 \ (345)$ & $< 10^{-2}$  & - & - & - & B23 \\
J0451$-$6951 & 04:51:52.7 & $-$69:51:41 & 2.8   & $< 10^{-2}$   & - & - & - & B23 \\
J0452$-$6638 & 04:52:42.2 & $-$66:38:43 & $4.6\times6.6 \ (0)$   & $(1.50 \pm 1.83)\times 10^{-2}$  & $-0.13 \pm	0.33$ & - & - & B23 \\
J0455$-$6830 & 04:55:36.8 & $-$68:30:35 & 1.3 & $< 10^{-2}$ & - & - & - & Y21 \\
J0457$-$6739 & 04:57:33.0 & $-$67:39:05 & 2.5   & $< 10^{-2}$ & - & - & - & B17 \\
J0457$-$6823 & 04:57:33.6 & $-$68:23:39 & $6.6\times3.8 \ (330)$ & $(2.50 \pm 1.28)\times 10^{-2}$  & $-0.69 \pm 0.37$ &- & - & B23 \\
J0457$-$6923 & 04:57:07.8 & $-$69:23:58 & $4.0\times3.4 \ (90)$  & $< 10^{-2}$ & - & - & - & B17 \\
J0459$-$6757 & 04:59:55.0 & $-$67:57:01 & $4.4\times3.8 \ (357)$ & $(1.20 \pm 0.98)\times 10^{-2}$  & $0.42 \pm 1.69$ & $0.27 \pm 0.63$ & - & B23 \\
J0459$-$7008b & 04:59:38.7 & $-$70:08:37 & $3.8\times3.2 \ (150)$ & $(1.73 \pm 0.14)\times 10^{-1}$ & $-0.32 \pm 0.07$ & $-0.80 \pm 0.12$ & - & B23 \\
J0500$-$6512 & 05:00:53.2 & $-$65:11:46 & $6.2\times3.8 \ (55)$  & $(7.10 \pm 1.30)\times 10^{-2}$  & $0.53 \pm 0.20$ & - & - & Y21 \\
J0502$-$6739 & 05:02:02.5 & $-$67:39:31 & $6.4\times5.6 \ (60)$  &$< 10^{-2}$   & - & - & - & Y21 \\
J0504$-$6901 & 05:04:04.8 & $-$69:01:12 & $8.6\times8.2 \ (22) $ & $(4.70 \pm 2.48)\times 10^{-2}$   & $-0.66 \pm 0.33$ & - & - & B23 \\
J0506$-$6509 & 05:06:49.1 & $-$65:09:19 & $6.4\times6.2 \ (170)$ & $< 10^{-2}$  & - & - & - & Y21 \\
J0507$-$7110 & 05:07:35.3 & $-$71:10:15 & $7.4\times6.0 \ (0)$   & $< 10^{-2}$   & - & - & - & B17 \\
J0508$-$6928 & 05:08:46.5 & $-$69:28:16 & 5.5   & $< 10^{-2}$  & - & - & - & Y21 \\
J0509$-$6402 & 05:09:15.5 & $-$64:02:07 & $5.4\times4.0 \ (50)$  & $(3.10 \pm 0.76)\times 10^{-2}$  & $-0.10 \pm 0.08$ & $-0.20 \pm 0.10$ & $-0.13 \pm 0.14$ & Y21 \\
J0513$-$6731 & 05:13:26.9 & $-$67:31:53 & $5.0\times3.6 \ (60)$ & $(4.40 \pm 0.96)\times 10^{-2}$  & $-0.61 \pm 0.17$ & - & - & B17 \\
J0517$-$6757 & 05:17:53.6 & $-$67:57:25 & 1.4 & $< 10^{-2}$  & - & - & - & Y21 \\
J0528$-$7018 & 05:28:46.0 & $-$70:17:57 & $9.2\times8.8 \ (140)$ & $(1.10 \pm 3.31)\times 10^{-2}$ & - & - & - & Y21 \\
J0534$-$6700 & 05:34:42.4 & $-$66:59:55 & $3.0\times2.6 \ (0)$   & $(1.70 \pm 0.44)\times 10^{-2}$ & $-0.73 \pm 0.26$ & $0.11 \pm	0.63$ & - & B23 \\
J0534$-$6720 & 05:34:04.9 & $-$67:20:51 & 2.7   & $(9.90 \pm 1.27)\times 10^{-2}$  & $-0.66 \pm 0.10$ & $-0.72 \pm 0.37$ & $-0.14 \pm 1.30$ & B23 \\
J0538$-$6921 & 05:38:14.7 & $-$69:21:24 & $5.6\times5.4 \ (0)$   & $(3.83 \pm 0.24)\times 10^{-1}$  & $0.24 \pm 0.06$ & $-0.57 \pm 0.07$ & $-0.70\pm 0.20$ & B17 \\
J0538$-$7004 & 05:38:44.9 & $-$70:04:24 & 1.3 & $< 10^{-2}$ & - & - & - & Y21 \\
J0539$-$7001\tablefootmark{a} & 05:39:35.5 & $-$70:01:52 & $6.0\times2.4 \ (135)$ & $(3.08 \pm 0.14)\times 10^{-1}$   & $0.71\pm0.04$ & $-0.88\pm0.04$ & $-0.82\pm0.47$  &  HP99       \\
J0542$-$6852 & 05:41:59.3 & $-$68:52:01 & $6.0\times5.8 \ (343)$ & $(7.50 \pm 1.58)\times 10^{-2}$  & $-0.02 \pm 0.12$ & - & - & B23 \\
J0543$-$6906 & 05:43:27.0 & $-$69:07:21 & 7.3  & $(3.97 \pm 0.25)\times 10^{-1}$ & $0.23 \pm 0.06$ & $-0.53 \pm 0.07$ & - & B23  \\
J0543$-$6923 & 05:43:16.5 & $-$69:23:27 & $9.6\times7.0 \ (0)$   & $< 10^{-2}$  & - & - & - & B23 \\
J0543$-$6928 & 05:43:06.3 & $-$69:28:42 & $4.8\times3.0 \ (35)$  & $(3.60 \pm 1.18)\times 10^{-2}$  & $-0.38 \pm 0.46$& $0.14 \pm 0.56$ & - & B23 \\
J0548$-$6941 & 05:48:49.2 & $-$69:41:22 & $5.2\times3.2 \ (150)$ & $< 10^{-2}$  & - & - & - & Y21 \\
J0549$-$6618 & 05:49:30.4 & $-$66:17:37 & $2.2\times2.0 \ (45)$  & $< 10^{-2}$  & - & - & - & Y21   \\ 
J0549$-$6633 & 05:49:25.6 & $-$66:33:46 & $5.8\times4.2 \ (45)$  & $< 10^{-2}$  &- & - & - & Y21      \\
J0624$-$6948 & 06:24:13.5 & $-$69:48:31 &  4.5 & $(3.70 \pm 0.73)\times 10^{-2}$  & $-0.07 \pm 0.22$ & $0.13\pm0.18$ &- & F22  \\\bottomrule  
\label{Tab:PreviousCandRemainCand}
\end{longtable}

\FloatBarrier
\section{MCSNR and candidate Images}
\label{appendixImages}
\begin{subappendices}
\label{appendixImaPreviousCandidateROSAT}
\def\names{{J0454-7003}/{MCSNR J0454--7003}}
\begin{figure*}[h]
        \centering
        \foreach \i/\j in \names{
            \centering
                 \includegraphics[width=0.49\textwidth]{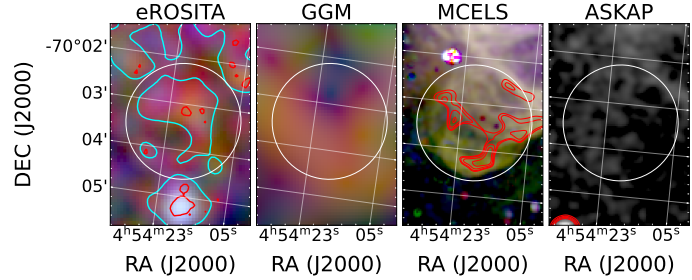}
                 \label{\i}
         }
    \caption{\textbf{Previous MCELS candidate confirmed with eROSITA (Sect. \ref{previousCandConf} and Table \ref{Tab:PreviousCandConfi}).} \textbf{MCSNR J0454--7003}. From left to right we show the eROSITA count rate three-colour image where red: $0.2\mbox{--}\SI{0.7}{\keV}$, green: $0.7\mbox{--}\SI{1.1}{\keV}$, and blue: $1.1\mbox{--}\SI{5.0}{\keV}$. The white circle shows the extraction region for determining the count rates (see Sect.\,\ref{section:Sizes}). The cyan (red) contours show the detection at $1\sigma \ (3\sigma)$ over the background in the energy band $0.2\mbox{--}\SI{1.1}{\keV}$. In the second image, the GGM filter has been applied to the eROSITA count rate images (GGM filter described in Sect.\,\ref{section:GGM}). The third image shows the MCELS survey three-colour image where red: H$\alpha$, green: [\ion{S}{II}], and blue: [\ion{O}{III}]. The contours in the optical image represent [\ion{S}{II}]/H$\alpha > 0.67$. The last image shows the ASKAP radio continuum. The contours in the radio image show the non-thermal emission calculated as described in Sect.\,\ref{radioSection}.}
    \label{J0454-7003}
\end{figure*}
\FloatBarrier
\label{appendixImaEROSITAcandidate}
\def\namesOne{{J0451-6717}/{4eRASSU J045145.7--671724},{J0456-6830}/{4eRASSU J045625.5--683052},{J0507-7143}/{4eRASSU J050750.8$-$714241},{J0510-6853}/{4eRASSU J051028.3--685329},{J0521-6707}/{4eRASSU J052136.6--670741},{J0521-6853}/{4eRASSU J052126.5--685245}}

\def\namesTwo{{J0521-6936}/{4eRASSU J052148.7--693649}/{6},{J0523-6804}/{4eRASSU J052330.7--680400}/{7},{J0525-6621}/{4eRASSU J052502.7--662125}/{8},{J0528-6719}/{4eRASSU J052849.7--671913}/{9},{J0532-6554}/{4eRASSU J053224.5--655411}/{10},{J0549-7001}/{4eRASSU J054949.7--700145}/{11}}
\def\namesThree{}
\begin{figure*}[h]
        \centering
        \foreach \i/\j in \namesOne{
            \begin{subfigure}{0.49\textwidth} 
            \centering
                 \includegraphics[width=0.97\textwidth]{Images/Subplots/Ima\i.pdf}
                 \includegraphics[width=0.7\textwidth]{Images/SFH/SFH_\i_100pc.pdf}
                 \caption{\textbf{\j}}
                 \label{\i}
        \end{subfigure}
         }
    \caption{ \textbf{SNR candidates detected with eROSITA (Sect. \ref{eROSITACandidate} and Table \ref{Tab:eROSITACandidate}).} For each source we show the eROSITA count rate three-colour image (left) with red: $0.2\mbox{--}\SI{0.7}{\keV}$, green: $0.7\mbox{--}\SI{1.1}{\keV}$, and blue: $1.1\mbox{--}\SI{5.0}{\keV}$,  the three-colour image with the GGM filter  (see Sect.\,\ref{section:GGM}) applied to the eROSITA count rate images (middle left),  the MCELS survey three-colour image (middle right) with red: H$\alpha$, green: [\ion{S}{II}], and blue: [\ion{O}{III}], and the ASKAP radio continuum (right) in the upper panel. The white circles (ellipses) show the extraction region for determining the count rates (see Sect.\,\ref{section:Sizes}). The cyan (red) contours in the eROSITA three-colour image show the detection at $1\sigma \ (3\sigma)$ over the background in the energy band $0.2\mbox{--}\SI{1.1}{\keV}$. The contours in the optical image represent [\ion{S}{II}]/H$\alpha > 0.67$. The contours in the radio image show the non-thermal emission calculated as described in Sect.\,\ref{radioSection}. In the lower panel we show the SFH as measured in \cite{SFHpaper} using $J-K_\text{s}$ and $Y-K_\text{s}$ (see Sect.\,\ref{section:SFH}).}
    \label{eROSITACandi_images}    
\end{figure*}

\setcounter{figure}{1}

\begin{figure*}[h]
        \centering
        \foreach \i/\j/\letters in \namesTwo{
            \begin{subfigure}{0.49\textwidth}
            \setcounter{subfigure}{\letters}
            \centering
                 \includegraphics[width=0.97\textwidth]{Images/Subplots/Ima\i.pdf}
                 \includegraphics[width=0.7\textwidth]{Images/SFH/SFH_\i_100pc.pdf}
                 \caption{\textbf{\j}}
                 \label{\i}
        \end{subfigure}
         }
    \caption{\textbf{continued. SNR candidates detected with eROSITA (Sect. \ref{eROSITACandidate} and Table \ref{Tab:eROSITACandidate}).} For each source we show the eROSITA count rate three-colour image (left) with red: $0.2\mbox{--}\SI{0.7}{\keV}$, green: $0.7\mbox{--}\SI{1.1}{\keV}$, and blue: $1.1\mbox{--}\SI{5.0}{\keV}$, the three-colour image with the GGM filter  (see Sect.\,\ref{section:GGM}) applied to the eROSITA count rate images  (middle left),  the MCELS survey three-colour image (middle right) with red: H$\alpha$, green: [\ion{S}{II}], and blue: [\ion{O}{III}], and the ASKAP radio continuum (right) in the upper panel. The white circles (ellipses) show the extraction region for determining the count rates (see Sect.\,\ref{section:Sizes}). The cyan (red) contours in the eROSITA three-colour image show the detection at $1\sigma \ (3\sigma)$ over the background in the energy band $0.2\mbox{--}\SI{1.1}{\keV}$. The contours in the optical image represent [\ion{S}{II}]/H$\alpha > 0.67$. The contours in the radio image show the non-thermal emission calculated as described in Sect.\,\ref{radioSection}. In the lower panel we show the SFH as measured in \cite{SFHpaper} using $J-K_\text{s}$ and $Y-K_\text{s}$ (see Sect.\,\ref{section:SFH}).}
\end{figure*}
\FloatBarrier
\def\namesOne{{J0447-6918}/{MCSNR J0447-6918},{J0449-6903}/{MCSNR J0449-6903},{J0456-6950}/{MCSNR J0456-6950},{J0504-6723}/{MCSNR J0504-6723},{J0506-6815}/{MCSNR J0506-6815},{J0507-6847}/{MCSNR J0507-6847},{J0510-6708}/{MCSNR J0510-6708},{J0512-6716}/{MCSNR J0512-6716},{J0513-6724}/{MCSNR J0513-6724},{J0522-6543}/{MCSNR J0522-6543}}

\def\namesTwo{{J0522-6740}/{MCSNR J0522-6740}/{10},{J0527-7134}/{MCSNR J0527-7134}/{11},{J0529-7004}/{MCSNR J0529-7004}/{12},{J0542-7104}/{MCSNR J0542-7104}/{13}}
\begin{figure*}[h]
        \centering
        \foreach \i/\j in \namesOne{
            \begin{subfigure}{0.49\textwidth}
            \centering
                 \includegraphics[width=\textwidth]{Images/Subplots/Ima\i.pdf}
                 \caption{\textbf{\j}}
                 \label{\i}
        \end{subfigure}
         }
    \caption{ \textbf{Known SNRs not included in \citet{Maggi16}, but observed with eROSITA (Sect. \ref{KnownSNRnotMaggi} and Table \ref{Tab:SNRNotInMaggi}).} For each source we show the eROSITA count rate three-colour image (left) with red: $0.2\mbox{--}\SI{0.7}{\keV}$, green: $0.7\mbox{--}\SI{1.1}{\keV}$, and blue: $1.1\mbox{--}\SI{5.0}{\keV}$, the three-colour image with the GGM filter  (see Sect.\,\ref{section:GGM}) applied to the eROSITA count rate images  (middle left),  the MCELS survey three-colour image (middle right) with red: H$\alpha$, green: [\ion{S}{II}], and blue: [\ion{O}{III}], and the ASKAP radio continuum (right). The white circles (ellipses) show the extraction region for determining the count rates (see Sect.\,\ref{section:Sizes}). The cyan (red) contours in the eROSITA three-colour image show the detection at $1\sigma \ (3\sigma)$ over the background in the energy band $0.2\mbox{--}\SI{1.1}{\keV}$. The contours in the optical image represent [\ion{S}{II}]/H$\alpha > 0.67$. The contours in the radio image show the non-thermal emission calculated as described in Sect.\,\ref{radioSection}.}
    \label{appendix:SNRNotInMaggi} 
\end{figure*}

\setcounter{figure}{2}
\begin{figure*}[h]
        \centering
        \foreach \i/\j/\letters in \namesTwo{
            \begin{subfigure}{0.49\textwidth}
           \setcounter{subfigure}{\letters}
            \centering
                 \includegraphics[width=\textwidth]{Images/Subplots/Ima\i.pdf}
                 \caption{\textbf{\j}}
                 \label{\i}
        \end{subfigure}
         }
    \caption{\textbf{continued. Known SNRs not included in \citet{Maggi16}, but observed with eROSITA (Sect. \ref{KnownSNRnotMaggi} and Table \ref{Tab:SNRNotInMaggi}).} For each source we show the eROSITA count rate three-colour image (left) with red: $0.2\mbox{--}\SI{0.7}{\keV}$, green: $0.7\mbox{--}\SI{1.1}{\keV}$, and blue: $1.1\mbox{--}\SI{5.0}{\keV}$, three-colour image with the GGM filter  (see Sect.\,\ref{section:GGM}) applied to the eROSITA count rate images (middle left),  the MCELS survey three-colour image (middle right) with red: H$\alpha$, green: [\ion{S}{II}], and blue: [\ion{O}{III}], and the ASKAP radio continuum (right) in the upper panel. The white circles (ellipses) show the extraction region for determining the count rates (see Sect.\,\ref{section:Sizes}). The cyan (red) contours in the eROSITA three-colour image show the detection at $1\sigma \ (3\sigma)$ over the background in the energy band $0.2\mbox{--}\SI{1.1}{\keV}$. The contours in the optical image represent [\ion{S}{II}]/H$\alpha > 0.67$. The contours in the radio image show the non-thermal emission calculated as described in Sect.\,\ref{radioSection}.}
      
\end{figure*}

\label{appendix:CandRemainCand}
\renewcommand\thesubfigure{\alphalph{\value{subfigure}}}
%
\def\namesZero{{J0444-6758},{J0450-6818},{J0451-6906},{J0451-6951}}
\def\namesOne{{J0452-6638}/{4},{J0455-6830}/{5},{J0457-6739}/{6},{J0457-6823}/{7},{J0457-6923}/{8},{J0459-6757}/{9},{J0459-7008b}/{10},{J0500-6512}/{11},{J0502-6739}/{12},{J0504-6901}/{13}}

\def\namesTwo{{J0506-6509}/{14},{J0507-7110}/{15},{J0508-6928}/{16},{J0509-6402}/{17},{J0513-6731}/{18},{J0517-6757}/{19},{J0528-7018}/{20},{J0534-6700}/{21},{J0534-6720}/{22},{J0538-6921}/{23}}
\def\namesThree{{J0538-7004}/{24},{J0539-7001}/{25},{J0542-6852}/{26},{J0543-6906}/{27},{J0543-6923}/{28},{J0543-6928}/{29},{J0548-6941}/{30},{J0549-6618}/{31},{J0549-6633}/{32},{J0624-6948}/{33}}
\def\namesFour{}

\begin{figure*}[h]
        \centering
        \foreach \i in \namesZero{
            \begin{subfigure}{0.49\textwidth}
            \centering
                 \includegraphics[width=\textwidth]{Images/Subplots/Ima\i.pdf}
                 \caption{\textbf{\i}}
                 \label{\i}
        \end{subfigure}
         }
    \caption{\textbf{Previous candidates which remain candidates (Sect. \ref{canNotConf} and Table \ref{Tab:PreviousCandRemainCand}).} For each source we show the eROSITA count rate three-colour image (left) with red: $0.2\mbox{--}\SI{0.7}{\keV}$, green: $0.7\mbox{--}\SI{1.1}{\keV}$, and blue: $1.1\mbox{--}\SI{5.0}{\keV}$, the three-colour image with the GGM filter  (see Sect.\,\ref{section:GGM}) applied to the eROSITA count rate images (middle left),  the MCELS survey three-colour image (middle right) with red: H$\alpha$, green: [\ion{S}{II}], and blue: [\ion{O}{III}], and the ASKAP radio continuum (right). The white circles (ellipses) show the extraction region for determining the count rates (see Sect.\,\ref{section:Sizes}). The cyan (red) contours in the eROSITA three-colour image show the detection at $1\sigma \ (3\sigma)$ over the background in the energy band $0.2\mbox{--}\SI{1.1}{\keV}$. The contours in the optical image represent [\ion{S}{II}]/H$\alpha > 0.67$. The contours in the radio image show the non-thermal emission calculated as described in Sect.\,\ref{radioSection}.}    
    \label{CandiRemainCandi_images}
    \end{figure*}
\setcounter{figure}{3}
\begin{figure*}[h]
        \centering
        \foreach \i/\letters in \namesOne{
            \begin{subfigure}{0.49\textwidth}
            \setcounter{subfigure}{\letters}
                       \centering
                 \includegraphics[width=\textwidth]{Images/Subplots/Ima\i.pdf}
                 \caption{\textbf{\i}}
                 \label{\i}
        \end{subfigure}
         }
    \caption{\textbf{continued. Previous candidates which remain candidates (Sect. \ref{canNotConf} and Table \ref{Tab:PreviousCandRemainCand}).} For each source we show the eROSITA count rate three-colour image (left) with red: $0.2\mbox{--}\SI{0.7}{\keV}$, green: $0.7\mbox{--}\SI{1.1}{\keV}$, and blue: $1.1\mbox{--}\SI{5.0}{\keV}$, the three-colour image with the GGM filter  (see Sect.\,\ref{section:GGM}) applied to the eROSITA count rate images (middle left),  the MCELS survey three-colour image (middle right) with red: H$\alpha$, green: [\ion{S}{II}], and blue: [\ion{O}{III}], and the ASKAP radio continuum (right). The white circles (ellipses) show the extraction region for determining the count rates (see Sect.\,\ref{section:Sizes}). The cyan (red) contours in the eROSITA three-colour image show the detection at $1\sigma \ (3\sigma)$ over the background in the energy band $0.2\mbox{--}\SI{1.1}{\keV}$. The contours in the optical image represent [\ion{S}{II}]/H$\alpha > 0.67$. The contours in the radio image show the non-thermal emission calculated as described in Sect.\,\ref{radioSection}.}    
    \label{CandiRemainCandi_images}
    \end{figure*}

\setcounter{figure}{3}
\begin{figure*}[h]
        \centering
        \foreach \i/\letters in \namesTwo{
            \begin{subfigure}{0.49\textwidth}
           \setcounter{subfigure}{\letters}
            \centering
                 \includegraphics[width=\textwidth]{Images/Subplots/Ima\i.pdf}
                 \caption{\textbf{\i}}
                 \label{\i}
        \end{subfigure}
         }
    \caption{\textbf{continued. Previous candidates which remain candidates (Sect. \ref{canNotConf} and Table \ref{Tab:PreviousCandRemainCand}).} For each source we show the eROSITA count rate three-colour image (left) with red: $0.2\mbox{--}\SI{0.7}{\keV}$, green: $0.7\mbox{--}\SI{1.1}{\keV}$, and blue: $1.1\mbox{--}\SI{5.0}{\keV}$, the three-colour image with the GGM filter  (see Sect.\,\ref{section:GGM}) applied to the eROSITA count rate images  (middle left),  the MCELS survey three-colour image (middle right) with red: H$\alpha$, green: [\ion{S}{II}], and blue: [\ion{O}{III}], and the ASKAP radio continuum (right). The white circles (ellipses) show the extraction region for determining the count rates (see Sect.\,\ref{section:Sizes}). The cyan (red) contours in the eROSITA three-colour image show the detection at $1\sigma \ (3\sigma)$ over the background in the energy band $0.2\mbox{--}\SI{1.1}{\keV}$. The contours in the optical image represent [\ion{S}{II}]/H$\alpha > 0.67$. The contours in the radio image show the non-thermal emission calculated as described in Sect.\,\ref{radioSection}.}
      
\end{figure*}
\setcounter{figure}{3}

\begin{figure*}[h]
        \centering
        \foreach \i/\letters in \namesThree{
            \begin{subfigure}{0.49\textwidth}
           \setcounter{subfigure}{\letters}
            \centering
                 \includegraphics[width=\textwidth]{Images/Subplots/Ima\i.pdf}
                 \caption{\textbf{\i}}
                 \label{\i}
        \end{subfigure}
         }
    \caption{\textbf{continued. Previous candidates which remain candidates (Sect. \ref{canNotConf} and Table \ref{Tab:PreviousCandRemainCand}).} For each source we show the eROSITA count rate three-colour image (left) with red: $0.2\mbox{--}\SI{0.7}{\keV}$, green: $0.7\mbox{--}\SI{1.1}{\keV}$, and blue: $1.1\mbox{--}\SI{5.0}{\keV}$, the three-colour image with the GGM filter  (see Sect.\,\ref{section:GGM}) applied to the eROSITA count rate images  (middle left),  the MCELS survey three-colour image (middle right) with red: H$\alpha$, green: [\ion{S}{II}], and blue: [\ion{O}{III}], and the ASKAP radio continuum (right). The white circles (ellipses) show the extraction region for determining the count rates (see Sect.\,\ref{section:Sizes}). The cyan (red) contours in the eROSITA three-colour image show the detection at $1\sigma \ (3\sigma)$ over the background in the energy band $0.2\mbox{--}\SI{1.1}{\keV}$. The contours in the optical image represent [\ion{S}{II}]/H$\alpha > 0.67$. The contours in the radio image show the non-thermal emission calculated as described in Sect.\,\ref{radioSection}. For J0624--6948 no MCELS images are available.}
      
\end{figure*}

\end{subappendices}
\end{appendix}
\end{document}